\documentclass{aa}
\usepackage[utf8]{inputenc}
\usepackage{natbib}
\usepackage{graphicx}
\usepackage{amsmath}
\usepackage{gensymb}
\usepackage{subcaption}
\usepackage{dblfloatfix}
\usepackage{makecell}
\usepackage{ragged2e}
\usepackage[colorlinks=true,allcolors=blue]{hyperref}
\usepackage[scientific-notation=true]{siunitx}
\usepackage{xcolor}
\usepackage{mhchem}
\usepackage{longtable}
\usepackage{lscape}
\usepackage{tablefootnote}
\usepackage{float}
\usepackage{tabularx}

\newcommand{\Msun}{M$_\odot$}
\newcommand{\Mearth}{M$_\oplus$}

\title{An analysis of the Herbig star population and their protoplanetary disks within 1~kpc}

\author{L. M. Stapper\inst{\ref{inst1}} \and M. Vioque \inst{\ref{inst2}} \and A. J. Winter\inst{\ref{inst3}} \and D. J. Wilner\inst{\ref{inst4}} \and J. P. Williams\inst{\ref{inst5}} \and M. Benisty\inst{\ref{inst1}} \and A. S. Booth\inst{\ref{inst4}} \and S.~L.~Grant\inst{\ref{inst6}} \and M. R. Hogerheijde\inst{\ref{inst7},\ref{inst8}}}

\institute{Max-Planck-Institut für Astronomie, Königstuhl 17, 69117 Heidelberg, Germany \\e-mail: \texttt{lustapper@mpia.de} \label{inst1} \and European Southern Observatory, Karl-Schwarzschild-Strasse 2, 85748 Garching bei München, Germany \label{inst2} \and Astronomy Unit, School of Physics and Astronomy, Queen Mary University of London, London E1 4NS, UK \label{inst3} \and Center for Astrophysics  Harvard \& Smithsonian, Cambridge, MA 02138, USA \label{inst4} \and Institute for Astronomy, University of Hawaii, Honolulu, HI 96822, USA \label{inst5} \and Earth and Planets Laboratory, Carnegie Institution for Science, 5241 Broad Branch Road, NW, Washington, DC 20015, USA \label{inst6} \and Leiden Observatory, Leiden University, 2300 RA Leiden, The Netherlands \label{inst7} \and Anton Pannekoek Institute for Astronomy, University of Amsterdam, The Netherlands \label{inst8}}

\date{\today}

\abstract
{ 
Herbig stars are intermediate mass pre-main sequence stars of 1.5 to $\sim15$~\Msun~and are in between low and high-mass star formation. Despite the giant exoplanet occurrence rate peaking at intermediate mass stars, a systematic analysis of the complete known Herbig star and protoplanetary disk population is lacking.
}{
This work provides a catalog of all 243 known bona-fide Herbig stars with clear infrared excess and in many cases accretion signatures within 1~kpc. It contains archival, recalibrated, and newly derived stellar parameters, accretion rate properties, and disk dust mass measurements. With these, we aim to examine the Herbig star relationship of the accretion rate against stellar and disk dust mass, get an estimate for the completeness of our catalog, and obtain an estimate for the Herbig star lifetime. These relationships and disk dust masses are put into the context of the T~Tauri stars and the giant exoplanet population.
}{
We revise the stellar masses, radii, and ages using a Hertzsprung-Russell diagram, and compute new accretion rates for part of our sample. We derive dust masses for all targets using a combination of newly obtained millimeter photometry (60\% of the targets) and literature values, from a combination of ALMA, ACA, SMA, and NOEMA data.
}{
We find that 50\% of Herbig dust disks are more massive than 10~\Mearth, while this is only true for 20\% and 5\% of the T~Tauri disks in Lupus and Upper~Scorpius respectively. Furthermore, the Herbig disk dust mass distribution is bimodal, it consists of a high and low disk mass population with a mean dust mass of 21~\Mearth~and 0.45~\Mearth~respectively. 
We find that the catalog is near complete for within 300~pc, however, this decreases to 24\% out to 1~kpc. Hence, a large fraction of stars which meet our selection criteria are still missing. We also find a decreasing Herbig disk lifetime with increasing stellar mass of the form $t=12 \times M_\star^{-1.4}$~Myr.
The bulk of the Herbig disks are above the T~Tauri accretion rate -- disk mass relationship. We observe a flat relation for Herbig disks due to objects with short disk lifetimes, caused by a combination of low disk masses and high accretion rates.
Lastly, we find the peak of the occurrence rate of massive (>10~\Mearth) disks to occur at $\sim$1-2~\Msun~stars, coinciding with the peak in occurrence rate of giant exoplanets. Furthermore, we find an increase of disk dust mass with stellar mass up to 1-2~\Msun, and then a decrease with stellar mass, likely related to an increase in UV irradiation, multiplicity, and/or radial drift efficiency.
}{
Our catalog provides stellar parameters, accretion properties, and disk masses for 243 Herbig stars within 1~kpc. This catalog describes the properties of intermediate mass star formation. We find a direct link to the occurrence rate of massive disks with that of giant exoplanets. The small number of Herbig stars with low accretion rates ($<10^{-8}$~\Msun~yr$^{-1}$) or low disk dust masses ($\lesssim1$~\Mearth), combined with the lack of an age dependence in the disk dust mass distribution, suggests that the observed Herbig star population represents the surviving disk-hosting phase in the optically visible late stages of intermediate-mass star formation. We argue that this is either due to the formation of deep dust traps in these disks or replenishment by late-infall, or a combination of the two.
}


\begin{document}

\maketitle

\section{Introduction}
\label{sec:introduction}
Herbig stars are pre-main sequence stars with intermediate masses of 1.5 to $\sim15$~\Msun \citep[e.g.,][]{Herbig1960, Waters1998, Brittain2023}, and are in between low and high-mass star formation. Spectral types range from A/B-type, the well-known Herbig Ae/Be stars, to G-type stars, which are also known as Intermediate Mass T~Tauri stars and are the precursors of the Herbig Ae/Be stars \citep{Calvet2004, Valegard2021}. In this paper, we refer to both of these as Herbig stars. Herbig stars were selected based on the presence of accretion signatures and infrared excess (e.g., \citealp{The1994, Malfait1998, Waters1998, Vieira2003, Chen2016, Vioque2018, Brittain2023}), and generally have higher accretion rates than T~Tauri stars, with typical values of $10^{-8}$-$10^{-6}$~\Msun~yr$^{-1}$ \citep[e.g.,][]{Wichittanakom2020}. The lower mass Herbig stars follow the same $\dot{M}$-$M_\star$ correlation as T~Tauri stars, but this breaks at a stellar mass of $\sim4$~\Msun, which is thought to be related to a transition from magnetospheric accretion to a different accretion mechanism (probably a direct disk-to-star hot boundary layer, \citealp{Wichittanakom2020, Mendigutia2020, Vioque2022, Grant2022}).

Herbig stars host some of the most well-known disks, and many have been part of detailed studies in both the infrared and millimeter \citep{Bae2023, Benisty2023}. In scattered light many Herbig disks are known to host spirals \citep[e.g.,][]{Benisty2015, Ren2020, Uyama2020, Pinilla2022b, Ren2024, Columba2024}, which are more common compared to disks around lower mass stars \citep{Garufi2018, Garufi2026}. Furthermore, a dichotomy is seen in the far-infrared brightness of disks \citep{Meeus2001}. The bright disks are generally associated with having large cavities \citep{Maaskant2013, Honda2015}, while the faint disks are self-shadowed \citep{Dullemond2004a}. At millimeter wavelengths, Herbig disks have been part of many detailed studies, from programs covering dust substructures and kinematics \citep[e.g.,][]{Andrews2018b, Teague2025}, to chemistry studies \citep[e.g.,][]{Oberg2021,Booth2024, Booth2025, Booth2026, Leemker2024}. In general, the millimeter observations show Herbig disks to be large and massive, and have rich gas-phase chemistry, quite different from T~Tauri disks \citep{Oberg2015, Booth2021a}. While many of these disks have been well known for quite some time \citep{The1994}, large millimeter interferometric surveys of these disks, as is done for the disks around T~Tauri stars (see \citealt{Manara2023} for an overview), have until recently been missing.

\citet{Stapper2022} published a survey of 36 Herbig disks observed with the Atacama Large Millimeter/submillimeter Array (ALMA). From this, they found that Herbig disks are generally more massive than T~Tauri disks. In particular by a factor of 3-7 compared to the disks in Lupus \citep{Ansdell2016} and Upper~Scorpius \citep{Barenfeld2016} respectively. Furthermore, while the median dust disk mass was higher, the range in dust disk masses were found to be similar to that of T~Tauri disks. Additionally, a dichotomy in disk mass between cavity-bearing (higher dust mass) and full disks (lower dust mass) was found, which may be related to giant planets being formed in the infrared bright disks \citep{Kama2015, GuzmanDiaz2023}. This work was later expanded upon by \citet{Stapper2025b} by utilizing ALMA archival data of later spectral type Herbig stars and with a NOEMA survey of Herbig stars in Orion \citep{Stapper2025a}. Both works draw similar conclusions as \citet{Stapper2022} did: Herbig disks are on average more massive than T~Tauri disks.

The finding of more massive disks around Herbig stars may be linked to the fact that giant exoplanets are more common around intermediate mass stars \citep[peaking at $1.5-3.0$~\Msun,][]{Johnson2010, Nielsen2019, Wittenmyer2020, Fulton2021, Wolthoff2022, Squicciarini2025}. This may be either a cause or an effect: the higher dust disk mass will form such massive exoplanets, but the massive exoplanets also produce deep dust traps and stop radial drift and keep the disk large resulting in a high inferred disk mass \citep{Stapper2022, GuzmanDiaz2023}. Famous directly imaged planetary systems such as HR~8799 \citep{Marois2008, Marois2010}, $\beta$~Pic \citep{Lagrange2010}, and 51~Eri \citep{Chauvin2017} have been formed in Herbig disks, and all host a debris disk beyond the orbits of their giant planets. In addition, debris disks around intermediate mass stars are also prime sites for planetary system studies (e.g. \citealp{2017ApJ...849..123M,2021A&A...651L..11V,2025A&A...703A.235M, Marino2026}). Furthermore, hints of these giant exoplanets forming inside disks have been readily found in multiple Herbig disks: AB~Aur \citep{Currie2022}, HD~169142 \citep{Hammond2023}, HD~100546 \citep{Booth2023}, HD~97048 \citep{Pinte2018}, HD~163296 \citep{Izquierdo2022}, and HD~135344B \citep{Latour2026}. These objects are the precursors to planets whose atmospheres are currently best suited for characterization \citep{ZhangY2021}, which can be tied to their formation locations within the natal disk. Therefore, as the main formation sites of giant exoplanets, Herbig disks are a vital component in understanding the formation and migration history of exoplanets.

Yet, around 60\% of all known pre-main sequence intermediate mass stars within 1~kpc lack millimeter interferometry observations. In this paper we aim to solve this by presenting new millimeter photometry observations of all known Herbig stars within 1~kpc. In Section~\ref{sec:target_selection} we present the target selection and the newly derived stellar parameters and accretion rates. In Section~\ref{sec:data_reduction} the millimeter interferometric data is reduced. Section~\ref{sec:results} presents the results in the form of continuum images, millimeter fluxes and dust disk masses, the latter of which is compared to that of T~Tauri disks. Section~\ref{sec:completeness} models the completeness of the sample and obtains a lifetime of Herbig stars. In Sec.~\ref{sec:discussion} we discuss the results in the context of the accretion rates (\S\ref{subsec:accretion}) and the long-lived nature of Herbig disk (\S\ref{subsec:long_lived_disks}). We close the discussion by comparing the occurrence rate of massive disks with that of giant exoplanets (\S\ref{subsec:connection_to_exoplanets}). Section~\ref{sec:conclusion} summarizes our conclusions.

\begin{figure*}[b!]
    \centering
    \includegraphics[width=\textwidth]{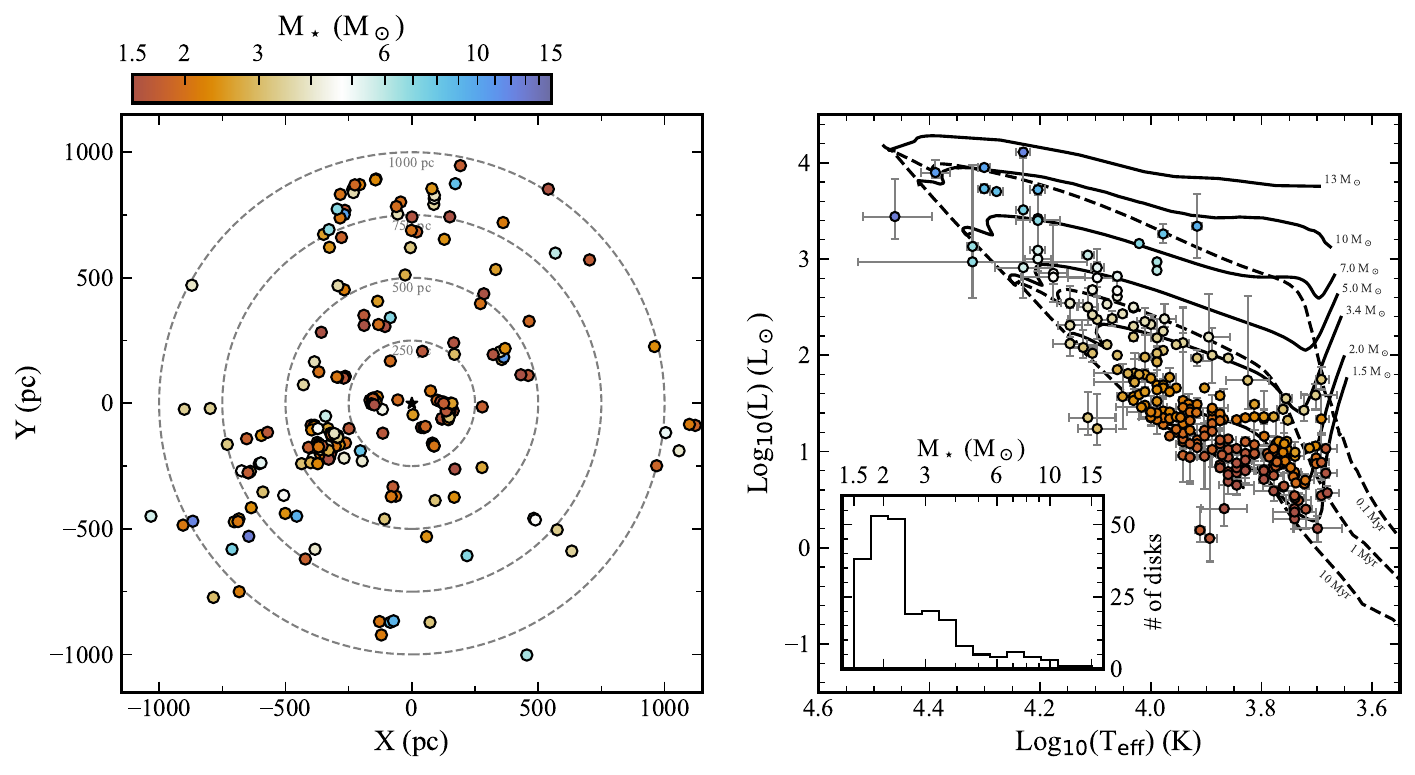}
    \caption{Positions of the 243 Herbig stars in the galactic plane ($l=0\degree$ on the right) and on the HR-diagram. The color indicates the stellar mass, and the distribution of stellar masses is shown in the inset in the right panel.}
    \label{fig:sky_and_HR_plot}
\end{figure*}

\section{Target selection, and stellar and accretion properties derivation}
\label{sec:target_selection}

We compile all spectroscopically confirmed Herbig stars in the literature with \textit{Gaia} DR3 data available and distances compatible with under 1 kpc (geometric distances of \citealp{2021AJ....161..147B}). These are those described in \citet{Vioque2018, GuzmanDiaz2021, Valegard2021, Vioque2022, Vioque2023} and references therein\footnote{See Appendix A of \citet{Vioque2020}, where some historically considered Herbig stars, which now better fall into the categories of Classical Be or FS CMa stars, are detailed.}. We consider a Herbig star to be any pre-main sequence star with a stellar mass compatible with $M_\star>1.5$ M$_{\odot}$. This definition ensures all optically visible stages of intermediate-mass pre-main sequence evolution are considered, irrespective of their spectral type or emission line properties\footnote{i.e., we are not bound to the classical ``Ae/Be'' definition}. Generally, these Herbig stars were selected to be part of these catalogs by having infrared excess and in many cases accretion signatures (often H$\alpha$ emission). In Section~\ref{sec:completeness} we discuss how complete this selection process has been. 

The compiled sample comprises 258 Herbig stars. The clustering and spatial properties of this sample were analyzed in \citet{Vioque2023}. We note that there are many other Herbig star candidates within 1 kpc that are not included in this selection (e.g. \citealp{Vioque2020, Kuhn2021, Shridharan2021, 2022ApJS..259...38Z, 2023MNRAS.524.5166N, Delfini2025,Liu2026}), as well as many hot stars in nearby star forming regions with no clear sign of disk emission or ongoing accretion (e.g. \citealp{Zari2018,Iglesias2022, Luhman2022}). Hence, we do not claim that this sample of Herbig stars is volume-complete, but it does contain most bona fide Herbig stars with disks studied in the literature.

For the 258 targets selected within 1 kpc, we compile the effective temperatures from the literature (mostly from \citealp{Vioque2018,Wichittanakom2020,GuzmanDiaz2021,Valegard2021,Vioque2022} and references therein, all references in Table~\ref{tab:herbig_stellar_params}). For reference, we convert the effective temperatures to spectral types using the relations from \citet{Pecaut2013}. For those Herbigs with luminosity determinations in \citet{GuzmanDiaz2021,Valegard2021,Vioque2022} we take the luminosities directly from those works (212 sources). For those sources only in \citet{Vioque2018} or \citet{Wichittanakom2020} we rescale their luminosities to \textit{Gaia} DR3 distances (20 sources). For the remaining 26 sources from \citet{Vioque2018} and \citet{Wichittanakom2020}, we convert the A\textsubscript{V} extinctions to A\textsubscript{RP} extinctions using the range of coefficients of \citet{2018MNRAS.479L.102C} and \citet{2023ApJS..264...14Z} and assuming a standard extinction law with $R_{V}=3.1$. We then use these A\textsubscript{RP} extinctions to de-redden the \textit{Gaia} G\textsubscript{RP} magnitude of each source. BT-NextGen (AGSS2009, \citealp{2009ARA&A..47..481A,2011ASPC..448...91A}) stellar spectra models of the right effective temperature are then scaled to the de-reddened G\textsubscript{RP} photometry of each target. Total fluxes are obtained from integrating the stellar spectra models, which are then converted to luminosities ($L=F\cdot4\pi d^2$). Uncertainties in distance, effective temperature, extinction, surface gravity, and metallicity are propagated throughout the calculations. 

\begin{table*}[t]
\caption{Stellar parameters of the (candidate) Herbig stars within 1~kpc.}
\tiny\centering
\resizebox{\textwidth}{!}{\begin{tabular}{ll|rrrcccrccc|c}
\hline\hline
\makecell{Name \\ \hspace{1mm}} & \makecell{Alt. Name \\ \hspace{1mm}} & \makecell{RA \\ (h:m:s)} & \makecell{Dec \\ (deg:m:s)} & \makecell{Dist. \\ (pc)} & \makecell{M$_\star$ \\ (M$_\odot$)} & \makecell{R$_\star$ \\ (R$_\odot$)} & \makecell{Log$_{10}$(L$_\star$) \\ (L$_\odot$)} & \makecell{T$_{\rm eff}$ \\ (K)} & \makecell{Age \\ (Myr)} & \makecell{Log$_{10}$(L$_{{acc}}$) \\ (L$_\odot$)} & \makecell{Log$_{10}$($\dot{M}$) \\ (M$_\odot$ yr$^{-1}$)} & \makecell{Ref. \\ \hspace{1mm}} \\ \hline
HBC 324 &  & 00:07:30.7 & +65:39:52.6 & $398.5^{+80.2}_{-48.5}$ & $1.64\pm0.05$ & $1.10\pm0.04$ & $0.10^{+0.85}_{-0.24}$ & $7830^{+160}_{-225}$ & $26.16\pm2.69$ & $-0.34^{+0.15}_{-0.12}$ & $-7.79^{+0.12}_{-0.11}$ & V18,W20 \\
MQ Cas &  & 00:09:37.6 & +58:13:10.7 & $758.2^{+47.2}_{-37.8}$ & $2.57\pm0.03$ & $2.53\pm0.11$ & $1.80^{+0.05}_{-0.06}$ & $10250^{+125}_{-125}$ & $3.55\pm0.11$ & $^{}_{}$ & $^{}_{}$ & GD21 \\
VX Cas &  & 00:31:30.7 & +61:58:50.9 & $525.3^{+6.4}_{-5.4}$ & $2.16\pm0.02$ & $1.70\pm0.06$ & $1.40^{+0.05}_{-0.05}$ & $10000^{+125}_{-125}$ & $18.79\pm4.46$ & $0.66^{+0.08}_{-0.07}$ & $-6.72^{+0.07}_{-0.07}$ & GD21,W20 \\
EM* GGR 195 & VOS 77 & 00:37:22.5 & +62:44:27.3 & $363.4^{+2.1}_{-2.1}$ & $1.43\pm0.13$ & $1.68\pm0.26$ & $0.20^{+0.13}_{-0.14}$ & $5000^{+500}_{-500}$ & $5.12\pm3.65$ & $^{}_{}$ & $^{}_{}$ & V22 \\
BD+61 154 & V594 Cas & 00:43:18.3 & +61:54:40.1 & $552.2^{+3.6}_{-4.3}$ & $3.60\pm0.14$ & $3.78\pm0.24$ & $2.38^{+0.08}_{-0.08}$ & $11750^{+215}_{-215}$ & $1.46\pm0.16$ & $2.07^{+0.05}_{-0.04}$ & $-5.19^{+0.09}_{-0.08}$ & GD21,W20 \\
PDS 2 & CD-53 251 & 01:17:43.5 & -52:33:30.8 & $399.0^{+3.0}_{-2.7}$ & $1.47\pm0.02$ & $1.79\pm0.05$ & $0.77^{+0.01}_{-0.01}$ & $6750^{+125}_{-125}$ & $12.54\pm0.16$ & $0.01^{+0.10}_{-0.08}$ & $-7.18^{+0.08}_{-0.08}$ & GD21,W20 \\
HD 9672 & * 49 Ceti & 01:34:37.9 & -15:40:34.9 & $57.1^{+0.2}_{-0.2}$ & $1.94\pm0.02$ & $1.64\pm0.05$ & $1.20^{+0.02}_{-0.02}$ & $9000^{+125}_{-125}$ & $21.65\pm13.99$ & $-0.26^{+0.11}_{-0.09}$ & $-7.60^{+0.09}_{-0.08}$ & GD21,W20 \\
HD 17081 & * pi. Cet & 02:44:07.3 & -13:51:31.7 & $119.1^{+3.0}_{-2.6}$ & $4.38\pm0.12$ & $4.54\pm0.21$ & $2.68^{+0.04}_{-0.05}$ & $12750^{+228}_{-228}$ & $0.87\pm0.07$ & $0.42^{+0.09}_{-0.08}$ & $-6.84^{+0.09}_{-0.08}$ & GD21,W20 \\
BX Ari A &  & 02:58:11.2 & +20:30:03.0 & $233.4^{+1.2}_{-1.1}$ & $1.42\pm0.05$ & $1.63\pm0.13$ & $0.21^{+0.07}_{-0.05}$ & $5040^{+240}_{-210}$ & $5.98\pm2.01$ & $^{}_{}$ & $^{}_{}$ & V21 (ext revised) \\
HD 19745 &  & 03:07:03.3 & -65:26:56.9 & $397.8^{+4.6}_{-5.6}$ & $3.13\pm0.40$ & $10.27\pm0.99$ & $1.75^{+0.12}_{-0.11}$ & $4920^{+110}_{-160}$ & $0.12\pm0.04$ & $^{}_{}$ & $^{}_{}$ & SIMBAD+PM2013 \\
HBC 338 &  & 03:25:49.8 & +31:10:23.7 & $284.6^{+1.8}_{-1.7}$ & $2.09\pm0.06$ & $2.68\pm0.11$ & $0.75^{+0.02}_{-0.03}$ & $5490^{+40}_{-150}$ & $2.83\pm0.54$ & $^{}_{}$ & $^{}_{}$ & V21 \\
BD+30 549 &  & 03:29:19.8 & +31:24:56.9 & $284.8^{+2.1}_{-1.8}$ & $2.36\pm0.05$ & $1.76\pm0.11$ & $1.57^{+0.11}_{-0.10}$ & $10750^{+201}_{-201}$ & $8.63\pm2.78$ & $^{}_{}$ & $^{}_{}$ & GD21 \\
TYC 4062-230-1 & VOS 1225 & 03:31:56.9 & +60:07:37.7 & $456.5^{+2.5}_{-3.1}$ & $1.47\pm0.06$ & $1.45\pm0.16$ & $0.65^{+0.13}_{-0.10}$ & $7000^{+500}_{-500}$ & $21.00\pm7.61$ & $^{}_{}$ & $^{}_{}$ & V22 \\
$\dotsb$ & $\dotsb$ & $\dotsb$ & $\dotsb$ & $\dotsb$ & $\dotsb$ & $\dotsb$ & $\dotsb$ & $\dotsb$ & $\dotsb$ & $\dotsb$ & $\dotsb$ & $\dotsb$  \\

\hline
\end{tabular}}\\
\tablefoot{Table in combination with Table~\ref{tab:herbig_fluxes} is available in its entirety on CDS\footnote{For now this can be found on \url{https://datashare.mpcdf.mpg.de/s/wqYx7D9boe5i4S3}}. The effective temperatures, luminosities, and A\textsubscript{V} extinctions are from \citet{Vioque2018} (V18), \citet{Wichittanakom2020} (W20), \citet{GuzmanDiaz2021} (GD21), \citet{Valegard2021} (V21), and \citet{Vioque2022} (V22). The distances are the Gaia DR3 geometric distances from \citet{2021AJ....161..147B}. Accretion rates are compiled from \citet{Fairlamb2015}, \citet{Wichittanakom2020}, \citet{Vioque2022}, \citet{Manara2023}, and \citet{Mendigutia2026}. Stellar masses, radii, and ages are derived in this work. Luminosities were updated to \textit{Gaia} DR3 distances when needed. A total of 26 luminosities and 23 mass accretion rates are first presented in this work.}
\label{tab:herbig_stellar_params}
\end{table*}

After obtaining effective temperatures and luminosities for the whole sample, we derive the stellar masses, radii, and ages from the Hertzsprung–Russell (HR) diagram by comparing with the PARSEC V2.0 stellar evolutionary tracks (\citealp{2022A&A...665A.126N}). This method provides accurate stellar masses (c.f. \citealp{Zallio2026}). Upon inspection, 25 sources appear in unexpected locations of the HR diagram. Nineteen are to the left of the ZAMS (and hence are not compatible with the pre-main sequence phase) and six, at the top-right of the diagram, are so massive and young it is surprising they were detected by the \textit{Gaia} passbands. We discuss these 25 sources in Appendix \ref{app:individual_sources}. Five of the sources left of the ZAMS are likely to have inaccurate \textit{Gaia} parallaxes and thus we use for them the parallax-independent distances and luminosities provided by \citet{Fairlamb2015} instead. For the massive source \object{MWC 297}, we use the effective temperature and luminosity from the detailed analysis of \citet{2022ApJ...941..189V}. Lastly, we note that for the more massive Herbig stars ($\gtrsim8$~\Msun) being to the right of the ZAMS does not necessarily mean that the star is still moving toward it, it could also be moving away from it while infall and disk-to-star accretion still occurs. This should be a topic for future work.

We remove 15 sources from the final sample because for six we cannot derive the stellar mass and more data is needed, and the remaining nine are not Herbig stars because their stellar mass is less than 1.5~\Msun~even when taking errors on their stellar mass into account. Hence, the final sample consists of 243 Herbig stars out of the initial sample of 258. Figure~\ref{fig:sky_and_HR_plot} shows the positions of the 243 Herbig stars in the galactic plane and on an HR-diagram. Table~\ref{tab:herbig_stellar_params} presents the stellar parameters for the 243 Herbig stars\footnote{\label{foot1}Including the 15 additional discarded sources}.

Accretion rates are compiled for 60 sources from the work of \citet{Fairlamb2015}, who measured accretion rates directly from UV-excess over the Balmer jump. We update \citet{Fairlamb2015} accretion rates by rescaling their accretion luminosities to \textit{Gaia} DR3 distances and re-deriving mass accretion rates assuming a disk truncation radius 2.5 times the radius of the star (\citealp{2011A&A...535A..99M}). Errors are propagated via bootstrapping. To this set we add the mass accretion rates from \citet{Manara2023}, also often derived from the Balmer jump (9 sources). \citet{Manara2023} lists no uncertainties, and thus for those 9 sources we take a generic 0.25~dex uncertainty in log($\dot{M}$). For the sources for which neither \citet{Fairlamb2015} nor \citet{Manara2023} provided accretion rate measurements, we check \citet{Wichittanakom2020} H$\alpha$ emission line measurements and use them to derive mass accretion rates for 89 more sources. We note this includes mass accretions rates for 23 sources not published in \citet{Wichittanakom2020} because \textit{Gaia} DR2 parallaxes were too uncertain at the time. To this list we add \citet{Vioque2022} mass accretion rates (19 sources, as determined from H$\alpha$ and H$\beta$), which we modify to adopt the truncation radius assumption used before to homogenize the sample. We note that \citet{Vioque2022} did not publish the accretion luminosities, so these were derived using their Eq.(2). Finally, we add mass accretion rates for the Intermediate-Mass T Tauris in our sample from \citet{Mendigutia2026}. We could not find more measurements in the literature to derive accurate accretion rates for more sources. In total, we have mass accretion rate estimates for 184 sources, 180 of which are in the final selection of 243 Herbig stars. We note that the mass accretion rates of very massive Herbig stars, in particular those over $10^{-5}$ M$_{\odot}$/yr, are to be taken with caution as no good boundary layer accretion models are currently available for them (e.g. \citealp{Mendigutia2020,2021A&A...652A..68M}). We report the accretion rate properties in Table~\ref{tab:herbig_stellar_params}.

\begin{table*}[b]
\caption{Millimeter fluxes and dust disk masses of the 243 Herbig stars within 1 kpc.}
\tiny\centering
\resizebox{\textwidth}{!}{\begin{tabular}{ll|cccc|ccc|ccc|ccc|c}
\hline\hline
\makecell{Name \\ \hspace{1mm}} & \makecell{Alt. Name \\ \hspace{1mm}} & \makecell{Flux \\ (mJy)} & \makecell{M$_{\rm dust}$ \\ (M$_\oplus$)} & \makecell{Freq. \\ (GHz)} & \makecell{Array \\ \hspace{1mm}} & \makecell{Flux$_{\rm ALMA}$ \\ (mJy)} & \makecell{M$_{\rm dust, ALMA}$ \\ (M$_\oplus$)} & \makecell{Freq. \\ (GHz)} & \makecell{Flux$_{\rm ACA}$ \\ (mJy)} & \makecell{M$_{\rm dust, ACA}$ \\ (M$_\oplus$)} & \makecell{Freq. \\ (GHz)} & \makecell{Flux$_{\rm SMA}$ \\ (mJy)} & \makecell{M$_{\rm dust, SMA}$ \\ (M$_\oplus$)} & \makecell{Freq. \\ (GHz)} & \makecell{Ref. \\ \hspace{1mm}} \\ \hline
HBC 324 &  & $2.90\pm0.68$ & $9.81\pm1.96$ & 225 & SMA &  &  &  &  &  &  & $2.9\pm0.7$ & $9.8\pm2.0$ & 225 &  \\
MQ Cas &  & $28.50\pm0.82$ & $114.53\pm22.91$ & 225 & SMA &  &  &  &  &  &  & $28.5\pm0.8$ & $114.5\pm22.9$ & 225 &  \\
VX Cas &  & $3.10\pm0.63$ & $7.68\pm1.54$ & 225 & SMA &  &  &  &  &  &  & $3.1\pm0.6$ & $7.7\pm1.5$ & 225 &  \\
EM* GGR 195 & VOS 77 & $8.70\pm0.78$ & $22.77\pm4.55$ & 225 & SMA &  &  &  &  &  &  & $8.7\pm0.8$ & $22.8\pm4.6$ & 225 &  \\
BD+61 154 & V594 Cas & $22.80\pm0.77$ & $34.03\pm6.81$ & 225 & SMA &  &  &  &  &  &  & $22.8\pm0.8$ & $34.0\pm6.8$ & 225 &  \\
PDS 2 & CD-53 251 & $4.13\pm0.75$ & $8.82\pm1.76$ & 225 & ACA &  &  &  & $4.13\pm0.75$ & $8.82\pm1.76$ & 225 &  &  &  &  \\
HD 9672 &  & $4.14\pm0.04$ & $0.13\pm0.03$ & 232 & ALMA &  &  &  &  &  &  &  &  &  & S22 \\
HD 17081 & * pi. Cet & $<1.93$ & $<0.11$ & 225 & ACA &  &  &  & $<1.93$ & $<0.11$ & 225 &  &  &  &  \\
BX Ari A &  & $<2.65$ & $<2.83$ & 225 & ACA &  &  &  & $<2.65$ & $<2.83$ & 225 &  &  &  &  \\
HD 19745 &  & $0.90\pm0.04$ & $1.02\pm0.20$ & 225 & ALMA & $0.90\pm0.04$ & $1.02\pm0.20$ & 225 &  &  &  &  &  &  &  \\
HBC 338 &  & $30.96\pm0.05$ & $34.16\pm6.83$ & 225 & ALMA & $30.96\pm0.05$ & $34.16\pm6.83$ & 225 & $31.06\pm0.61$ & $34.13\pm6.83$ & 225 &  &  &  &  \\
BD+30 549 &  & $<5.00$ & $<3.26$ & 225 & ACA &  &  &  & $<5.00$ & $<3.26$ & 225 &  &  &  &  \\
TYC 4062-230-1 & VOS 1225 & $<3.03$ & $<9.22$ & 225 & SMA &  &  &  &  &  &  & $<3.0$ & $<9.2$ & 225 &  \\
$\dotsb$ & $\dotsb$ & $\dotsb$ & $\dotsb$ & $\dotsb$ & $\dotsb$ & $\dotsb$ & $\dotsb$ & $\dotsb$ & $\dotsb$ & $\dotsb$ & $\dotsb$ & $\dotsb$ & $\dotsb$ & $\dotsb$ & $\dotsb$  \\
\hline
\end{tabular}}\\
\tablefoot{Table in combination with Table~\ref{tab:herbig_stellar_params} is available in its entirety on CDS\footnote{For now this can be found on \url{https://datashare.mpcdf.mpg.de/s/wqYx7D9boe5i4S3}}. For 40 targets fluxes have been taken from \citet{Stapper2022} (S22), \citet{Stapper2024} (S24), \citet{Stapper2025a} (S25a), and \citet{Stapper2025b} (S25b). Due to overlap between the ALMA, ACA, and SMA data, the fluxes and resulting dust masses are given for all three. We use the ALMA data over the ACA data, and the ACA data over the SMA data. The used fluxes and the resulting dust masses are presented in the third and fourth columns together with the fluxes resulting from the NOEMA data. The observing frequencies are also mentioned for each flux measurement.}
\label{tab:herbig_fluxes}
\end{table*}

\begin{figure*}[h!]
    \centering
    \includegraphics[width=0.95\textwidth]{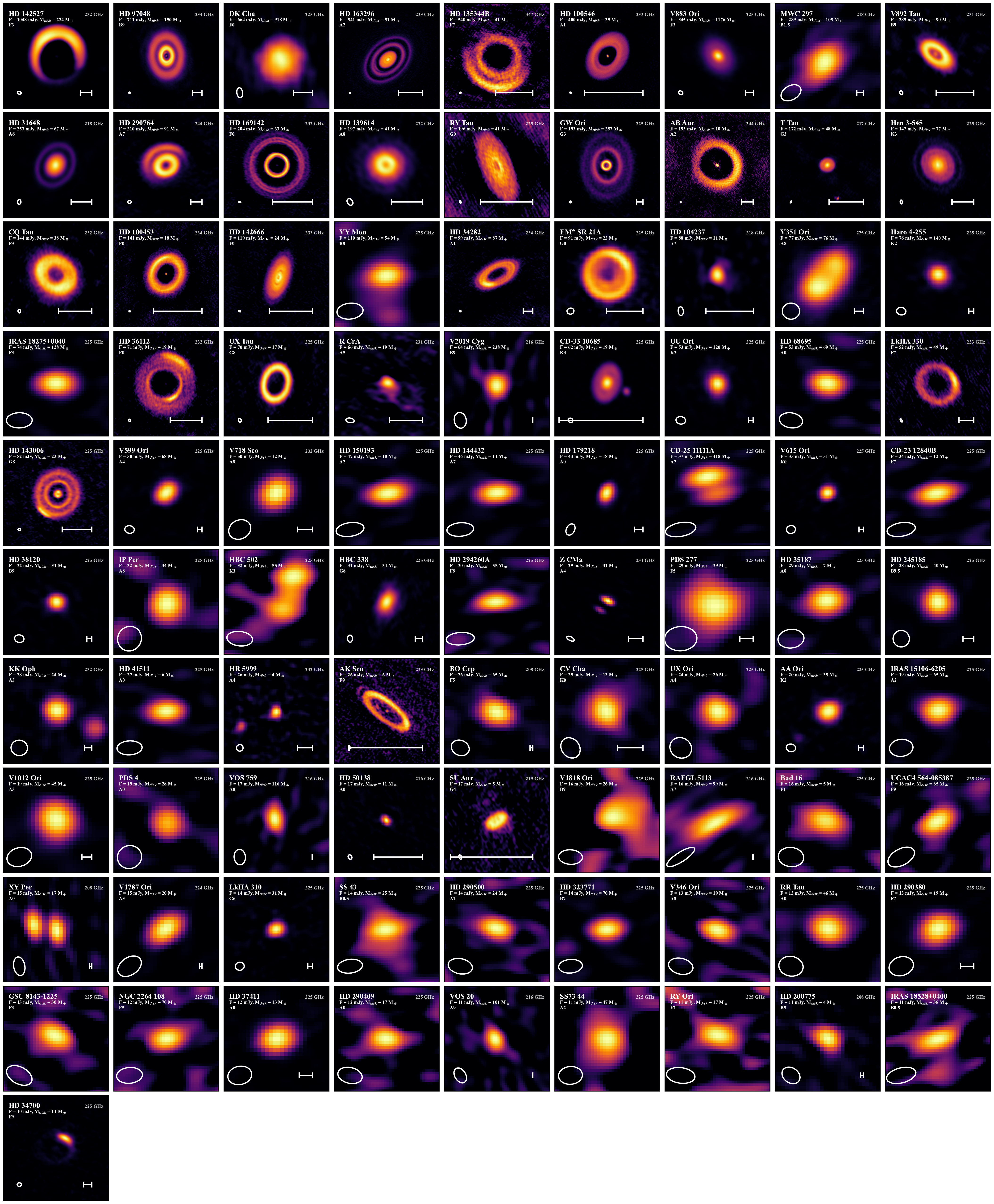}
    \caption{Images of all Herbig disks within 1 kpc with an integrated flux of at least 10~mJy, ordered by integrated flux density. An asinh normalization is used with the maximum set at the peak brightness of the disk. The size of the beam is indicated in the bottom left and a scale bar of 100~au is shown in the bottom right. For the ACA images no scale bar is shown. Additionally, the millimeter flux, resulting dust mass, observing frequency, and the stellar spectral type are also indicated.}
    \label{fig:massive_disks_gallery}
\end{figure*}

\section{Millimeter interferometric data}
\label{sec:data_reduction}
We report millimeter fluxes for all 243 disks around the Herbig stars in our catalog\footref{foot1} in Table~\ref{tab:herbig_fluxes}. Of these, we present new millimeter photometry data for 207 Herbig disks. For the remaining 36 Herbig disks, we use fluxes from previously published works by \citet{Stapper2022}, \citet{Stapper2024}, \citet{Stapper2025a}, and \citet{Stapper2025b}. We note that around 40\% of the 243 Herbig stars in our catalog was already covered by the aforementioned works. However, in most cases where both our new data and prior published fluxes are available, we prioritize our new observations. In four cases (XY~Per, T~Tau, HD~35929, and T~Ori), we use the previously published fluxes from \citet{Stapper2024, Stapper2025a, Stapper2025b} due to contamination by a companion or a non-detection in the new data.

The new interferometric photometry data are gathered from ten different projects, observed with the Atacama Large (sub)Millimeter Array (ALMA), the Atacama Compact Array (ACA), the Submillimeter Array (SMA), and the Northern Extended Millimeter Array (NOEMA). For the observations done with the ACA the data are from projects 2021.2.00005.S (PI: J.~Williams), 2022.1.01460.S (PI: J.~Williams), 2023.1.00937.S (PI: M.~Vioque), and 2024.1.00408.S (PI: M.~Vioque), while the ALMA data are from projects 2022.1.01155.S (PI: M.~Vioque) and 2023.1.00561.S (PI: M.~Vioque). For one target, HD~50138, we use ALMA archival product data from project 2019.1.01693.S (PI: J.~Varga). Additionally, we have SMA data from 2024A-S016 (PI: M.~Vioque) and 2024A-H001 (PI: M.~Vioque), and NOEMA data from W25BM (PI: L.~Stapper). Some targets have overlapping observations between the ACA, ALMA, and SMA data (we note that all fluxes are consistent between the observations). We report the measured fluxes from each telescope separately and for our analysis we favor the ALMA data over the ACA data, and the ACA data over the SMA data. The NOEMA observations do not overlap with the other observations. A general overview of the new data (e.g., observing conditions, configuration, tuning), how the imaging of the data was done,  and how the fluxes were determined can be found in Appendix~\ref{app:new_mm_data}. In total, the survey resulted in a detection rate of 70\%.

An overview of the continuum images resulting from the data reduction can be found for the ALMA data in Fig.~\ref{fig:ALMA_gallery}, for the ACA data in Fig.~\ref{fig:ACA_gallery}, and for the NOEMA data in Fig.~\ref{fig:NOEMA_gallery}. The images are ordered in decreasing integrated flux. For the SMA data the $uv$-coverage is relatively poor and therefore no images are presented.

\begin{figure*}[b]
    \centering
    \includegraphics[width=1.0\textwidth]{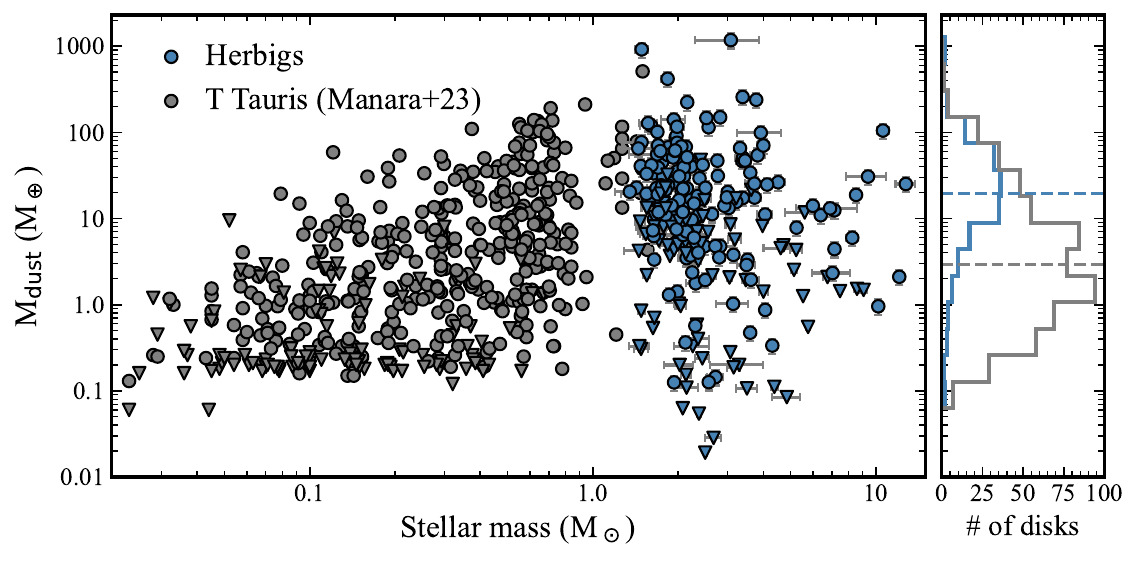}
    \caption{The stellar mass versus dust disk mass of T~Tauri disks (gray, \citealt{Manara2023}), and Herbig disks (blue, this work). Upper limits are indicated by the triangles. A histogram of the detected disks is shown in the right panel with the corresponding median dust disk mass as indicated by the horizontal dashed line. This shows that the distribution of Herbig disk dust masses are essentially contained within that of the T~Tauri disks, but skewed toward higher disk masses.}
    \label{fig:Mdust_Mstar}
\end{figure*}

\section{Results}
\label{sec:results}
\subsection{Continuum images}

Figure~\ref{fig:massive_disks_gallery} presents the continuum images of all Herbig disks with an integrated flux of more than 10~mJy, 91 disks in total, and includes the currently highest resolution Band~6 or 7 observations available in the ALMA archive\footnote{\url{https://almascience.eso.org/aq/}}. These high resolution data were not used for the determination of the total flux, those were obtained from the data and references as mentioned in Section~\ref{sec:data_reduction}. The high resolution observations in Fig.~\ref{fig:massive_disks_gallery} belong to some of the most famous disks in the field. Many of these disks show (strong) asymmetries: HD~142527 \citep{Temmink2023}, HD~290764 \citep{Kraus2017}, HD~34282 \citep{vanderPlas2017a}, LkH$\alpha$~330 \citep{Pinilla2022b}, HD~135344B \citep{Casassus2021}, EM$^*$~SR~21A \citep{Yang2023}, HD~36112 \citep{Dong2018}, HD~34700 \citep{Stadler2026}, and AB~Aur \citep{Tang2017}, and others have (multiple) rings: GW~Ori \citep{Kraus2020}, HD~97048 \cite{vanderPlas2017b}, V892~Tau \citep[][see also App.~\ref{app:individual_sources}]{Alaguero2024, Alaguero2025}, Hen~3-545 \citep[CR~Cha,][]{Kim2020}, HD~31648 \citep{Liu2019}, HD~163296 \citep{Isella2016}, T~Tau, HD~139614, RY~Tau \citep{Valegard2022}, HD~100546 \citep{Pineda2019}, CQ~Tau \citep{Wolfer2021}, HD~169142 \citep{Perez2019}, HD~142666 \citep{Andrews2018b}, CD-33~10685 \citep[HT~Lup,][]{Andrews2018b}, HD~100453 \citep{Rosotti2020}, UX~Tau \citep{Menard2020}, and AK~Sco. In addition, the famous outbursting source V883~Ori \citep{Cieza2016, Leemker2025}, and the Class~I/II object DK~Cha \citep[e.g.,][]{vanKempen2010} are also part of the sample. While many of the brightest disks already have high resolution observations, there are still many which do not have resolved observations. The least bright disk with high resolution observations is HD~34700, which has an arc-like feature similar to that of HD~142527 \citep{Stadler2026, Fasano2026}, and has many spiral structures in scattered light \citep{Monnier2019, Columba2024}. Many of the unresolved disks in Fig.~\ref{fig:massive_disks_gallery} may therefore reveal very substructured and interesting disks with higher resolution observations. In the new ALMA observations there are eight newly resolved disks, which are shown as the first eight disks in Fig.~\ref{fig:ALMA_gallery}, and are also shown in Fig.~\ref{fig:massive_disks_gallery}. No particular structures are visible in the dust, but some extended emission is visible in most of these eight bright disks. In particular the disks of Haro~4-255, UU~Ori, and HBC~338 show extended faint continuum emission.

In a significant fraction of sources, around 30 in total, emission offset from the position of the Herbig star is visible within the field of view of the data. Specifically, in the ACA data of VY~Mon, CD-25~11111A, HBC~502, and CV~Cha, there is emission in addition to that of the Herbig disk (see Fig.~\ref{fig:ACA_gallery}). Additionally, ten other cases in the ACA data show emission in the field of view where the Herbig disk is not detected, typically at the edge of the field and likely not bound objects. For the new ALMA data (Fig.~\ref{fig:ALMA_gallery}), two sources (AA~Ori and CO~Ori~A) display emission besides the Herbig disk. Notably, the additional emission to that of CO Ori A originates from a disk around a secondary star, as it is a known binary system \citep{Mason2001}. Other well-known binaries or multiple systems with multiple components showing continuum emission include KK~Oph, HR~5999 \citep{Panic2021}, T~Tau \citep{Kohler2016, Rota2022}, CD-33~10685 (HT~Lup) \citep{Andrews2018b}, and Z~CMa \citep{Dong2022}. In some cases, only one component has detected millimeter emission. Examples include HD~100453 \citep{Rosotti2020}, which has a low stellar mass companion \citep{Collins2009}, and HD~135344B, which is one of the most studied disks \citep{Stolker2016, Cazzoletti2018}, while the A component lacks a disk but hosts a planet \citep{Stolker2025}. Furthermore, our new ALMA data reveal instances where no flux is detected around the Herbig star, yet nearby emission is present, such as in HD~288313, PR~Ori, IRAS~05393-0838, and V373~Cep (Figs.\ref{fig:ALMA_gallery} and \ref{fig:NOEMA_gallery}). Additionally, slight deviations in the emission peak from the position of the Herbig star are observed in HBC~502, HD~259431, and V1818~Ori (Fig.\ref{fig:ACA_gallery}). These detections could indicate potential companions, though they may also be foreground or background sources requiring further investigation. 

To estimate whether the coarser beam of the ACA observations catches more diffuse emission not associated with the disk, such as an envelope or a binary component, we can make use of the overlapping observations between the ACA and ALMA data. A comparison between these observations shows that there are no significant differences between the respective integrated fluxes (see the fluxes in Table~\ref{tab:herbig_fluxes}). Nonetheless, to properly distinguish between disk emission from a primary and a possible secondary disk in the ACA observations, a foreground/background object, or extended envelope emission, higher resolution observations are necessary.

\begin{figure*}[t]
    \centering
    \includegraphics[width=\textwidth]{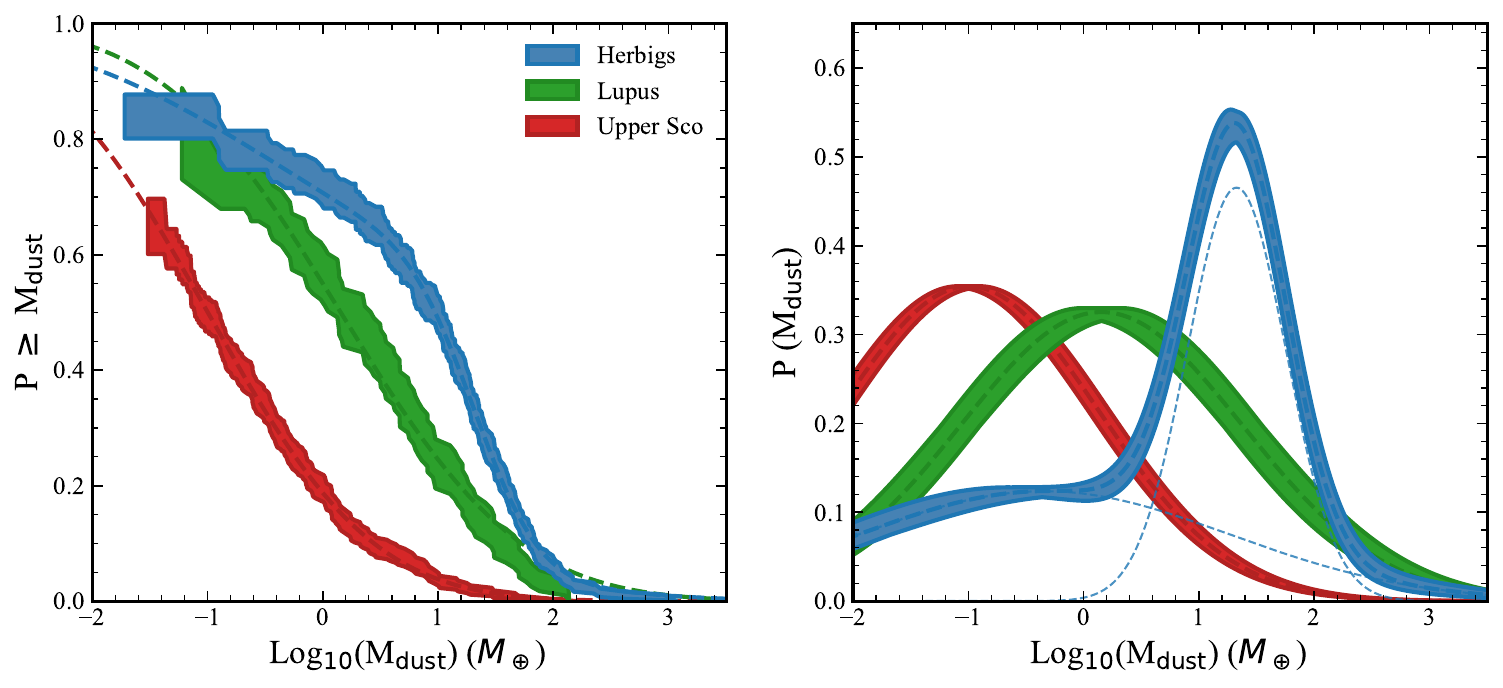}
    \caption{The cumulative and probability dust mass distributions of the Herbig stars within 1~kpc, the Lupus star-forming region \citep{Manara2023} and the Upper~Scorpius star-forming region \citep{Carpenter2025}. To obtain the probability distributions in the right panel, a lognormal Gaussian distribution has been fitted through the cumulative distributions in the left panel, as indicated by the dashed lines. For the Herbig sample a bimodal lognormal distribution is needed to fit the cumulative dust mass distribution. The two contributions have been plotted separately.}
    \label{fig:cdf_Mdust}
\end{figure*}

\subsection{Obtaining dust masses}
\label{subsec:get_dust_masses}
To obtain the dust disk masses from the millimeter continuum flux emission, we use the scaling relationship from \citet{Hildebrand1983}, which assumes optically thin emission,

\begin{equation}
    M_\text{dust} = \frac{F_\nu d^2}{\kappa_\nu B_\nu(T_\text{dust})}.
    \label{eq:Mdust}
\end{equation}

Here, $F_\nu$ and $\kappa_\nu$ are respectively the continuum millimeter flux and dust opacity at frequency $\nu$, $d$ the distance to the object, and $B_\nu(T_\text{dust})$ the value of the Planck function given a dust temperature $T_\text{dust}$ and a frequency $\nu$. The dust opacity is given by a power-law of the form $\kappa_\nu\propto\nu^\beta$, which is set to 10~cm$^2$~g$^{-1}$ at a frequency of 1000~GHz \citep{Beckwith1990}. We follow previous works by assuming that the power-law index $\beta$ equals 1. Lastly, Herbig stars are more luminous than T~Tauri stars, hence the general assumption of a dust temperature of 20~K \citep[see, e.g.,][]{Ansdell2016} does not hold. As was done in the works of \citet{Stapper2022, Stapper2025a, Stapper2025b}, we scale the dust temperature depending on the stellar luminosity following the scaling relationship used in \citet{Andrews2013}, $T_\text{dust}=25\text{K}\times(L_*/L_\odot)^{1/4}$. Hence, the dust temperatures range from 26~K to 266~K with a median temperature of 54~K. To compute the error on the dust mass, we quadratically add the absolute flux calibration error and the rms noise estimated from each image. In addition, we propagate the error on the distances. In most cases the flux calibration error is the main contributing source of uncertainty.

The resulting dust masses are plotted against their stellar mass in Figure~\ref{fig:Mdust_Mstar}, together with the T~Tauri sample of \citet{Manara2023}. From this figure, it is clear that the median dust mass is higher for Herbig disks than for T~Tauri disks, yet the range in dust masses is relatively similar. The histograms in the right panel show this clearly: the distribution of the Herbig disk dust masses is essentially contained within that of the T~Tauri disks, but it is significantly skewed toward higher disk masses. Furthermore, for Herbig stars more massive than 4~\Msun, which includes 34 stars in total, we find generally lower disk masses, resulting in a decrease of disk mass with an increase in stellar mass. This will be further discussed quantitatively in Sec.~\ref{subsec:connection_to_exoplanets}. Lastly, we note that there is a lack of dust mass measurements for disks around solar mass stars in the sample of \citet{Manara2023}, resulting in low-number statistics at that particular stellar mass regime.

\subsection{Dust mass cumulative and probability distributions}
\label{subsec:mdust_cdf_pdf}
A cumulative dust mass distribution can be made using the \texttt{lifelines} package \citep{DavidsonPilon2021}, using the \texttt{fit\_left\_censoring} of the \texttt{Kaplan-Meier} (KM) estimator to properly account for the upper limits in the data. We note that the KM estimator should not be used when the upper limits are covariant with the disk mass. The main source of such a covariance would be caused by a relation between stellar mass and the upper limits. Given that we do not see such a relation, we use the KM estimator for our analysis. The resulting cumulative distributions can be found in Figure~\ref{fig:cdf_Mdust}, where the vertical spread in the distribution indicates the $1\sigma$ confidence intervals. The Herbig disk dust mass distribution is shown as the blue curve which is compared to the cumulative distributions of the Lupus star-forming region (\citealt{Manara2023} who combined the works of \citealt{Cleeves2016}, \citealt{Ansdell2018}, and \citealt{Sanchis2020}) and the Upper~Scorpius star-forming region \citep{Carpenter2025}.  The latter distribution has been obtained by using the fluxes and distances provided in \citet{Carpenter2025} and scaling those to a dust mass using Eq.~(\ref{eq:Mdust}). For the disks in both Lupus and Upper~Sco, a dust temperature of 20~K was assumed, as scaling the temperature is unnecessary given the scarcity of intermediate-mass stars in these regions. In Appendix~\ref{app:flux_distributions}, a version of Fig.~\ref{fig:cdf_Mdust} is presented based on the fluxes, scaled to a common distance and frequency.

It is clear that the distributions are different between the different star-forming regions and the Herbig disks, as was reported in \citet{Stapper2022}. Both Figs.~\ref{fig:Mdust_Mstar} and \ref{fig:cdf_Mdust} show that the median dust disk mass is higher for Herbig disks than for the disks around lower mass stars. The cumulative distribution of the Herbig disks has a steeper slope than for Lupus and Upper~Sco, resulting in 50\% of the Herbig dust disks being more massive than 10~\Mearth, which is only 20\% and 5\% for Lupus and Upper~Scorpius respectively. 

Following the previous work of \citet[][]{Williams2019}, we can fit a Gaussian lognormal distribution through the cumulative dust mass distributions shown in the left panel of Fig.~\ref{fig:cdf_Mdust} to obtain probability distributions. The cumulative distribution is made such that the sum from zero to infinity is normalized to 1. We find that the distributions are well reproduced with a lognormal distribution, especially for Lupus and Upper~Sco. The fit results in a mean dust disk mass of $1.43^{+0.69}_{-0.46}$~\Mearth~and $0.10^{+0.02}_{-0.02}$~\Mearth~for Lupus and Upper~Scorpius respectively, with standard deviations of $1.23^{+0.02}_{-0.03}$ and $1.13^{+0.01}_{-0.01}$ in units of Log$_{10}$(M$_{\rm dust}$/\Mearth). The errors were propagated by fitting the upper and lower bounds of cumulative distribution confidence interval. For the Herbig disk dust mass distribution, we find that the distribution is not well described by a single lognormal distribution. This is due to a significant fraction of low mass disks in the sample, which gives rise to a low mass tail deviating from a single lognormal distribution. We therefore fit a bimodal lognormal distribution, again made such that the sum from zero to infinity is normalized to 1. Comparing the goodness-of-fit $\chi^2$ of a single lognormal distribution fit ($\chi^2=1639$) to the bimodal lognormal distribution fit ($\chi^2=41$), it is clear that the fit has significantly been improved by fitting a bimodal lognormal distribution. We therefore find that the Herbig disk dust mass distribution seems to consist of two populations. One population has relatively high disk masses, with a mean dust mass of $21.19^{+1.48}_{-1.44}$~\Mearth~and a standard deviation of $0.43^{+0.01}_{-0.01}$. Additionally, there is a low dust mass population with a mean dust mass of only $0.45^{+0.042}_{-0.024}$~\Mearth\;with a much broader standard deviation of $1.61^{+0.07}_{-0.05}$. Section~\ref{subsec:long_lived_disks} will discuss the interpretation of these high and low dust disk mass Herbig populations.

\section{Completeness and lifetime of Herbig stars}
\label{sec:completeness}
In this section our goal is to quantify the completeness of our sample, i.e., estimate the number of Herbig stars we might currently be missing. This will ultimately also result in an estimate of the lifetime of a pre-main sequence intermediate mass star being recognized as a Herbig star. In other words, the Herbig lifetime is how long a star would comply with the selection criteria set out in Section~\ref{sec:target_selection} and be put in our catalog. We can achieve an estimate of the completeness if we are almost complete for nearby regions, and we can determine a Herbig lifetime by measuring the distribution of stars as a function of mass and distance compared to what would be expected from a complete sample. We do this by using the constraints on the star formation rate in the solar neighborhood assuming a known initial mass function.

\subsection{Modeling the observed sample of Herbigs}
Our Herbig sample is defined as in Section~\ref{sec:target_selection}. For each star we compute a Sun–centered Cartesian position and determine its cylindrical radius via $R_\star = \sqrt{x_\star^2 + y_\star^2}$. We split the sample into $N_M$ bins in stellar mass, $[M_{j,\min},M_{j,\max})$ with $j=1,\dots,N_M$, using the $M_\star$ estimates in Table~\ref{tab:herbig_stellar_params}. To connect the local SFR to the number of Herbigs formed in each mass bin, we assume a \citet{Chabrier_ea_2003} initial mass function (IMF).

We start by constructing a Sun-centered map of the star-formation rate surface density $\dot{\Sigma}_{\rm SFR}(x,y)$, which is rescaled to match the star-formation rate by \citet{Quintana_ea_2025} (see for a detailed derivation App.~\ref{app:sfr_from_dust}). Then, the mean star-formation rate per annulus SFR$_{\rm ring}(R_i)$ is determined by averaging $\dot{\Sigma}_{\rm SFR}(x,y)$ in concentric annuli of width $\Delta R$. For a given parametrization of the Herbig lifetime as a function of stellar mass we then predict the expected number of Herbig systems in annulus $i$ and mass bin $j$ as
\begin{equation}
  \lambda^{\rm (0)}_{ij}
  \;=\;
  \mathrm{SFR}_{\rm ring}(R_i)\;
  \frac{f_j}{\bar{M}}\;
  t_{\mathrm{eff},j},
  \label{eq:n_model_simple}
\end{equation}
where $f_j$ is the IMF number fraction in bin $j$, $t_{\mathrm{eff},j}$ is an effective lifetime appropriate for that bin, and $\bar{M}$ the IMF-averaged stellar mass (see App.~\ref{app:sfr_from_dust}).

We have so far assumed both that we know the exact star formation rate, and that we detect all stars. Neither of these things is true, and we must account for them in the Bayesian model. The mean expected number of \emph{detected} Herbig stars in annulus $i$ and mass bin $j$ is then:

\begin{equation}
\begin{aligned}
      \lambda_{ij} &= \lambda^{\rm (0)}_{ij} \times f_\mathrm{SFR} \times f_\mathrm{det}\\
 &=  f_\mathrm{SFR} \mathrm{SFR}_{\rm ring}(R_i)\frac{f_j}{\bar{M}}\tau(M_j)f_{\mathrm{det}}(R_i,M_j),
  \label{eq:lambda_ij_def}
\end{aligned}
\end{equation}
where:

\begin{itemize}
    \item $f_\mathrm{SFR}$ is a global normalization parameter that absorbs any residual mismatch between the Quintana SFR map and the true underlying SFR. Since $\dot{\Sigma}_{\rm SFR}(x,y)$ has already been rescaled to match the \citet{Quintana_ea_2025} SFR inside 1~kpc, we place a narrow normal prior on $\ln f_\mathrm{SFR}$. This distribution is centered on zero and truncated below zero with an upper dispersion $0.034$ corresponding to the uncertainty quoted by \citet{Quintana_ea_2025};
    \item The effective lifetime $t_{\mathrm{eff},j}$ is set to $\tau(M_j)$, which is the mass--dependent Herbig lifetime, parametrized as
    \begin{equation}
      \tau(M_j) \;=\; \tau_0
      \left(\frac{M_j}{M_0}\right)^{\alpha_\tau},
      \label{eq:tau_mass}
    \end{equation}
    with both the normalization $\tau_0$ and the slope $\alpha_\tau$
    treated as free parameters, while we fix $M_0 = 1\, M_\odot$. In the inference we assign a lognormal prior to $\tau_0/1 \, \mathrm{Myr}$ with mean $\mu = \ln 5$ and dispersion $\sigma =0.4$. We assume a  normal prior for $\alpha_\tau$ with $\mu =-1.5$ and $\sigma =0.5$;
    \item $f_{\mathrm{det}}(R,M)$ is a detection probability that accounts for radial and mass--dependent incompleteness. We adopt a simple logistic form,
    \begin{equation}
    \begin{aligned}
      \mathrm{logit}\,f_{\mathrm{det}}(R,M) &\equiv \ln\!\left[\frac{f_{\mathrm{det}}}{1-f_{\mathrm{det}}}\right]\\
      &= a_0 + a_R \ln\!\left(\frac{R}{R_0}\right) + a_M \ln\!\left(\frac{M}{M_0}\right),
      \label{eq:logit_fdet}
    \end{aligned}
    \end{equation}
    where $R_0$ and $M_0$ are fixed reference values (we adopt $R_0 = 500$~pc and $M_0 = 1~\mathrm{M_\odot}$), and $a_0$, $a_R$, and $a_M$ are free parameters with weakly constrained priors. All are normal with means $0$, $-1$ and $1$, and dispersions $2$, $1$, and $1$ respectively.
\end{itemize}

We fit the model using \texttt{Pymc} \citep{pymc2023}. We divide our sample into rings of radius $\Delta R=50$~pc and include $6$ evenly spaced stellar mass bins between $1.5$ and $12 \, M_\odot$. We then consider the likelihood of observing the number of Herbig stars in each bin $(i, j)$ as a Poisson distribution with mean $\lambda_{ij}$. The outcome of the fitting procedure is presented in Appendix~\ref{app:corner_plot}.

\subsection{Modeling results}
\begin{figure}[t]
    \centering
    \includegraphics[width=0.5\textwidth]{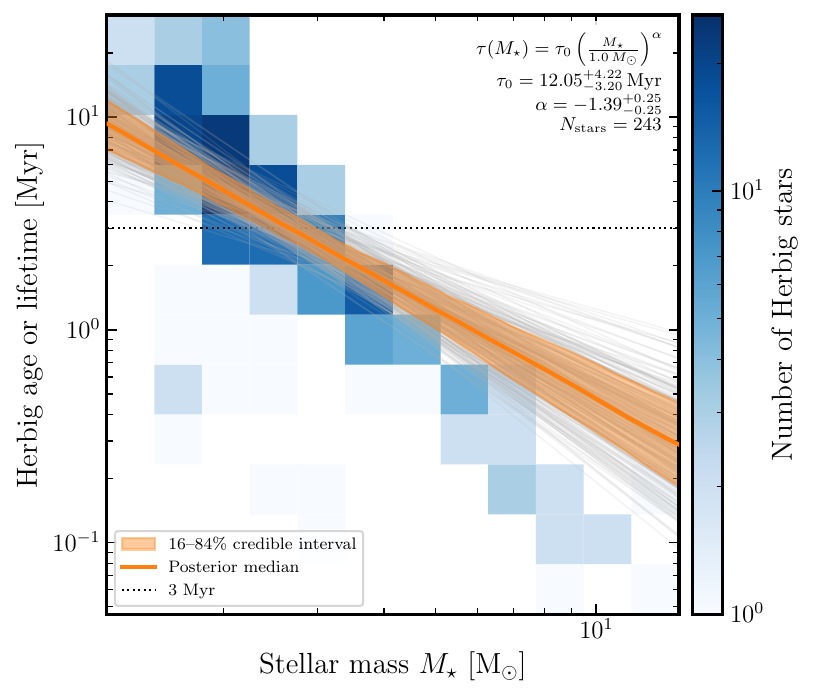}
    \caption{The orange lines are the posterior of the Herbig lifetimes with respect to stellar mass based on our Bayesian model. The blue background shows a histogram of the stellar (Herbig) ages obtained from the PARSEC evolutionary tracks as a function of stellar mass. The Herbig lifetime is found to decrease as stellar mass increases.}
    \label{fig:lifetimes}
\end{figure}

In Figure~\ref{fig:lifetimes} we show the posterior of the Herbig lifetimes based on our Bayesian model compared to the stellar ages obtained from the pre-main sequence evolutionary tracks. Again, we note that the lifetime here means how long we would recognize the star to be a Herbig star, i.e., that the intermediate mass pre-main sequence star has (detectable) infrared excess and in many cases accretion signatures. We have the strongest constraints for low mass Herbig stars, which are most numerous. However, it is clear that the data supports a Herbig lifetime that decreases with stellar mass in the Herbig regime, with index $\alpha_\tau = -1.39$. We estimate that dust-disks around stars with masses $3-4\,M_\odot$ have the canonical $3$~Myr lifetimes. Such a decrease is broadly consistent with the findings of other authors \citep{Yasui_ea_2014, Ribas_ea_2015, Pinilla2022a, Ronco2024}. Generally,  we find long disk lifetimes, close to $\sim 12$~Myr when extrapolated to solar mass stars \citep[similar to the lifetimes suggested by][]{Pfalzner_ea_2022, Polnitzky2025}. Furthermore, the Herbig disk lifetimes and the stellar ages from the HR diagram seem to be consistent with each other. Yet, trusting the found relationship for the Herbig lifetime (Eq.(\ref{eq:tau_mass})), this comparison does suggest that the ages from the HR diagram result in too old ages for the lower mass stars and too young ages for the higher mass stars. Dynamical stellar mass estimates will help in better determining the ages of these stars \citep{Zallio2026}.

\begin{figure}[t]
    \centering
    \includegraphics[width=0.5\textwidth]{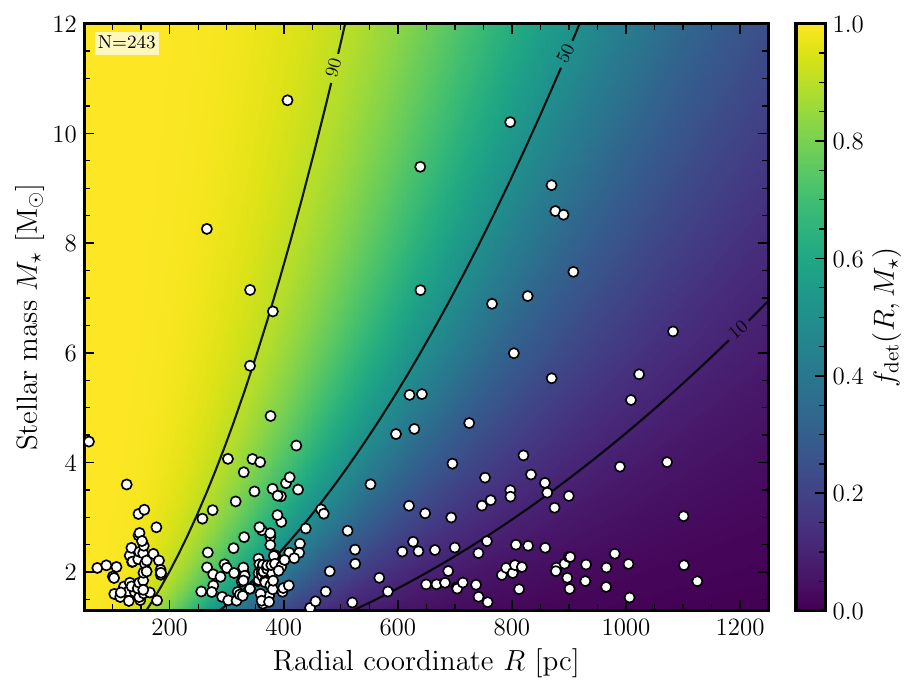}
    \caption{The completeness of our survey as a function of stellar mass and distance from Earth. Our sample of Herbig stars are indicated by the markers. The sample is close to complete within 300~pc, but this decreases steeply with an increase in distance. We are more complete for higher than for lower mass stars.}
    \label{fig:completeness}
\end{figure}

\begin{figure}[t]
    \centering
    \includegraphics[width=0.5\textwidth]{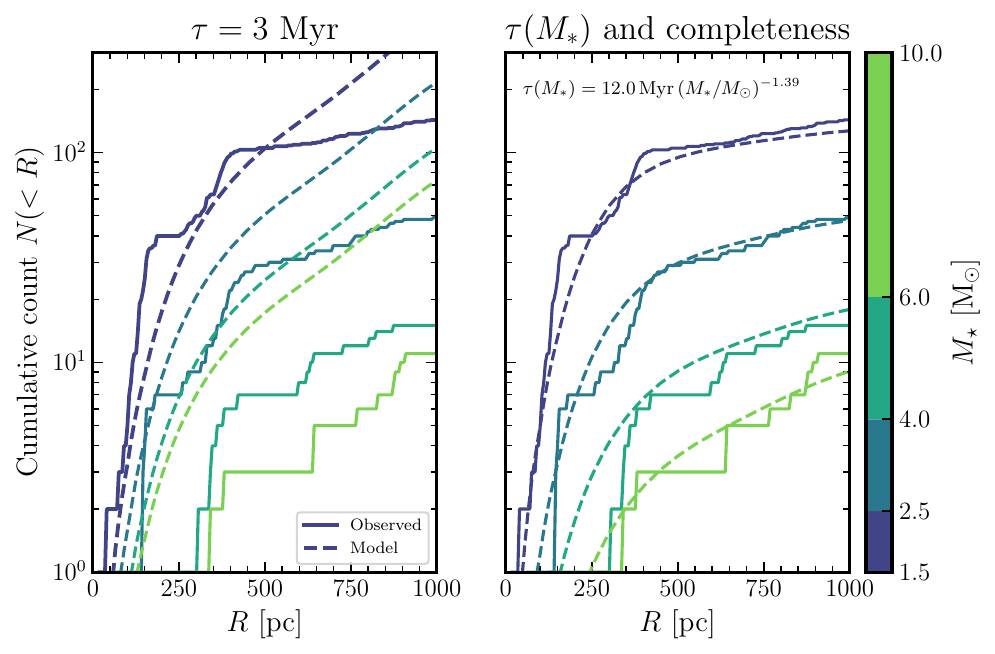}
    \caption{Model predictions for the number of Herbig stars contained within a cylindrical radius $R$.  In all panels we show the cumulative observed counts $N^{\mathrm{obs}}_j(<R)$ (solid lines) and the corresponding model predictions (dashed lines), with different colors denoting the different mass bins. The color bar on the right indicates the mass range associated with each color. We assume a star formation rate as estimated in Section~\ref{app:sfr_from_dust}. Our assumptions vary from: $100$~percent completeness and all stars would have been selected as Herbigs for $3$~Myr (left), and a power-law completeness in $M_*$ and $R$ and a power-law lifetime $\tau(M_*)$ from Figs.~\ref{fig:lifetimes} and \ref{fig:completeness} (right). The left panel illustrates the onset of strong radial incompleteness at $R \approx 500$~pc.}
    \label{fig:counts}
\end{figure}

Using Eq.~(\ref{eq:logit_fdet}) and filling in the results from our fit (see Eq.~\ref{eq:logit_fdet_filled_in}) we can estimate the completeness $f_{\rm det}$ of our Herbig sample per stellar mass and radial bin. This gives the result shown in Figure~\ref{fig:completeness}, where we see that the completeness increases with stellar mass, and decreases with distance. The sample is close to $100$~percent complete inside $300$~pc across the full stellar mass range. Within $400$~pc (the distance of Orion), we expect approximately $162$ Herbig stars, which means that with the 126 Herbig stars in our catalog within 400~pc we include 78\% of the total number of Herbig stars. However, this can become significantly lower for specific stellar mass bins: for $\sim 2\, M_\odot$ stars, we are $50$~percent complete at that distance. Still, we are quite complete for relatively local regions when considering the full stellar mass range. Within $1$~kpc we would expect based on our model approximately 1000 Herbig stars between $1.5 - 12\, M_\odot$ if we were $100$~percent complete. Hence, regarding the whole 1~kpc volume we are only 24\% complete. We may now ask how well we reproduce the total number of Herbigs when we put our entire model together. Figure~\ref{fig:counts} summarizes the comparison between observed and predicted Herbig counts as a function of cylindrical radius and stellar mass. The two panels show a simple model with a constant lifetime of 3~Myr for all mass bins on the left and the fitted radial completeness and lifetime model on the right.

Figure~\ref{fig:counts} shows that the fitted radial completeness and lifetime model reproduces the observed cumulative counts in each mass bin remarkably well. Given that the underlying SFR field has been normalized to be very close to the \citet{Quintana_ea_2025} value inside 1~kpc, the agreement between the right–hand panel model and the observed Herbig counts is reassuring: a Quintana–consistent local SFR, combined with a simple parametrization of $\tau(M_\star)$ and $f_{\rm det}(R,M_\star)$, is sufficient to match both the normalization and the mass dependence of our Herbig sample.

The comparison between the two panels of Fig.~\ref{fig:counts} also highlights that the shape of the mass–segregated cumulative distribution functions (CDF) requires a decreasing Herbig lifetime with increasing stellar mass. The constant--lifetime model ($\tau = 3$~Myr; left panel) fails to reproduce the relative separation between the curves in different mass bins: it cannot simultaneously match the low--mass counts and the rapid rise of the high--mass CDF at small radii. The fitted power--law relation $\tau(M_\star) = \tau_0 (M_\star/M_\odot)^{\alpha_\tau}$ in the right–hand panel captures the observed mass dependence of the CDFs well.

\begin{figure*}[t]
    \centering
    \includegraphics[width=\textwidth]{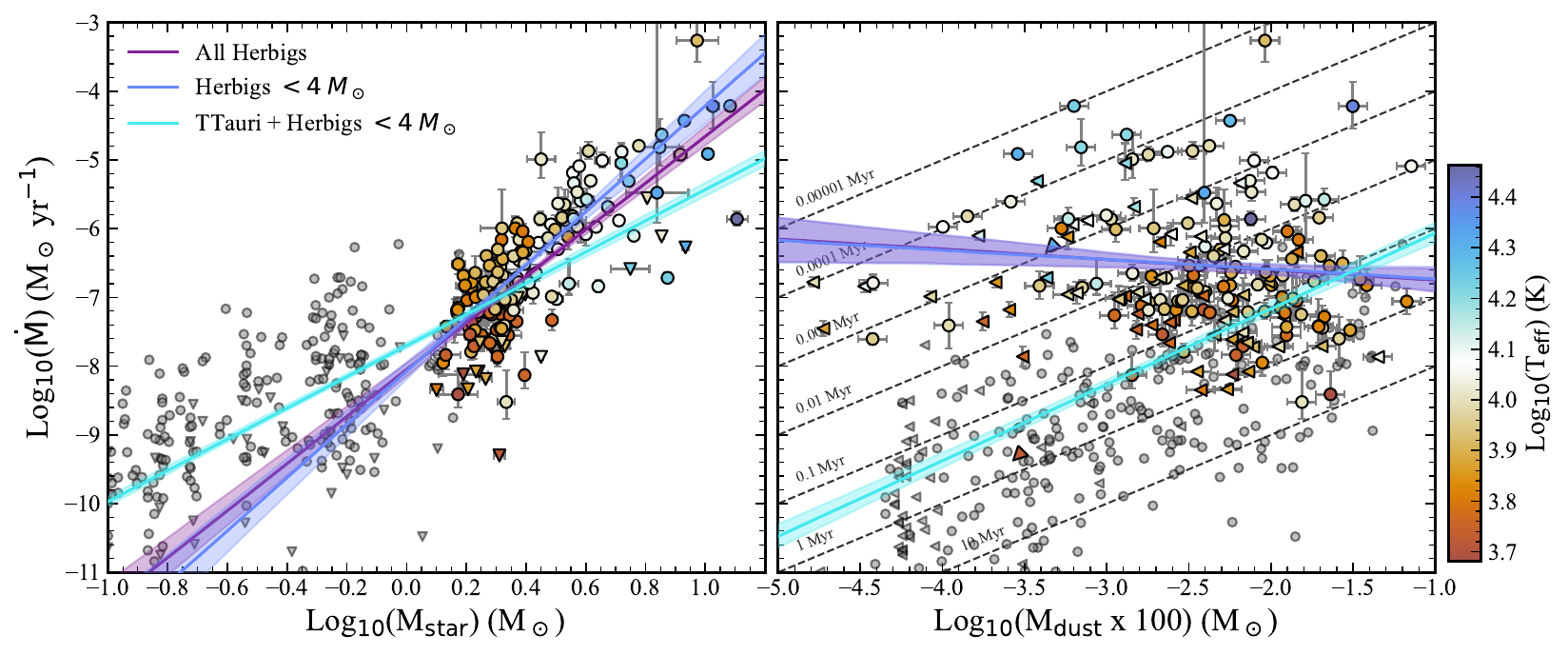}
    \caption{The accretion rate plotted against the stellar mass (left) and the disk mass (right). The Herbig disks are colored by the effective temperature of their host star. The gray markers are the T~Tauri disks from \citet{Manara2023}. The dashed lines in the right panel indicate different disk lifetimes assuming a viscously accreting disk. Three fits are shown in both panels, with a $1\sigma$ range highlighted around the median value. We find a steeper accretion rate stellar mass relationship for Herbig stars compared to T~Tauri stars. Above 4~\Msun the relationship flattens. For the accretion rate disk mass relationship we find an almost flat relationship for Herbig stars.}
    \label{fig:mdust_mdot}
\end{figure*}

The figure also illustrates the onset of strong radial incompleteness. The observed cumulative counts flatten beyond $R \approx 500$~pc in all mass bins, whereas the simple ``100\% complete'' model in the left panel continues to rise too quickly with radius. This behavior requires a detection probability that declines with $R$, as encoded in our logistic completeness function $f_{\rm det}(R,M_\star)$ (as shown in Fig.~\ref{fig:completeness}).

Finally, the low--mass bins ($M_\star \approx 1.5$–$2~{\rm M}_\odot$) clearly favor dust disk lifetimes longer than 3~Myr. In the left panel of Fig.~\ref{fig:counts}, the $\tau=3$~Myr model already underpredicts the number of Herbigs at small radii, where the sample should be close to complete. The fitted model remedies this by adopting a longer lifetime at low mass, with an inferred normalization $\tau_0 \sim 7$~Myr at $1~{\rm M}_\odot$, consistent with the behavior required by the innermost part of the CDF.

\section{Discussion}
\label{sec:discussion}

\subsection{The $\dot{M}$-$M_{dust \rm}$ and $\dot{M}$-$M_{\odot}$ relationships}
\label{subsec:accretion}
One way to understand the evolution of stars and their protoplanetary disks is by comparing accretion rates to stellar and disk masses. A relationship between the mass accretion rate and the stellar mass has been well established in T~Tauri stars for quite some time \citep[e.g.,][]{Natta2006}, with a typical relationship of $\dot{M}\propto M_\star^2$ and a large spread of a couple orders of magnitude \citep[e.g.,][]{Testi2022, Manara2023}. It remains unknown what the exact origin is of this correlation \citep{Ercolano2017}. This correlation has been found to extend toward the Herbig star regime, up to 4~\Msun, after which the relationship flattens \citep{Fairlamb2015,Wichittanakom2020,Grant2022,Vioque2022}. The lower-mass Herbig stars are known to accrete as T~Tauri stars via magnetospheric accretion (\citealp{Hartmann2016}). On the other hand, accretion in Herbig stars more massive than around 4 M$_{\odot}$, with weak or negligible magnetic fields, probably occurs directly from the disk to the star through a hot boundary layer. This results in a less steep relationship (see \citealt{Mendigutia2020} for a review). Additionally, a clear relationship has been found between the stellar accretion rate and the total disk mass as determined from dust millimeter observations in multiple star-forming regions with a spread of around 1~dex \citep{Manara2016, Mulders2017, Testi2022, Manara2023}, although recent work for Upper~Scorpius finds a lack of correlation in that region \citep{Empey2026}. This relationship is expected from viscous evolution, with internal photo-evaporation, disk truncation, and optically thick dust increasing the spread in the correlation \citep{Rosotti2017, Sellek2020, Zagaria2022, Anania2025AGEPRO, Anania2025}, but can also arise from magnetohydrodynamic (MHD) disk evolution \citep{Tabone2022, Somigliana2024, Tabone2025}. For Herbig stars initially this relationship was also tentatively found \citep{Mendigutia2012}. More recent work by \citet{GrantStapper2023} showed that while there is a significant fraction of Herbig stars which are indeed the scaled-up version of T~Tauri stars and follow the $\dot{M}$--$M_{disk \rm}$ correlation, there is a group of Herbig stars which have high accretion rates and low disk masses. This results in a practically flat relationship between the two quantities in the Herbig regime. 

Figure~\ref{fig:mdust_mdot} presents the accretion rates of our catalog, both as a function of stellar mass (left panel) and disk mass (right panel). In both panels the Herbig stars are compared to the T~Tauri stars from the sample of \citet{Manara2023}. In the left panel the break in the stellar mass accretion rate relationship around a stellar mass of 4~\Msun is reproduced. In the Herbig regime the accretion rates range from $10^{-9}$ to $10^{-3}$~\Msun~yr$^{-1}$. In the right panel, isochrones assuming a viscously accreting disk at a constant accretion rates are plotted as dashed lines. We find that the disk mass accretion rate relationship of the Herbig disks are slightly above that of the T~Tauri disks. While the relationship for T~Tauri stars is centered around a 1~Myr lifetime, only the tail end of the Herbigs coincides with the T~Tauris. A large number of Herbig stars in our sample have high accretion rates, but relatively low disk masses. This results in 40\% of the Herbig stars having an expected disk lifetime shorter than 10~kyr. This is only the case for less than 1\% for the T~Tauri stars. This is in stark contrast with the Herbig lifetime found in Section~\ref{sec:completeness}, where the majority of the Herbig stars are expected to survive more than 1~Myr.

We fit the $\dot{M}$-$M_{star \rm}$ and  $\dot{M}$-$M_{dust \rm}$ relationships using \texttt{linmix} \citep{Kelly2007}\footnote{\url{https://linmix.readthedocs.io/}} for three different cases: all Herbig stars, Herbig stars with a stellar mass of less than 4~\Msun, and the latter combined with T~Tauri stars. Using a power-law relationship of the form log$_{10}$($\dot{M}$/$M_\odot$~yr$^{-1}$)=$\alpha$+$\beta\times$log$_{10}$($M_{star \rm}/M_\odot$) we obtain intercepts of $-8.05\pm0.10$, $-8.07\pm0.13$, and $-7.70\pm0.04$, and slopes of $3.40\pm0.22$, $3.87\pm0.35$, and $2.27\pm0.07$, respectively. The resulting fits are also plotted in Fig.~\ref{fig:mdust_mdot}. We find a generally steeper relationship for the Herbig stars compared to the T~Tauri stars. Excluding the more massive Herbig stars slightly decreases the slope of the relationship, indicating that the relationship indeed changes for the higher mass stars. For the disk mass and accretion rate relationship we use a power-law relationship of the form log$_{10}$($\dot{M}$/$M_\odot$~yr$^{-1}$)=$\alpha$+$\beta\times$log$_{10}$($100\times M_{dust \rm}/M_\odot$). We obtain intercepts of $-6.90\pm0.29$, $-6.88\pm0.29$, and $-4.93\pm0.22$, and slopes of $-0.15\pm0.12$, $-0.15\pm0.12$, and $1.11\pm0.08$, for respectively all Herbigs, only including the Herbig stars with a mass of $<4$~\Msun, and combining the latter with the T~Tauris, see the right panel of Fig.~\ref{fig:mdust_mdot}. Similar to \citet{GrantStapper2023}, we find a flat relationship for the Herbig stars, caused by a fraction of Herbig stars with relatively low disk masses. No difference is found for the relationship when including or excluding the higher mass Herbig stars, showing that the flat relationship is robust across the Herbig star population, and not only caused by the highest mass stars in our sample.

The dust masses are likely underestimating the total amount of material available in the disks. A comparison between the gas masses and dust masses of Herbig disks showed that the dust masses are likely underestimated by at least a factor of a few \citep{Stapper2024, Longarini2025}. This can even be as high as an order of magnitude in some disks \citep{Kaeufer2023, Liu2022, Radley2026}. However, in order to explain the Herbig disks with very short disk lifetimes, we note that the accretion rates may also be overestimated, as the accretion tracers can indeed originate from winds or outflows instead of accretion onto the star \citep[e.g.,][]{Kraus2008, Mendigutia2017}. However, as of now it is unclear how much impact this has on the determination of the real accretion rates. Furthermore, binarity can overestimate the accretion rate as well. Lastly, \citet{Brittain2026} and \citet{Mendigutia2026} showed that the accretion rates increase as Herbig stars evolve toward the main sequence. =While this does not solve our current problem, this would mean that a lower initial disk mass is necessary to account for these high accretion rates.

\begin{figure*}[t]
    \centering
    \includegraphics[width=\textwidth]{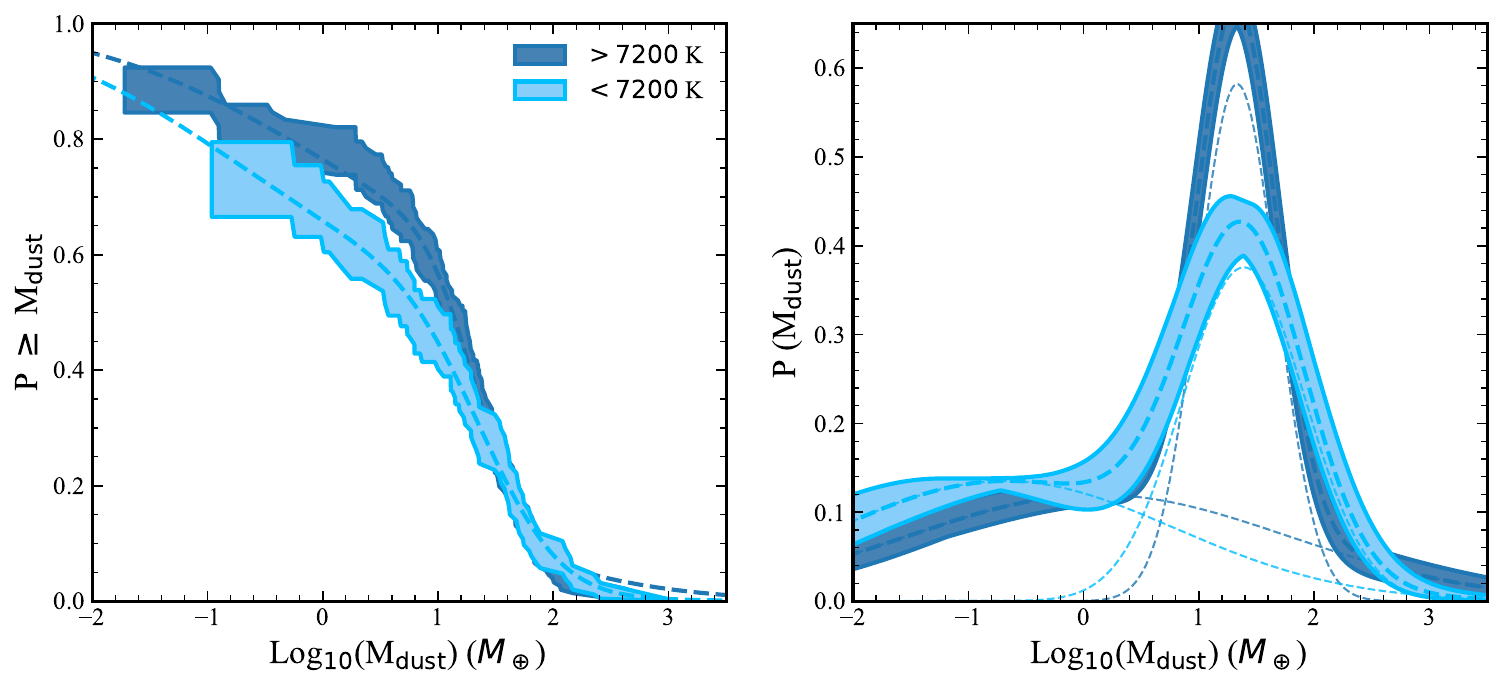}
    \caption{The cumulative and probability dust mass distributions of the disks around Herbig stars with an effective temperature below (also known as Intermediate Mass T~Tauri stars, \citealt{Calvet2004}) or above 7200~K, representing young and old pre-main sequence intermediate mass stars. All stars were selected to have masses of less than 4~\Msun. In the right panel, the two contributions have been plotted separately as dashed lines for each distribution. We see no significant differences in dust disk mass with age for the Herbig star population.}
    \label{fig:cdf_IMTTs}
\end{figure*}

Ultimately, as described in Sect. \ref{sec:target_selection}, our Herbig star sample was selected based on the presence of accretion signatures and infrared excess. Hence, our sample of Herbig stars is biased toward the highest accretors and most massive disks, which in part originates from the emission line criterion used for Herbig Ae/Be stars by \citet{Herbig1960}.  Additionally, low accretion rates for intermediate mass stars are notoriously difficult to measure due to their strong intrinsic ultraviolet emission, which coincides with the wavelength range in which accretion-shock emission peaks, and their deep Balmer lines \citep{Donehew2011, Brittain2023}. This results in the high accretion rates prevalent across our selected Herbig star population. Ideally, a sample of pre-main sequence intermediate mass stars, regardless of their accretion rate or IR excess level, should be constructed.

\subsection{The long-lived disks of Herbig stars}
\label{subsec:long_lived_disks}
Figure~\ref{fig:cdf_IMTTs} presents the disk dust mass cumulative and probability distributions for our sample with stellar masses of $<4$~\Msun, and divided by stellar effective temperature (above or below 7200~K, or a spectral type of F0). These intermediate mass stars with spectral types later than F0 are also known as Intermediate Mass T~Tauri stars \citep{Calvet2004, Valegard2021, Brittain2026}. Due the relatively horizontal evolutionary tracks for intermediate mass stars (see Fig.~\ref{fig:sky_and_HR_plot}), this division also indicates age; the later spectral type intermediate mass stars are younger compared to earlier spectral types. As was found by \citet{Stapper2025b}, Fig.~\ref{fig:cdf_IMTTs} shows that no significant differences are found between the dust masses of the disks around the young and old Herbig stars.

We hypothesize that our sample therefore represents the surviving disk-hosting population in the optically visible late stages of intermediate-mass star formation, as proposed by \citet{Stapper2025b}. Forming intermediate-mass stars will start with higher disk masses \citep{Longarini2025_infall}, and either retain bright disks until being relatively old, or dissipate them early on and are therefore already invisible to our selection criteria, which is supported by Fig.~\ref{fig:cdf_IMTTs}. This dichotomy is seen in both Figs.~\ref{fig:Mdust_Mstar} and \ref{fig:mdust_mdot}: we predominantly observe Herbig stars with high dust disk masses, which follow the upper end of the T~Tauri disk-mass distribution, and we are missing those with low disk masses ($\lesssim1$~\Mearth) and low accretion rates ($<10^{-8}$~\Msun~yr$^{-1}$). A volume limited study within 300~pc of intermediate mass objects with infrared excess done by \citet{Iglesias2022} found such objects: it resulted in 135 new pre-main sequence intermediate mass star candidates out of which only six could be considered hosting protoplanetary disks. Furthermore, there are many O, B, and A stars in nearby star forming regions \citep[e.g.,][]{vanTerwisga2022, Anania2025} that are most likely as young as the rest of the young stellar objects in that region that do not fit our Herbig selection criteria of emission lines or infrared excess. Together, these two observations show that the surviving disks we observe represent only a small fraction of a much larger underlying population of `boring' intermediate-mass young stars that have already dispersed most of their disks. This dichotomy could be due to the early development of strong dust traps in some sources, or the replenishment of material by late-infall in others, as will be discussed below.

\subsubsection{Containment via dust traps}

For a dust disk of a Herbig star to survive for a significant amount of time, formation of deep dust traps may be necessary. Indeed, many of the high mass disks in the Herbig disk population have large dust cavities \citep{Stapper2022}, and the disks around more massive stars tend to show more substructures in general than those around lower mass stars, which could be linked to the high prevalence of giant exoplanets \citep{vanderMarel2021, Bosschaart2026}. Such substructures influence the dust dynamics in disks around intermediate mass stars by trapping the dust and keeping the disk large compared to disks around lower mass stars \citep{Pinilla2020}. Indeed, some substructures like large inner dust cavities may be related to the presence of giant planets in Herbig stars (\citealp{Kama2015,GuzmanDiaz2023}). However, dust trapping does not necessarily need to be completely driven by giant exoplanets, as (sub)-stellar companions are common for Herbig stars \citep[$\sim50-70\%$, ][likely going up to 100\% for the highest mass stars]{Baines2006, Wheelwright2010, 2023AJ....165..135T,Garufi2026}. Such binaries, especially eccentric ones, can create large cavities in circumbinary disks \citep{Price2018, Calcino2019,2025A&A...698A.102R}.

Regardless of whether strong dust traps are created by giant exoplanets or stellar companions, such traps lead to a clear dichotomy in disk properties. Disks with strong dust traps retain higher masses, as the traps prevent efficient radial drift and allow large amounts of dust to remain at wide orbital radii. In contrast, disks without such traps become compact due to rapid radial drift, resulting in lower inferred disk masses and faster disk dispersal. This dichotomy is now clearly reflected in the dust mass cumulative and probability distribution shown in Fig.~\ref{fig:cdf_Mdust}. Recent studies support this scenario. \citet{Stapper2022} demonstrated that cavity-bearing disks generally have a high dust mass, whereas full or compact disks have lower dust masses. A compelling explanation is that giant planets have formed in cavity-bearing disks, preserving dust at larger radii, while full disks either still need to form such a cavity, or undergo unimpeded drift, steadily depleting their solid reservoir onto the star. Indeed, the only two directly imaged candidate planets in Herbig disks both reside in cavity-bearing disks (AB~Aur, \citealp{Currie2022, Bowler2025}; HD~169142, \citealp{Hammond2023}) lending direct support to this scenario. Further support comes from \citet{Stapper2025b}, who found that full disks around younger Herbig stars (the Intermediate Mass T~Tauris) have dust masses comparable to their cavity-bearing counterparts, consistent with drift having not yet run its course. Together, these results paint a coherent evolutionary picture in which the low and high disk mass dichotomy reflects two distinct pathways of Herbig disk evolution, set in motion by the early formation of strong dust traps, which result into the relatively old Herbig disk population of our catalog. A similar evolutionary scenario has been proposed for the disks in Ophiuchus \citep{Cieza2021, Orcajo2025, Bhowmik2026}. These objects may then provide the possibly primordial origin of the material in gas-rich debris disks, which are primarily found around A-type stars and can have ages up to 50~Myr \citep{Nakatani2023}.

The disks which cannot form the dust traps are likely part of the low disk mass component of the bimodal dust mass distribution presented in Fig.~\ref{fig:cdf_Mdust}, or do not fall within our selection criteria anymore. However, enhanced grain growth may also play a role here. In this scenario, the millimeter-sized grains, instead of having drifted onto the star, have been incorporated into much larger pebbles/planetesimals, thereby reducing the millimeter opacity. Hence, these disks may be transitioning from the Class~II to the Class~III phase \citep{Lada1987}. For example the debris disk HD~9672 (49~Ceti; \citealt{Marino2026}) is part of these low dust mass disks. Therefore, the low disk mass component of the bimodal distribution may partly consist of more evolved Class~III disks or even debris disks. However, this grain growth scenario likely explains at most a factor of a few change in the apparent flux at fixed disk mass.

\subsubsection{Replenishment via infall}

Another way to make Herbig disks survive for a long time is via replenishment by late-infall. In recent years multiple cases of late-infall streamers have been found, many of which are around intermediate mass stars such as AB~Aur \citep{Speedie2025, Calcino2025, Jiang2026}, SU~Aur \citep{Ginski2021}, S~CrA \citep{Gupta2024}, HL~Tau \citep{Garufi2022, Gupta2024}, GW~Ori \citep{GallowaySprietsma2026}, and indirect evidence in HD~142527 \citep{Temmink2026} and HD~34700A \citep{Stadler2026}.
Other signatures of late-infall were found around stars with a mass of at least $1$~\Msun \citep{Huang2021, Mesa2022}. Furthermore, in the all-sky census of disks with near-IR high-contrast images of \citet{Garufi2026}, ambient material is visible in more than 20\% of the sample, with a mild prevalence for intermediate-mass stars and higher accretors. This high incidence rate of late-infall around more massive stars should not be surprising. In dynamical systems the smaller objects get ejected first, while the more massive stars stay left behind in their natal cloud \citep{Bate2009, Price2009, Bate2012}. This is reflected in the original definition of Herbig stars, as they were associated with nearby nebulosity \citep{Herbig1960}. When this natal cloud is again accreted, \citet{Kuffmeier2023} found, based on MHD simulations of a molecular cloud, that the fraction of the stellar mass accreted by late-stage infall increases with stellar mass. Additionally, Bondi-Hoyle (BH) accretion of a turbulent and dense interstellar medium has recently been shown to create streamers similar to those observed \citep{Huhn2025}, and can even explain the observed sizes of protoplanetary disks \citep{Padoan2025}. Given that BH accretion scales with the stellar mass squared \citep{Bondi1944}, late-infall is expected to be relatively common around intermediate mass stars. This may also solve the mass budget problem for planet formation found in both Herbig disks \citep{Stapper2025b} and planet-forming disks in general \citep{Manara2018, Mulders2021}. Indeed, late-infall may facilitate the formation of distant giant exoplanets \citep{Zhao2026}. However, for the earlier spectral type stars it remains unclear whether late-stage infall can proceed efficiently, as they launch substantial photoevaporative winds that may suppress accretion flows.

Late-infall may also help in explaining the low disk mass, high accretion rate objects found in Fig.~\ref{fig:mdust_mdot}. As mentioned before, we expect our catalog to consist of the small fraction of pre-main sequence intermediate mass stars with a surviving disk, with the disk-less pre-main sequence stars dominating the population. If only a small fraction of these disk-less stars were to be stochastically replenished via infall, this could give rise to the sources at short disk lifetimes. In addition, the turbulent energy injected by such infall could also drive short-lived episodes of enhanced accretion, potentially helping to explain accretion timescales that are much shorter than nominal disk ages \citep[e.g.,][]{Winter2024}. 

To conclude, the fact that the Herbig population, following the definition of our sample, includes only objects with a high accretion rate, suggests that there may be an undiscovered population of intermediate mass pre-main sequence stars with low accretion rates and little or no disk remaining. This is supported by surveys identifying many intermediate-mass stars in nearby star forming regions that were not identified as Herbig stars. The Herbig star population in our catalog consists of the surviving massive disks, which are either retained by the formation of deep dust traps or replenished by late-infall, and do not seem to evolve over time. The sources for which we find short disk lifetimes may be transitioning to the Class~III phase, and/or are on the verge of dissipating their disks, or were part of the large fraction of disk-less stars and are currently being replenished by late-stage infall.

\subsection{Connecting to the exoplanet population}
\label{subsec:connection_to_exoplanets}

One of the profound results regarding disk populations in recent years is the observed trend between disk mass and stellar mass \citep[e.g.,][]{Andrews2013, Pascucci2016, Ansdell2016, Barenfeld2016, Ansdell2017}. For an increase in stellar mass, there is an increase in disk mass. While this is not particularly well visible in Fig.~\ref{fig:Mdust_Mstar} due to a relatively large spread, for each region separately there is a trend \citep{Ansdell2017,2022A&A...663A..98T}. 

\begin{figure*}[t]
    \centering
    \includegraphics[width=\textwidth]{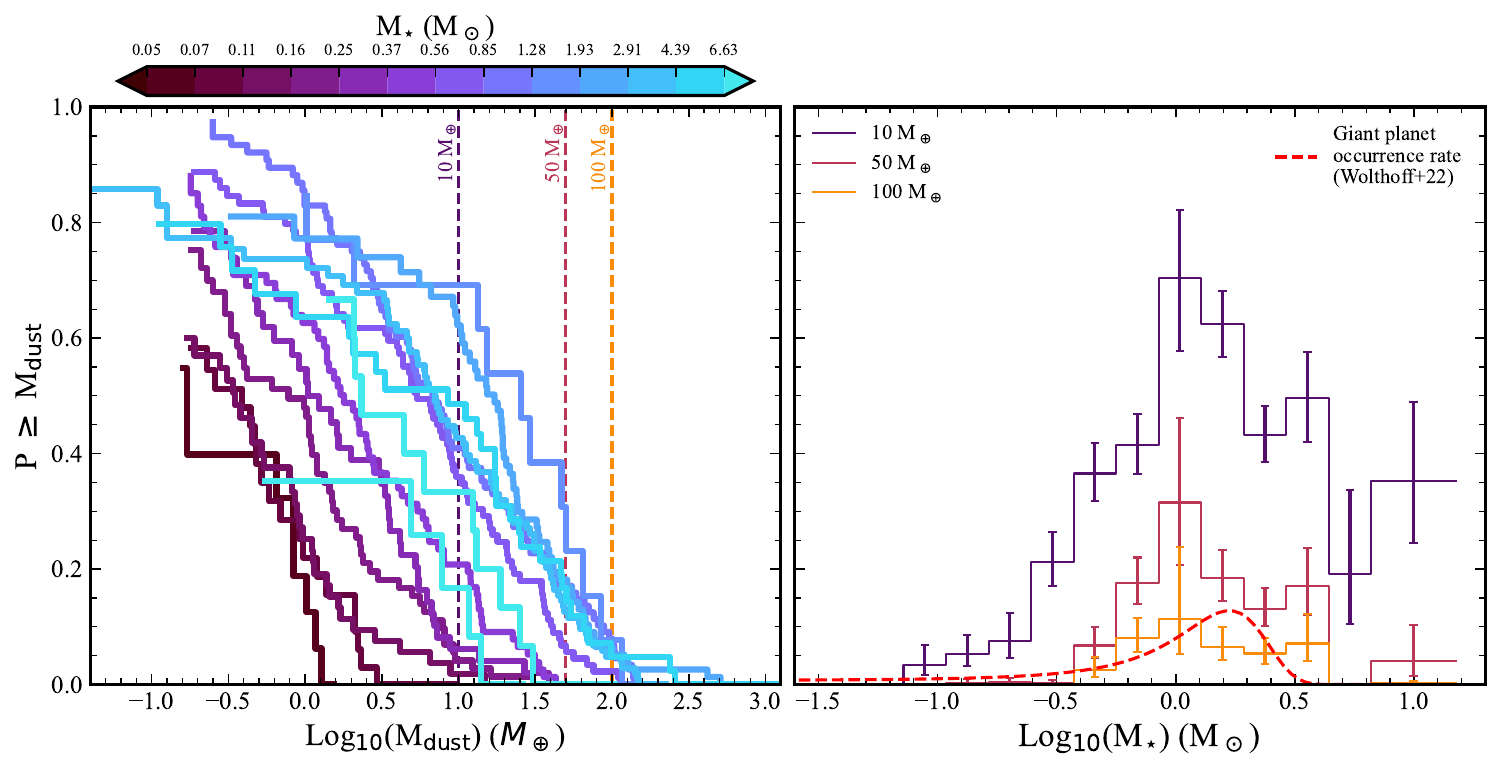}
    \caption{Cumulative dust disk mass distributions for different stellar mass bins (left), and the resulting fraction of disks per stellar mass bin larger than the disk mass as indicated in the left figure (right). The giant planet occurrence rate distribution of \citet{Wolthoff2022} is plotted for solar metallicity. The peak of the high disk mass occurrence rate coincides with that of the giant planet occurrence rate.}
    \label{fig:Mdisk_probability}
\end{figure*}

\begin{figure}[t]
    \centering
    \includegraphics[width=0.5\textwidth]{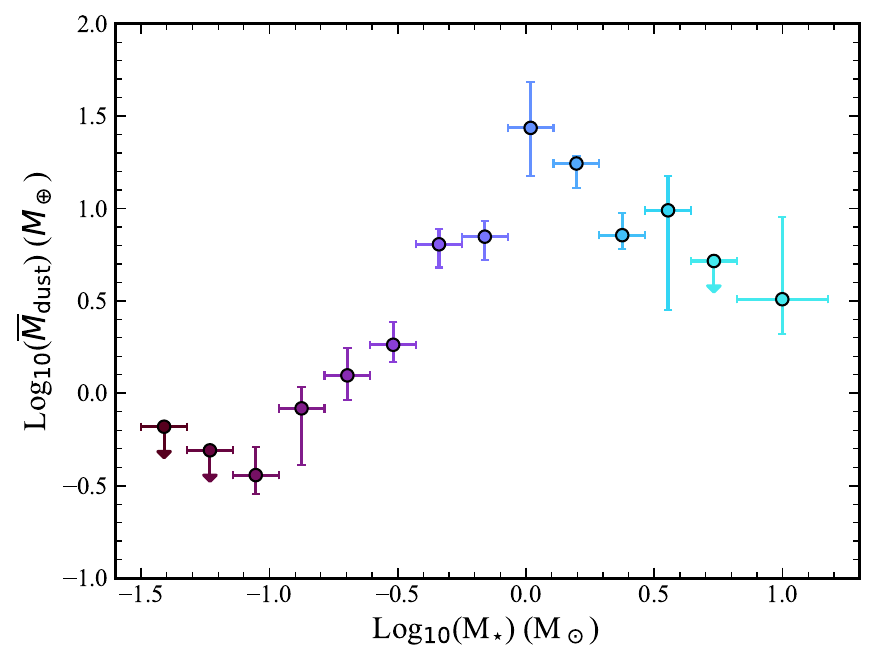}
    \caption{The median dust disk mass compared to the stellar mass determined from the dust mass distributions in the left panel of Fig.~\ref{fig:Mdisk_probability}. Colors correspond to the stellar mass bins in Fig.~\ref{fig:Mdisk_probability}, the width of each bin is indicated by the horizontal errorbar. The vertical errorbar gives the $1\sigma$ confidence intervals on the median dust mass. The median dust mass increases with stellar mass up until 1-2~\Msun, after which the median dust mass decreases.}
    \label{fig:median_Mdust_Mstar}
\end{figure}

This increase in disk mass is expected to stop for higher mass stars due to multiple different processes. First, the disk lifetime decreases due to an increase of internal photo-evaporation due to far ultraviolet (FUV) photons coming from the central star \citep{Kunitomo2021} or from nearby hot stars (\citealp{Anania2025,2026arXiv260222050P}). Regarding the inner disks, \citet{Vioque2018} and \citet{Cody2025} found a transition in inner disk properties at $\sim$7~\Msun\, based on optical variability, likely indicating an efficient inner disk dispersal at that stellar mass regime. Furthermore, the efficiency of radial drift from 1~Myr onward is also more efficient for higher mass stars \citep{Pinilla2022a}, which additionally has an effect on the ability to form planets via the streaming instability \citep{Das2025}. Lastly, the multiplicity also increases with stellar mass \citep{Offner2023}, which can lead to disk truncation and thus lower disk masses, and shorter lived disks \citep{Zagaria2023, Panic2021}. The more massive Herbig stars (>3-4~\Msun) without a magnetosphere may also be more efficient at accreting their newborn planets (\citealp{2024A&A...686L...1M}). Furthermore, recent modeling by \citet{Johnston2026} suggests that the peak in giant planet occurrence rates \citep{Wolthoff2022} arises due to a timescale problem: while higher stellar mass leads to higher accretion rates, and therefore higher pebble flux (which favors giant planet formation) the finite lifetime of the protoplanetary disk sets a limit. Consequently, the enhanced accretion rates around Herbig stars may boost giant planet formation, but only up to a point, as rapid disk dispersal halts further growth for the higher mass stars. All these effects could explain the dearth of exoplanet detections seen in the exoplanet population for stars >3~\Msun\;(e.g., \citealp{Reffert2015})

To our knowledge \citet{AlonsoAlbi2009} is the only previous work which tentatively shows a decrease in dust mass for disks around stars more massive than $\sim$3-4~\Msun. Relatively low disk masses were indeed found for four pre-main sequence stars with stellar masses ranging from 4-10~\Msun~in the giant HII region M17 at a distance of 1.7~kpc \citep[][]{Poorta2025}. On the other hand, SMA observations presented in \citet{Wilner2026} resulted in the detection of five relatively massive Herbig disks around B-type stars, on the order of 10s to 100s of Earth masses, out of a sample of 24 sources. The wealth of data gathered in our work can be used to assess whether this decrease in disk mass is also present in our data. At $\sim4$~\Msun there is a clear jump to lower disk dust masses in Fig.~\ref{fig:Mdust_Mstar}, after which the disk masses are typically on the order of a couple of Earth masses.

To quantify the dust mass dependency on stellar mass, we generate cumulative dust mass distributions using \texttt{lifelines} for different logarithmically spaced stellar mass bins for a dataset which combines the data of \citet{Manara2023} with ours. This results in a total of 749 disks around stars with masses ranging from 0.023~\Msun~to 12.8~\Msun. The resulting cumulative distributions are shown in the left panel of Fig.~\ref{fig:Mdisk_probability}. The errors are not shown as the plot would become unreadable, but the typical $1\sigma$ error can be seen in the right panel of the same figure. Based on the generated distributions, we can see that the median dust mass of each distribution differs depending on the stellar mass. We plot the median dust disk mass of each distribution in Fig.~\ref{fig:median_Mdust_Mstar}. We find an increase of disk dust mass with stellar mass up to 1-2~\Msun, and then a decrease with stellar mass. We note that the peak at 1~\Msun~in Fig.~\ref{fig:median_Mdust_Mstar} is likely due to a scarcity of measurements around that stellar mass. In addition we note that the decrease in dust mass beyond 2~\Msun~could be partially attributed to the assumption that the dust temperature scales with the stellar luminosity (see Section~\ref{subsec:get_dust_masses}). However, using less steep relationships such as those in \citet{vanderPlas2016} or \citet{Deng2026}, both based on radiative transfer models, results in a similar decrease. Only under the incorrect assumption of a constant 20~K dust temperature across the sample does the median dust mass appear constant after 2~\Msun, rather than decreasing.

With the stellar mass dependent cumulative distributions in the left panel of Fig.~\ref{fig:Mdisk_probability}, we compute the fraction of disks with a particular minimum mass. These are shown in the right panel of Fig.~\ref{fig:Mdisk_probability} for masses of 10~\Mearth, 50~\Mearth, and 100~\Mearth. The peak of the distribution of disks over 10~\Mearth\, in mass lies at 1-2~\Msun, before and after which there is a decrease in the occurrence rate of these disks. Hence, the 1-2~\Msun\, stellar mass range is therefore likely the optimal range for efficient giant planet formation. The colored distributions in the right panel of Fig.~\ref{fig:Mdisk_probability} indicate the giant exoplanet occurrence rate from \citet{Wolthoff2022} for solar metallicity stars. The peak of the giant exoplanet occurrence rate coincides with the peak of the occurrence rate of massive disks. This can mean two things. Either the giant exoplanets have already formed in these disks, keeping the disks large and therefore resulting in a higher inferred disk mass, or giant planets are currently forming because of the high disk masses. The latter scenario assumes that the disk masses are underestimated, as the typical disk masses are not sufficient to form the observed giant exoplanet systems \citep{Stapper2025b}. Hence, this would support the scenario where the giant planets have already formed. To distinguish between these two scenarios, better disk mass estimates would be necessary. This could be achieved via multi-wavelength observations (see, e.g., \citealt{Painter2025}).

\section{Conclusion}
\label{sec:conclusion}
In this work we present a catalog of all well-known 243 Herbig stars within 1~kpc, containing archival, recalibrated, and newly derived stellar parameters, accretion rate properties, and disk masses. To obtain the latter, we also present new millimeter photometry observations made with ALMA, ACA, the SMA, and NOEMA. Our results can be summarized as follows:

\begin{enumerate}
    \item Disks around Herbig stars are on average more massive than the disks around T~Tauri stars, but do cover the same dust mass range. We find that 50\% of the Herbig disks are more massive than 10~\Mearth, while this is only true for 20\% and 5\% of the T~Tauri disks in Lupus and Upper~Scorpius respectively.
    \item We find that a bimodal dust mass distribution fits the Herbig disk population well. One part of the population consists of high mass disks with a mean dust mass of 21~\Mearth, and the other consists of low mass disks with a mean dust mass of only 0.45~\Mearth.
    \item The secondary peak at low disk masses in the dust mass distribution of the Herbig disks may be caused by disks transitioning toward the Class~III phase due to grain growth and radial drift. On the other hand, late-infall may also stochastically replenish these low mass disks, which may originate from the large population of disk-less pre-main sequence intermediate mass stars.
    \item Using constraints on the star formation rate in the solar neighborhood, we obtain the completeness of our catalog by estimating the time that a pre-main sequence intermediate mass star is recognized as a Herbig star, its so-called `lifetime'. We find that our catalog, while close to complete inside 300~pc, misses approximately 76\% of the expected number of Herbig stars within 1~kpc. Hence, a large fraction of stars which meet our selection criteria (i.e., with clear infrared excess and often accretion signatures) are still missing in our catalog. We find that the lifetime of a Herbig star inversely depends on its mass (slope of $-1.4$), and is estimated to be 3~Myr for stars with masses of 3-4~\Msun, or close to 7~Myr extrapolated to 1.5~\Msun.
    \item We compare the 180 accretion rates in our catalog to the stellar masses and observe the break at 4~\Msun~found by previous works. We find a slightly steeper relationship for the Herbig stars compared to T~Tauri stars. 
    \item We find a flat relationship for the Herbig stars when comparing the stellar accretion rates with their dust disk masses. The Herbig stars generally lie above the T~Tauri star relationship, and the flatness of the relationship is driven by a subset of Herbig stars surrounded by low dust mass disks, but still have high accretion rates. This results in estimated disk lifetimes of less than 10~kyr.
    \item As very few Herbig stars with low accretion rates ($<10^{-8}$~\Msun~yr$^{-1}$) or disk masses ($\lesssim1$~\Mearth) are found, and no significant changes in the disk dust mass distributions with age was found leads us to the following hypothesis. Our sample represents the surviving disk-hosting population of intermediate-mass pre-main sequence stars, drawn from a much larger population of intermediate-mass stars with no significant disks which fall outside our Herbig selection criteria. We argue that this is either due to the formation of deep dust traps in these disks or replenishment by late-infall, or a combination of the two.
    \item We find an increase of disk dust mass with stellar mass up to 1-2~\Msun, and then a decrease with stellar mass. This may be caused by the increase in internal UV irradiation, efficient radial drift, and/or multiplicity.
    \item The peak of the occurrence rate of massive disks ($>10$~\Mearth) is at $1-2$~\Msun, which coincides with the peak of the occurrence rate of giant exoplanets.
\end{enumerate}

Further efforts need to be made toward characterizing the low disk mass, high accretion rate objects in the Herbig disk population to properly understand their origin. Additionally, the missing Herbig stars within 1~kpc should be identified and added to the current Herbig star catalogs. Expanding the currently available sample by a factor of four would greatly improve statistics and help identify the influence of age and environment on the evolution of disks around intermediate mass stars. Lastly, to better understand the evolution of intermediate mass stars, comparisons with embedded objects would be vital to determine when the disk-hosting and disk-less dichotomy arises in the disks around intermediate mass stars.

\begin{acknowledgements}
LMS and MB have received funding from the European Research Council (ERC) under the European Union’s Horizon 2020 research and innovation programme (PROTOPLANETS, grant agreement No. 101002188). 
This paper makes use of the following ALMA data: ADS/JAO.ALMA\#2012.1.00195.S, \#2012.1.00631.S, \#2015.1.00222.S, \#2015.1.00350.S, \#2015.1.01600.S, \#2016.1.00110.S, \#2016.1.00344.S, \#2016.1.00484.L, \#2016.1.00826.S, \#2016.1.01344.S, \#2017.1.00286.S, \#2017.1.00466.S, \#2017.1.00492.S, \#2017.1.01404.S, \#2017.1.01460.S, \#2017.1.01578.S, \#2017.1.01678.S, \#2018.1.00689.S, \#2018.1.00814.S, \#2018.1.01055.L, \#2018.1.01066.S, \#2018.1.01302.S, \#2018.1.01309.S, \#2019.1.00703.S, \#2019.1.01167.S, \#2019.1.01210.S, \#2019.1.01693.S, \#2019.1.01792.S, \#2019.1.01813.S, \#2021.1.00994.S, \#2021.1.01137.S, \#2021.1.01661.S, \#2021.1.01705.S, \#2021.2.00005.S, \#2022.1.00313.S, \#2022.1.00760.S, \#2022.1.01155.S, \#2022.1.01302.S, \#2022.1.01365.S, \#2022.1.01460.S, \#2023.1.00561.S, \#2023.1.00937.S, \#2023.1.01149.S, \#2023.1.01379.S, \#2024.1.00408.S. ALMA is a partnership of ESO (representing its member states), NSF (USA) and NINS (Japan), together with NRC (Canada), MOST and ASIAA (Taiwan), and KASI (Republic of Korea), in cooperation with the Republic of Chile. The Joint ALMA Observatory is operated by ESO, AUI/NRAO and NAOJ. 
The Submillimeter Array is a joint project between the Smithsonian Astrophysical Observatory and the Academia Sinica Institute of Astronomy and Astrophysics and is funded by the Smithsonian Institution and the Academia Sinica. We recognize that Maunakea is a culturally important site for the indigenous Hawaiian people; we are privileged to study the cosmos from its summit.
This work is based on observations carried out under project numbers S21AS, S22AU, and W25BM with the IRAM NOEMA Interferometer. IRAM is supported by INSU/CNRS (France), MPG (Germany) and IGN (Spain). We would like to thank Chiara Circosta as our local contact at IRAM.
This work has made use of data from the European Space Agency (ESA) mission Gaia (\url{https://www.cosmos.esa.int/gaia}), processed by the Gaia Data Processing and Analysis Consortium (DPAC, \url{https://www.cosmos.esa.int/web/gaia/dpac/consortium}). 
We acknowledge project support from the Max Planck Computing and Data Facility.
This work makes use of the following software: The Common Astronomy Software Applications (CASA) package \citep{CASA2022}, Python version 3.14.3, astropy \citep{astropy2013, astropy2018}, astroquery \citep{Ginsburg2019}, cmasher \citep{cmasher}, lifelines \citep{DavidsonPilon2021}, linmix \citep{Kelly2007}, matplotlib \citep{Hunter2007}, numpy \citep{Harris2020}, pandas \citep{reback2020pandas}, pymc \citep{pymc2023}, and scipy \citep{2020SciPy-NMeth}.

\end{acknowledgements}

\bibliographystyle{aa}
\bibliography{references.bib}

\begin{thebibliography}{242}
\expandafter\ifx\csname natexlab\endcsname\relax\def\natexlab#1{#1}\fi

\bibitem[{Abril-Pla {et~al.}(2023)Abril-Pla, Andreani, Carroll, Dong, Fonnesbeck, Kochurov, Kumar, Lao, Luhmann, Martin, Osthege, Vieira, Wiecki, \& Zinkov}]{pymc2023}
Abril-Pla, O., Andreani, V., Carroll, C., {et~al.} 2023, \href{http://dx.doi.org/10.7717/peerj-cs.1516}{\color{blue}{PeerJ} Computer Science}, 9, 9

\bibitem[{{Alaguero} {et~al.}(2024){Alaguero}, {Cuello}, {M{\'e}nard}, {Ceppi}, {Ribas}, {Nealon}, {Vioque}, {Izquierdo}, {Miley}, {Mac{\'\i}as}, \& {Price}}]{Alaguero2024}
{Alaguero}, A., {Cuello}, N., {M{\'e}nard}, F., {et~al.} 2024, \href{http://dx.doi.org/10.1051/0004-6361/202449683}{\color{blue}\aap}, \href{https://ui.adsabs.harvard.edu/abs/2024A&A...687A.311A}{687, A311}

\bibitem[{{Alaguero} {et~al.}(2025){Alaguero}, {M{\'e}nard}, {Cuello}, {Ribas}, {Viscardi}, {Mac{\'\i}as}, {Vioque}, \& {Miley}}]{Alaguero2025}
{Alaguero}, A., {M{\'e}nard}, F., {Cuello}, N., {et~al.} 2025, \href{http://dx.doi.org/10.1051/0004-6361/202556077}{\color{blue}\aap}, \href{https://ui.adsabs.harvard.edu/abs/2025A&A...703A.210A}{703, A210}

\bibitem[{{Allard} {et~al.}(2011){Allard}, {Homeier}, \& {Freytag}}]{2011ASPC..448...91A}
{Allard}, F., {Homeier}, D., \& {Freytag}, B. 2011, in Astronomical Society of the Pacific Conference Series, Vol. 448, 16th Cambridge Workshop on Cool Stars, Stellar Systems, and the Sun, ed. C.~{Johns-Krull}, M.~K. {Browning}, \& A.~A. {West}, \href{https://ui.adsabs.harvard.edu/abs/2011ASPC..448...91A}{91}

\bibitem[{{Alonso-Albi} {et~al.}(2009){Alonso-Albi}, {Fuente}, {Bachiller}, {Neri}, {Planesas}, {Testi}, {Bern{\'e}}, \& {Joblin}}]{AlonsoAlbi2009}
{Alonso-Albi}, T., {Fuente}, A., {Bachiller}, R., {et~al.} 2009, \href{http://dx.doi.org/10.1051/0004-6361/200810401}{\color{blue}\aap}, \href{https://ui.adsabs.harvard.edu/abs/2009A&A...497..117A}{497, 117}

\bibitem[{{Anania} {et~al.}(2025{\natexlab{a}}){Anania}, {Rosotti}, {G{\'a}rate}, {Pinilla}, {Vioque}, {Trapman}, {Carpenter}, {Zhang}, {Pascucci}, {Cieza}, {Sierra}, {Kurtovic}, {Miley}, {P{\'e}rez}, {Tabone}, {Hogerheijde}, {Deng}, {Agurto-Gangas}, {Ruiz-Rodriguez}, {Gonz{\'a}lez-Ruilova}, \& {TorresVillanueva}}]{Anania2025AGEPRO}
{Anania}, R., {Rosotti}, G.~P., {G{\'a}rate}, M., {et~al.} 2025{\natexlab{a}}, \href{http://dx.doi.org/10.3847/1538-4357/adb587}{\color{blue}\apj}, \href{https://ui.adsabs.harvard.edu/abs/2025ApJ...989....8A}{989, 8}

\bibitem[{{Anania} {et~al.}(2025{\natexlab{b}}){Anania}, {Winter}, {Rosotti}, {Vioque}, {Zari}, {Pantaleoni Gonz{\'a}lez}, \& {Testi}}]{Anania2025}
{Anania}, R., {Winter}, A.~J., {Rosotti}, G., {et~al.} 2025{\natexlab{b}}, \href{http://dx.doi.org/10.1051/0004-6361/202453011}{\color{blue}\aap}, \href{https://ui.adsabs.harvard.edu/abs/2025A&A...695A..74A}{695, A74}

\bibitem[{{Andrews} {et~al.}(2018){Andrews}, {Huang}, {P{\'e}rez}, {Isella}, {Dullemond}, {Kurtovic}, {Guzm{\'a}n}, {Carpenter}, {Wilner}, {Zhang}, {Zhu}, {Birnstiel}, {Bai}, {Benisty}, {Hughes}, {{\"O}berg}, \& {Ricci}}]{Andrews2018b}
{Andrews}, S.~M., {Huang}, J., {P{\'e}rez}, L.~M., {et~al.} 2018, \href{http://dx.doi.org/10.3847/2041-8213/aaf741}{\color{blue}\apjl}, \href{https://ui.adsabs.harvard.edu/abs/2018ApJ...869L..41A}{869, L41}

\bibitem[{{Andrews} {et~al.}(2013){Andrews}, {Rosenfeld}, {Kraus}, \& {Wilner}}]{Andrews2013}
{Andrews}, S.~M., {Rosenfeld}, K.~A., {Kraus}, A.~L., \& {Wilner}, D.~J. 2013, \href{http://dx.doi.org/10.1088/0004-637X/771/2/129}{\color{blue}\apj}, \href{https://ui.adsabs.harvard.edu/abs/2013ApJ...771..129A}{771, 129}

\bibitem[{{Ansdell} {et~al.}(2017){Ansdell}, {Williams}, {Manara}, {Miotello}, {Facchini}, {van der Marel}, {Testi}, \& {van Dishoeck}}]{Ansdell2017}
{Ansdell}, M., {Williams}, J.~P., {Manara}, C.~F., {et~al.} 2017, \href{http://dx.doi.org/10.3847/1538-3881/aa69c0}{\color{blue}\aj}, \href{https://ui.adsabs.harvard.edu/abs/2017AJ....153..240A}{153, 240}

\bibitem[{{Ansdell} {et~al.}(2018){Ansdell}, {Williams}, {Trapman}, {van Terwisga}, {Facchini}, {Manara}, {van der Marel}, {Miotello}, {Tazzari}, {Hogerheijde}, {Guidi}, {Testi}, \& {van Dishoeck}}]{Ansdell2018}
{Ansdell}, M., {Williams}, J.~P., {Trapman}, L., {et~al.} 2018, \href{http://dx.doi.org/10.3847/1538-4357/aab890}{\color{blue}\apj}, \href{https://ui.adsabs.harvard.edu/abs/2018ApJ...859...21A}{859, 21}

\bibitem[{{Ansdell} {et~al.}(2016){Ansdell}, {Williams}, {van der Marel}, {Carpenter}, {Guidi}, {Hogerheijde}, {Mathews}, {Manara}, {Miotello}, {Natta}, {Oliveira}, {Tazzari}, {Testi}, {van Dishoeck}, \& {van Terwisga}}]{Ansdell2016}
{Ansdell}, M., {Williams}, J.~P., {van der Marel}, N., {et~al.} 2016, \href{http://dx.doi.org/10.3847/0004-637X/828/1/46}{\color{blue}\apj}, \href{https://ui.adsabs.harvard.edu/abs/2016ApJ...828...46A}{828, 46}

\bibitem[{{Asplund} {et~al.}(2009){Asplund}, {Grevesse}, {Sauval}, \& {Scott}}]{2009ARA&A..47..481A}
{Asplund}, M., {Grevesse}, N., {Sauval}, A.~J., \& {Scott}, P. 2009, \href{http://dx.doi.org/10.1146/annurev.astro.46.060407.145222}{\color{blue}\araa}, \href{https://ui.adsabs.harvard.edu/abs/2009ARA&A..47..481A}{47, 481}

\bibitem[{{Astropy Collaboration} {et~al.}(2018){Astropy Collaboration}, {Price-Whelan}, {Sip{\H{o}}cz}, {G{\"u}nther}, {Lim}, {Crawford}, {Conseil}, {Shupe}, {Craig}, {Dencheva}, {Ginsburg}, {Vand erPlas}, {Bradley}, {P{\'e}rez-Su{\'a}rez}, {de Val-Borro}, {Aldcroft}, {Cruz}, {Robitaille}, {Tollerud}, {Ardelean}, {Babej}, {Bach}, {Bachetti}, {Bakanov}, {Bamford}, {Barentsen}, {Barmby}, {Baumbach}, {Berry}, {Biscani}, {Boquien}, {Bostroem}, {Bouma}, {Brammer}, {Bray}, {Breytenbach}, {Buddelmeijer}, {Burke}, {Calderone}, {Cano Rodr{\'\i}guez}, {Cara}, {Cardoso}, {Cheedella}, {Copin}, {Corrales}, {Crichton}, {D'Avella}, {Deil}, {Depagne}, {Dietrich}, {Donath}, {Droettboom}, {Earl}, {Erben}, {Fabbro}, {Ferreira}, {Finethy}, {Fox}, {Garrison}, {Gibbons}, {Goldstein}, {Gommers}, {Greco}, {Greenfield}, {Groener}, {Grollier}, {Hagen}, {Hirst}, {Homeier}, {Horton}, {Hosseinzadeh}, {Hu}, {Hunkeler}, {Ivezi{\'c}}, {Jain}, {Jenness}, {Kanarek}, {Kendrew}, {Kern}, {Kerzendorf}, {Khvalko}, {King}, {Kirkby}, {Kulkarni},
  {Kumar}, {Lee}, {Lenz}, {Littlefair}, {Ma}, {Macleod}, {Mastropietro}, {McCully}, {Montagnac}, {Morris}, {Mueller}, {Mumford}, {Muna}, {Murphy}, {Nelson}, {Nguyen}, {Ninan}, {N{\"o}the}, {Ogaz}, {Oh}, {Parejko}, {Parley}, {Pascual}, {Patil}, {Patil}, {Plunkett}, {Prochaska}, {Rastogi}, {Reddy Janga}, {Sabater}, {Sakurikar}, {Seifert}, {Sherbert}, {Sherwood-Taylor}, {Shih}, {Sick}, {Silbiger}, {Singanamalla}, {Singer}, {Sladen}, {Sooley}, {Sornarajah}, {Streicher}, {Teuben}, {Thomas}, {Tremblay}, {Turner}, {Terr{\'o}n}, {van Kerkwijk}, {de la Vega}, {Watkins}, {Weaver}, {Whitmore}, {Woillez}, {Zabalza}, \& {Astropy Contributors}}]{astropy2018}
{Astropy Collaboration}, {Price-Whelan}, A.~M., {Sip{\H{o}}cz}, B.~M., {et~al.} 2018, \href{http://dx.doi.org/10.3847/1538-3881/aabc4f}{\color{blue}\aj}, \href{https://ui.adsabs.harvard.edu/abs/2018AJ....156..123A}{156, 123}

\bibitem[{{Astropy Collaboration} {et~al.}(2013){Astropy Collaboration}, {Robitaille}, {Tollerud}, {Greenfield}, {Droettboom}, {Bray}, {Aldcroft}, {Davis}, {Ginsburg}, {Price-Whelan}, {Kerzendorf}, {Conley}, {Crighton}, {Barbary}, {Muna}, {Ferguson}, {Grollier}, {Parikh}, {Nair}, {Unther}, {Deil}, {Woillez}, {Conseil}, {Kramer}, {Turner}, {Singer}, {Fox}, {Weaver}, {Zabalza}, {Edwards}, {Azalee Bostroem}, {Burke}, {Casey}, {Crawford}, {Dencheva}, {Ely}, {Jenness}, {Labrie}, {Lim}, {Pierfederici}, {Pontzen}, {Ptak}, {Refsdal}, {Servillat}, \& {Streicher}}]{astropy2013}
{Astropy Collaboration}, {Robitaille}, T.~P., {Tollerud}, E.~J., {et~al.} 2013, \href{http://dx.doi.org/10.1051/0004-6361/201322068}{\color{blue}\aap}, \href{http://adsabs.harvard.edu/abs/2013A%26A...558A..33A}{558, A33}

\bibitem[{{Bae} {et~al.}(2023){Bae}, {Isella}, {Zhu}, {Martin}, {Okuzumi}, \& {Suriano}}]{Bae2023}
{Bae}, J., {Isella}, A., {Zhu}, Z., {et~al.} 2023, in Astronomical Society of the Pacific Conference Series, Vol. 534, Protostars and Planets VII, ed. S.~{Inutsuka}, Y.~{Aikawa}, T.~{Muto}, K.~{Tomida}, \& M.~{Tamura}, \href{https://ui.adsabs.harvard.edu/abs/2023ASPC..534..423B}{423}

\bibitem[{{Bailer-Jones} {et~al.}(2021){Bailer-Jones}, {Rybizki}, {Fouesneau}, {Demleitner}, \& {Andrae}}]{2021AJ....161..147B}
{Bailer-Jones}, C.~A.~L., {Rybizki}, J., {Fouesneau}, M., {Demleitner}, M., \& {Andrae}, R. 2021, \href{http://dx.doi.org/10.3847/1538-3881/abd806}{\color{blue}\aj}, \href{https://ui.adsabs.harvard.edu/abs/2021AJ....161..147B}{161, 147}

\bibitem[{{Baines} {et~al.}(2006){Baines}, {Oudmaijer}, {Porter}, \& {Pozzo}}]{Baines2006}
{Baines}, D., {Oudmaijer}, R.~D., {Porter}, J.~M., \& {Pozzo}, M. 2006, \href{http://dx.doi.org/10.1111/j.1365-2966.2006.10006.x}{\color{blue}\mnras}, \href{https://ui.adsabs.harvard.edu/abs/2006MNRAS.367..737B}{367, 737}

\bibitem[{{Barenfeld} {et~al.}(2016){Barenfeld}, {Carpenter}, {Ricci}, \& {Isella}}]{Barenfeld2016}
{Barenfeld}, S.~A., {Carpenter}, J.~M., {Ricci}, L., \& {Isella}, A. 2016, \href{http://dx.doi.org/10.3847/0004-637X/827/2/142}{\color{blue}\apj}, \href{https://ui.adsabs.harvard.edu/abs/2016ApJ...827..142B}{827, 142}

\bibitem[{{Bate}(2009)}]{Bate2009}
{Bate}, M.~R. 2009, \href{http://dx.doi.org/10.1111/j.1365-2966.2008.14106.x}{\color{blue}\mnras}, \href{https://ui.adsabs.harvard.edu/abs/2009MNRAS.392..590B}{392, 590}

\bibitem[{{Bate}(2012)}]{Bate2012}
{Bate}, M.~R. 2012, \href{http://dx.doi.org/10.1111/j.1365-2966.2011.19955.x}{\color{blue}\mnras}, \href{https://ui.adsabs.harvard.edu/abs/2012MNRAS.419.3115B}{419, 3115}

\bibitem[{{Beckwith} {et~al.}(1990){Beckwith}, {Sargent}, {Chini}, \& {Guesten}}]{Beckwith1990}
{Beckwith}, S. V.~W., {Sargent}, A.~I., {Chini}, R.~S., \& {Guesten}, R. 1990, \href{http://dx.doi.org/10.1086/115385}{\color{blue}\aj}, \href{https://ui.adsabs.harvard.edu/abs/1990AJ.....99..924B}{99, 924}

\bibitem[{{Benisty} {et~al.}(2023){Benisty}, {Dominik}, {Follette}, {Garufi}, {Ginski}, {Hashimoto}, {Keppler}, {Kley}, \& {Monnier}}]{Benisty2023}
{Benisty}, M., {Dominik}, C., {Follette}, K., {et~al.} 2023, in Astronomical Society of the Pacific Conference Series, Vol. 534, Protostars and Planets VII, ed. S.~{Inutsuka}, Y.~{Aikawa}, T.~{Muto}, K.~{Tomida}, \& M.~{Tamura}, \href{https://ui.adsabs.harvard.edu/abs/2023ASPC..534..605B}{605}

\bibitem[{{Benisty} {et~al.}(2015){Benisty}, {Juhasz}, {Boccaletti}, {Avenhaus}, {Milli}, {Thalmann}, {Dominik}, {Pinilla}, {Buenzli}, {Pohl}, {Beuzit}, {Birnstiel}, {de Boer}, {Bonnefoy}, {Chauvin}, {Christiaens}, {Garufi}, {Grady}, {Henning}, {Huelamo}, {Isella}, {Langlois}, {M{\'e}nard}, {Mouillet}, {Olofsson}, {Pantin}, {Pinte}, \& {Pueyo}}]{Benisty2015}
{Benisty}, M., {Juhasz}, A., {Boccaletti}, A., {et~al.} 2015, \href{http://dx.doi.org/10.1051/0004-6361/201526011}{\color{blue}\aap}, \href{https://ui.adsabs.harvard.edu/abs/2015A&A...578L...6B}{578, L6}

\bibitem[{{Bhowmik} {et~al.}(2026){Bhowmik}, {Cieza}, {Miley}, {Nogueira}, {Gonz{\'a}lez-Ruilova}, {Chavan}, {Sierra}, {Dasgupta}, {Casassus}, {Batalla-Falcon}, {Di Lernia}, {Hales}, {Jennings}, {Orcajo}, {Perez}, {Ru{\'\i}z-Rodriguez}, {Shi}, {Williams}, {Zhang}, \& {Zurlo}}]{Bhowmik2026}
{Bhowmik}, T., {Cieza}, L., {Miley}, J.~M., {et~al.} 2026, \href{https://ui.adsabs.harvard.edu/abs/2026arXiv260419246B}{\href{http://dx.doi.org/10.48550/arXiv.2604.19246}{\color{blue}arXiv e-prints}, arXiv:2604.19246}

\bibitem[{{Bondi} \& {Hoyle}(1944)}]{Bondi1944}
{Bondi}, H. \& {Hoyle}, F. 1944, \href{http://dx.doi.org/10.1093/mnras/104.5.273}{\color{blue}\mnras}, \href{https://ui.adsabs.harvard.edu/abs/1944MNRAS.104..273B}{104, 273}

\bibitem[{{Booth} {et~al.}(2026){Booth}, {Calahan}, {Temmink}, {W{\"o}lfer}, {Pegues}, {Law}, {Evans}, {Leemker}, {Notsu}, {{\"O}berg}, {Walsh}, \& {van Dishoeck}}]{Booth2026}
{Booth}, A.~S., {Calahan}, J., {Temmink}, M., {et~al.} 2026, \href{http://dx.doi.org/10.3847/1538-3881/ae286b}{\color{blue}\aj}, \href{https://ui.adsabs.harvard.edu/abs/2026AJ....171..128B}{171, 128}

\bibitem[{{Booth} {et~al.}(2023){Booth}, {Ilee}, {Walsh}, {Kama}, {Keyte}, {van Dishoeck}, \& {Nomura}}]{Booth2023}
{Booth}, A.~S., {Ilee}, J.~D., {Walsh}, C., {et~al.} 2023, \href{http://dx.doi.org/10.1051/0004-6361/202244472}{\color{blue}\aap}, \href{https://ui.adsabs.harvard.edu/abs/2023A&A...669A..53B}{669, A53}

\bibitem[{{Booth} {et~al.}(2024){Booth}, {Leemker}, {van Dishoeck}, {Evans}, {Ilee}, {Kama}, {Keyte}, {Law}, {van der Marel}, {Nomura}, {Notsu}, {{\"O}berg}, {Temmink}, \& {Walsh}}]{Booth2024}
{Booth}, A.~S., {Leemker}, M., {van Dishoeck}, E.~F., {et~al.} 2024, \href{http://dx.doi.org/10.3847/1538-3881/ad2700}{\color{blue}\aj}, \href{https://ui.adsabs.harvard.edu/abs/2024AJ....167..164B}{167, 164}

\bibitem[{{Booth} {et~al.}(2021){Booth}, {Walsh}, {Terwisscha van Scheltinga}, {van Dishoeck}, {Ilee}, {Hogerheijde}, {Kama}, \& {Nomura}}]{Booth2021a}
{Booth}, A.~S., {Walsh}, C., {Terwisscha van Scheltinga}, J., {et~al.} 2021, \href{http://dx.doi.org/10.1038/s41550-021-01352-w}{\color{blue}Nature Astronomy}, \href{https://ui.adsabs.harvard.edu/abs/2021NatAs...5..684B}{5, 684}

\bibitem[{{Booth} {et~al.}(2025){Booth}, {W{\"o}lfer}, {Temmink}, {Calahan}, {Evans}, {Law}, {Leemker}, {Notsu}, {{\"O}berg}, \& {Walsh}}]{Booth2025}
{Booth}, A.~S., {W{\"o}lfer}, L., {Temmink}, M., {et~al.} 2025, \href{http://dx.doi.org/10.3847/2041-8213/adc7b2}{\color{blue}\apjl}, \href{https://ui.adsabs.harvard.edu/abs/2025ApJ...986L...9B}{986, L9}

\bibitem[{{Bosschaart} {et~al.}(2026){Bosschaart}, {Guerra-Alvarado}, {van der Marel}, \& {Mulders}}]{Bosschaart2026}
{Bosschaart}, Q., {Guerra-Alvarado}, O.~M., {van der Marel}, N., \& {Mulders}, G.~D. 2026, \href{http://dx.doi.org/10.1051/0004-6361/202556059}{\color{blue}\aap}, \href{https://ui.adsabs.harvard.edu/abs/2026A&A...708A.143B}{708, A143}

\bibitem[{{Bowler} {et~al.}(2025){Bowler}, {Zhou}, {Biddle}, {Jiang}, {Bae}, {Close}, {Follette}, {Franson}, {Kraus}, {Sanghi}, {Tran}, {Ward-Duong}, {Wu}, \& {Zhu}}]{Bowler2025}
{Bowler}, B.~P., {Zhou}, Y., {Biddle}, L.~I., {et~al.} 2025, \href{http://dx.doi.org/10.3847/1538-3881/adb6a1}{\color{blue}\aj}, \href{https://ui.adsabs.harvard.edu/abs/2025AJ....169..258B}{169, 258}

\bibitem[{{Brittain} {et~al.}(2023){Brittain}, {Kamp}, {Meeus}, {Oudmaijer}, \& {Waters}}]{Brittain2023}
{Brittain}, S.~D., {Kamp}, I., {Meeus}, G., {Oudmaijer}, R.~D., \& {Waters}, L.~B.~F.~M. 2023, \href{http://dx.doi.org/10.1007/s11214-023-00949-z}{\color{blue}\ssr}, \href{https://ui.adsabs.harvard.edu/abs/2023SSRv..219....7B}{219, 7}

\bibitem[{{Brittain} {et~al.}(2026){Brittain}, {Kern}, {Meeus}, \& {Oudmaijer}}]{Brittain2026}
{Brittain}, S.~D., {Kern}, J.~W., {Meeus}, G., \& {Oudmaijer}, R.~D. 2026, \href{http://dx.doi.org/10.3847/1538-3881/ae1a42}{\color{blue}\aj}, \href{https://ui.adsabs.harvard.edu/abs/2026AJ....171....4B}{171, 4}

\bibitem[{{Calcino} {et~al.}(2025){Calcino}, {Price}, \& {Ormel}}]{Calcino2025}
{Calcino}, J., {Price}, D.~J., \& {Ormel}, C.~W. 2025, \href{https://ui.adsabs.harvard.edu/abs/2025arXiv251005601C}{\href{http://dx.doi.org/10.48550/arXiv.2510.05601}{\color{blue}arXiv e-prints}, arXiv:2510.05601}

\bibitem[{{Calcino} {et~al.}(2019){Calcino}, {Price}, {Pinte}, {van der Marel}, {Ragusa}, {Dipierro}, {Cuello}, \& {Christiaens}}]{Calcino2019}
{Calcino}, J., {Price}, D.~J., {Pinte}, C., {et~al.} 2019, \href{http://dx.doi.org/10.1093/mnras/stz2770}{\color{blue}\mnras}, \href{https://ui.adsabs.harvard.edu/abs/2019MNRAS.490.2579C}{490, 2579}

\bibitem[{{Calvet} {et~al.}(2004){Calvet}, {Muzerolle}, {Brice{\~n}o}, {Hern{\'a}ndez}, {Hartmann}, {Saucedo}, \& {Gordon}}]{Calvet2004}
{Calvet}, N., {Muzerolle}, J., {Brice{\~n}o}, C., {et~al.} 2004, \href{http://dx.doi.org/10.1086/422733}{\color{blue}\aj}, \href{https://ui.adsabs.harvard.edu/abs/2004AJ....128.1294C}{128, 1294}

\bibitem[{{Carpenter} {et~al.}(2025){Carpenter}, {Esplin}, {Luhman}, {Mamajek}, \& {Andrews}}]{Carpenter2025}
{Carpenter}, J.~M., {Esplin}, T.~L., {Luhman}, K.~L., {Mamajek}, E.~E., \& {Andrews}, S.~M. 2025, \href{http://dx.doi.org/10.3847/1538-4357/ad8ebc}{\color{blue}\apj}, \href{https://ui.adsabs.harvard.edu/abs/2025ApJ...978..117C}{978, 117}

\bibitem[{{CASA Team} {et~al.}(2022){CASA Team}, {Bean}, {Bhatnagar}, {Castro}, {Donovan Meyer}, {Emonts}, {Garcia}, {Garwood}, {Golap}, {Gonzalez Villalba}, {Harris}, {Hayashi}, {Hoskins}, {Hsieh}, {Jagannathan}, {Kawasaki}, {Keimpema}, {Kettenis}, {Lopez}, {Marvil}, {Masters}, {McNichols}, {Mehringer}, {Miel}, {Moellenbrock}, {Montesino}, {Nakazato}, {Ott}, {Petry}, {Pokorny}, {Raba}, {Rau}, {Schiebel}, {Schweighart}, {Sekhar}, {Shimada}, {Small}, {Steeb}, {Sugimoto}, {Suoranta}, {Tsutsumi}, {van Bemmel}, {Verkouter}, {Wells}, {Xiong}, {Szomoru}, {Griffith}, {Glendenning}, \& {Kern}}]{CASA2022}
{CASA Team}, {Bean}, B., {Bhatnagar}, S., {et~al.} 2022, \href{http://dx.doi.org/10.1088/1538-3873/ac9642}{\color{blue}\pasp}, \href{https://ui.adsabs.harvard.edu/abs/2022PASP..134k4501C}{134, 114501}

\bibitem[{{Casagrande} \& {VandenBerg}(2018)}]{2018MNRAS.479L.102C}
{Casagrande}, L. \& {VandenBerg}, D.~A. 2018, \href{http://dx.doi.org/10.1093/mnrasl/sly104}{\color{blue}\mnras}, \href{https://ui.adsabs.harvard.edu/abs/2018MNRAS.479L.102C}{479, L102}

\bibitem[{{Casassus} {et~al.}(2021){Casassus}, {Christiaens}, {C{\'a}rcamo}, {P{\'e}rez}, {Weber}, {Ercolano}, {van der Marel}, {Pinte}, {Dong}, {Baruteau}, {Cieza}, {van Dishoeck}, {Jordan}, {Price}, {Absil}, {Arce-Tord}, {Faramaz}, {Flores}, \& {Reggiani}}]{Casassus2021}
{Casassus}, S., {Christiaens}, V., {C{\'a}rcamo}, M., {et~al.} 2021, \href{http://dx.doi.org/10.1093/mnras/stab2359}{\color{blue}\mnras}, \href{https://ui.adsabs.harvard.edu/abs/2021MNRAS.507.3789C}{507, 3789}

\bibitem[{{Cazzoletti} {et~al.}(2018){Cazzoletti}, {van Dishoeck}, {Pinilla}, {Tazzari}, {Facchini}, {van der Marel}, {Benisty}, {Garufi}, \& {P{\'e}rez}}]{Cazzoletti2018}
{Cazzoletti}, P., {van Dishoeck}, E.~F., {Pinilla}, P., {et~al.} 2018, \href{http://dx.doi.org/10.1051/0004-6361/201834006}{\color{blue}\aap}, \href{https://ui.adsabs.harvard.edu/abs/2018A&A...619A.161C}{619, A161}

\bibitem[{{Chabrier}(2003)}]{Chabrier_ea_2003}
{Chabrier}, G. 2003, \href{http://dx.doi.org/10.1086/376392}{\color{blue}\pasp}, \href{https://ui.adsabs.harvard.edu/abs/2003PASP..115..763C}{115, 763}

\bibitem[{{Chauvin} {et~al.}(2017){Chauvin}, {Desidera}, {Lagrange}, {Vigan}, {Gratton}, {Langlois}, {Bonnefoy}, {Beuzit}, {Feldt}, {Mouillet}, {Meyer}, {Cheetham}, {Biller}, {Boccaletti}, {D'Orazi}, {Galicher}, {Hagelberg}, {Maire}, {Mesa}, {Olofsson}, {Samland}, {Schmidt}, {Sissa}, {Bonavita}, {Charnay}, {Cudel}, {Daemgen}, {Delorme}, {Janin-Potiron}, {Janson}, {Keppler}, {Le Coroller}, {Ligi}, {Marleau}, {Messina}, {Molli{\`e}re}, {Mordasini}, {M{\"u}ller}, {Peretti}, {Perrot}, {Rodet}, {Rouan}, {Zurlo}, {Dominik}, {Henning}, {Menard}, {Schmid}, {Turatto}, {Udry}, {Vakili}, {Abe}, {Antichi}, {Baruffolo}, {Baudoz}, {Baudrand}, {Blanchard}, {Bazzon}, {Buey}, {Carbillet}, {Carle}, {Charton}, {Cascone}, {Claudi}, {Costille}, {Deboulbe}, {De Caprio}, {Dohlen}, {Fantinel}, {Feautrier}, {Fusco}, {Gigan}, {Giro}, {Gisler}, {Gluck}, {Hubin}, {Hugot}, {Jaquet}, {Kasper}, {Madec}, {Magnard}, {Martinez}, {Maurel}, {Le Mignant}, {M{\"o}ller-Nilsson}, {Llored}, {Moulin}, {Orign{\'e}}, {Pavlov}, {Perret}, {Petit},
  {Pragt}, {Puget}, {Rabou}, {Ramos}, {Rigal}, {Rochat}, {Roelfsema}, {Rousset}, {Roux}, {Salasnich}, {Sauvage}, {Sevin}, {Soenke}, {Stadler}, {Suarez}, {Weber}, {Wildi}, {Antoniucci}, {Augereau}, {Baudino}, {Brandner}, {Engler}, {Girard}, {Gry}, {Kral}, {Kopytova}, {Lagadec}, {Milli}, {Moutou}, {Schlieder}, {Szul{\'a}gyi}, {Thalmann}, \& {Wahhaj}}]{Chauvin2017}
{Chauvin}, G., {Desidera}, S., {Lagrange}, A.~M., {et~al.} 2017, \href{http://dx.doi.org/10.1051/0004-6361/201731152}{\color{blue}\aap}, \href{https://ui.adsabs.harvard.edu/abs/2017A&A...605L...9C}{605, L9}

\bibitem[{{Chen} {et~al.}(2016){Chen}, {Shan}, \& {Zhang}}]{Chen2016}
{Chen}, P.~S., {Shan}, H.~G., \& {Zhang}, P. 2016, \href{http://dx.doi.org/10.1016/j.newast.2015.09.001}{\color{blue}\na}, \href{https://ui.adsabs.harvard.edu/abs/2016NewA...44....1C}{44, 1}

\bibitem[{{Cieza} {et~al.}(2016){Cieza}, {Casassus}, {Tobin}, {Bos}, {Williams}, {Perez}, {Zhu}, {Caceres}, {Canovas}, {Dunham}, {Hales}, {Prieto}, {Principe}, {Schreiber}, {Ruiz-Rodriguez}, \& {Zurlo}}]{Cieza2016}
{Cieza}, L.~A., {Casassus}, S., {Tobin}, J., {et~al.} 2016, \href{http://dx.doi.org/10.1038/nature18612}{\color{blue}\nat}, \href{https://ui.adsabs.harvard.edu/abs/2016Natur.535..258C}{535, 258}

\bibitem[{{Cieza} {et~al.}(2021){Cieza}, {Gonz{\'a}lez-Ruilova}, {Hales}, {Pinilla}, {Ru{\'\i}z-Rodr{\'\i}guez}, {Zurlo}, {Casassus}, {P{\'e}rez}, {C{\'a}novas}, {Arce-Tord}, {Flock}, {Kurtovic}, {Marino}, {Nogueira}, {Perez}, {Price}, {Principe}, \& {Williams}}]{Cieza2021}
{Cieza}, L.~A., {Gonz{\'a}lez-Ruilova}, C., {Hales}, A.~S., {et~al.} 2021, \href{http://dx.doi.org/10.1093/mnras/staa3787}{\color{blue}\mnras}, \href{https://ui.adsabs.harvard.edu/abs/2021MNRAS.501.2934C}{501, 2934}

\bibitem[{{Cleeves} {et~al.}(2016){Cleeves}, {{\"O}berg}, {Wilner}, {Huang}, {Loomis}, {Andrews}, \& {Czekala}}]{Cleeves2016}
{Cleeves}, L.~I., {{\"O}berg}, K.~I., {Wilner}, D.~J., {et~al.} 2016, \href{http://dx.doi.org/10.3847/0004-637X/832/2/110}{\color{blue}\apj}, \href{https://ui.adsabs.harvard.edu/abs/2016ApJ...832..110C}{832, 110}

\bibitem[{{Cody} {et~al.}(2025){Cody}, {Hillenbrand}, {Chandragiri}, \& {Morgan}}]{Cody2025}
{Cody}, A.~M., {Hillenbrand}, L.~A., {Chandragiri}, S., \& {Morgan}, M. 2025, \href{http://dx.doi.org/10.3847/1538-4357/ae119a}{\color{blue}\apj}, \href{https://ui.adsabs.harvard.edu/abs/2025ApJ...994..253C}{994, 253}

\bibitem[{{Collins} {et~al.}(2009){Collins}, {Grady}, {Hamaguchi}, {Wisniewski}, {Brittain}, {Sitko}, {Carpenter}, {Williams}, {Mathews}, {Williger}, {van Boekel}, {Carmona}, {Henning}, {van den Ancker}, {Meeus}, {Chen}, {Petre}, \& {Woodgate}}]{Collins2009}
{Collins}, K.~A., {Grady}, C.~A., {Hamaguchi}, K., {et~al.} 2009, \href{http://dx.doi.org/10.1088/0004-637X/697/1/557}{\color{blue}\apj}, \href{https://ui.adsabs.harvard.edu/abs/2009ApJ...697..557C}{697, 557}

\bibitem[{{Columba} {et~al.}(2024){Columba}, {Rigliaco}, {Gratton}, {Mesa}, {D'Orazi}, {Ginski}, {Engler}, {Williams}, {Bae}, {Benisty}, {Birnstiel}, {Delorme}, {Dominik}, {Facchini}, {Menard}, {Pinilla}, {Rab}, {Ribas}, {Squicciarini}, {van Holstein}, \& {Zurlo}}]{Columba2024}
{Columba}, G., {Rigliaco}, E., {Gratton}, R., {et~al.} 2024, \href{http://dx.doi.org/10.1051/0004-6361/202347109}{\color{blue}\aap}, \href{https://ui.adsabs.harvard.edu/abs/2024A&A...681A..19C}{681, A19}

\bibitem[{{Currie} {et~al.}(2022){Currie}, {Lawson}, {Schneider}, {Lyra}, {Wisniewski}, {Grady}, {Guyon}, {Tamura}, {Kotani}, {Kawahara}, {Brandt}, {Uyama}, {Muto}, {Dong}, {Kudo}, {Hashimoto}, {Fukagawa}, {Wagner}, {Lozi}, {Chilcote}, {Tobin}, {Groff}, {Ward-Duong}, {Januszewski}, {Norris}, {Tuthill}, {van der Marel}, {Sitko}, {Deo}, {Vievard}, {Jovanovic}, {Martinache}, \& {Skaf}}]{Currie2022}
{Currie}, T., {Lawson}, K., {Schneider}, G., {et~al.} 2022, \href{http://dx.doi.org/10.1038/s41550-022-01634-x}{\color{blue}Nature Astronomy}, \href{https://ui.adsabs.harvard.edu/abs/2022NatAs...6..751C}{6, 751}

\bibitem[{{Das} {et~al.}(2025){Das}, {Vorobyov}, \& {Basu}}]{Das2025}
{Das}, I., {Vorobyov}, E., \& {Basu}, S. 2025, \href{http://dx.doi.org/10.3847/1538-4357/adb8ee}{\color{blue}\apj}, \href{https://ui.adsabs.harvard.edu/abs/2025ApJ...983..163D}{983, 163}

\bibitem[{Davidson-Pilon {et~al.}(2021)Davidson-Pilon, Kalderstam, Jacobson, Reed, Kuhn, Zivich, Williamson, AbdealiJK, Datta, Fiore-Gartland, Parij, WIlson, Gabriel, Moneda, Moncada-Torres, Stark, Gadgil, Jona, Singaravelan, Besson, Peña, Anton, Klintberg, GrowthJeff, Noorbakhsh, Begun, Kumar, Hussey, Seabold, \& Golland}]{DavidsonPilon2021}
Davidson-Pilon, C., Kalderstam, J., Jacobson, N., {et~al.} 2021, CamDavidsonPilon/lifelines: 0.25.10

\bibitem[{{Delfini} {et~al.}(2025){Delfini}, {Vioque}, {Ribas}, \& {Hodgkin}}]{Delfini2025}
{Delfini}, L., {Vioque}, M., {Ribas}, {\'A}., \& {Hodgkin}, S. 2025, \href{http://dx.doi.org/10.1051/0004-6361/202453539}{\color{blue}\aap}, \href{https://ui.adsabs.harvard.edu/abs/2025A&A...699A.145D}{699, A145}

\bibitem[{{Deng} {et~al.}(2026){Deng}, {Liu}, \& {Fang}}]{Deng2026}
{Deng}, H., {Liu}, Y., \& {Fang}, M. 2026, \href{https://ui.adsabs.harvard.edu/abs/2026arXiv260714850D}{\href{http://dx.doi.org/10.48550/arXiv.2607.14850}{\color{blue}arXiv e-prints}, arXiv:2607.14850}

\bibitem[{{Donehew} \& {Brittain}(2011)}]{Donehew2011}
{Donehew}, B. \& {Brittain}, S. 2011, \href{http://dx.doi.org/10.1088/0004-6256/141/2/46}{\color{blue}\aj}, \href{https://ui.adsabs.harvard.edu/abs/2011AJ....141...46D}{141, 46}

\bibitem[{{Dong} {et~al.}(2022){Dong}, {Liu}, {Cuello}, {Pinte}, {{\'A}brah{\'a}m}, {Vorobyov}, {Hashimoto}, {K{\'o}sp{\'a}l}, {Chiang}, {Takami}, {Chen}, {Dunham}, {Fukagawa}, {Green}, {Hasegawa}, {Henning}, {Pavlyuchenkov}, {Pyo}, \& {Tamura}}]{Dong2022}
{Dong}, R., {Liu}, H.~B., {Cuello}, N., {et~al.} 2022, \href{http://dx.doi.org/10.1038/s41550-021-01558-y}{\color{blue}Nature Astronomy}, \href{https://ui.adsabs.harvard.edu/abs/2022NatAs...6..331D}{6, 331}

\bibitem[{{Dong} {et~al.}(2018){Dong}, {Liu}, {Eisner}, {Andrews}, {Fung}, {Zhu}, {Chiang}, {Hashimoto}, {Liu}, {Casassus}, {Esposito}, {Hasegawa}, {Muto}, {Pavlyuchenkov}, {Wilner}, {Akiyama}, {Tamura}, \& {Wisniewski}}]{Dong2018}
{Dong}, R., {Liu}, S.-y., {Eisner}, J., {et~al.} 2018, \href{http://dx.doi.org/10.3847/1538-4357/aac6cb}{\color{blue}\apj}, \href{https://ui.adsabs.harvard.edu/abs/2018ApJ...860..124D}{860, 124}

\bibitem[{{Dullemond} \& {Dominik}(2004)}]{Dullemond2004a}
{Dullemond}, C.~P. \& {Dominik}, C. 2004, \href{http://dx.doi.org/10.1051/0004-6361:20031768}{\color{blue}\aap}, \href{https://ui.adsabs.harvard.edu/abs/2004A&A...417..159D}{417, 159}

\bibitem[{{Edenhofer} {et~al.}(2024){Edenhofer}, {Zucker}, {Frank}, {Saydjari}, {Speagle}, {Finkbeiner}, \& {En{\ss}lin}}]{Edenhofer_ea_2024}
{Edenhofer}, G., {Zucker}, C., {Frank}, P., {et~al.} 2024, \href{http://dx.doi.org/10.1051/0004-6361/202347628}{\color{blue}\aap}, \href{https://ui.adsabs.harvard.edu/abs/2024A&A...685A..82E}{685, A82}

\bibitem[{{Empey} {et~al.}(2026){Empey}, {Manara}, {Garcia Lopez}, {Natta}, {Claes}, {Zagaria}, {Alcal{\'a}}, {Anania}, {Beccari}, {Carpenter}, {Facchini}, {Fedele}, {Lodato}, {Mauco}, {Miotello}, {Nisini}, {Pascucci}, {Piscarreta}, {Rosotti}, {Scholz}, {Testi}, \& {Vioque}}]{Empey2026}
{Empey}, A., {Manara}, C.~F., {Garcia Lopez}, R., {et~al.} 2026, \href{https://ui.adsabs.harvard.edu/abs/2026arXiv260619455E}{\href{http://dx.doi.org/10.48550/arXiv.2606.19455}{\color{blue}arXiv e-prints}, arXiv:2606.19455}

\bibitem[{{Ercolano} \& {Pascucci}(2017)}]{Ercolano2017}
{Ercolano}, B. \& {Pascucci}, I. 2017, \href{http://dx.doi.org/10.1098/rsos.170114}{\color{blue}Royal Society Open Science}, \href{https://ui.adsabs.harvard.edu/abs/2017RSOS....470114E}{4, 170114}

\bibitem[{{Fairlamb} {et~al.}(2015){Fairlamb}, {Oudmaijer}, {Mendigut{\'\i}a}, {Ilee}, \& {van den Ancker}}]{Fairlamb2015}
{Fairlamb}, J.~R., {Oudmaijer}, R.~D., {Mendigut{\'\i}a}, I., {Ilee}, J.~D., \& {van den Ancker}, M.~E. 2015, \href{http://dx.doi.org/10.1093/mnras/stv1576}{\color{blue}\mnras}, \href{https://ui.adsabs.harvard.edu/abs/2015MNRAS.453..976F}{453, 976}

\bibitem[{{Fasano} {et~al.}(2026){Fasano}, {Benisty}, {Stadler}, {Zagaria}, {Ziampras}, {Winter}, {Bae}, {Facchini}, {Kurtovic}, {Ragusa}, \& {Teague}}]{Fasano2026}
{Fasano}, D., {Benisty}, M., {Stadler}, J., {et~al.} 2026, \href{https://ui.adsabs.harvard.edu/abs/2026arXiv260325541F}{\href{http://dx.doi.org/10.48550/arXiv.2603.25541}{\color{blue}arXiv e-prints}, arXiv:2603.25541}

\bibitem[{{Fulton} {et~al.}(2021){Fulton}, {Rosenthal}, {Hirsch}, {Isaacson}, {Howard}, {Dedrick}, {Sherstyuk}, {Blunt}, {Petigura}, {Knutson}, {Behmard}, {Chontos}, {Crepp}, {Crossfield}, {Dalba}, {Fischer}, {Henry}, {Kane}, {Kosiarek}, {Marcy}, {Rubenzahl}, {Weiss}, \& {Wright}}]{Fulton2021}
{Fulton}, B.~J., {Rosenthal}, L.~J., {Hirsch}, L.~A., {et~al.} 2021, \href{http://dx.doi.org/10.3847/1538-4365/abfcc1}{\color{blue}\apjs}, \href{https://ui.adsabs.harvard.edu/abs/2021ApJS..255...14F}{255, 14}

\bibitem[{Galloway-Sprietsma {et~al.}(2026)Galloway-Sprietsma, Bae, Phillips, Huang, Benisty, Porter, Ginski, \& Winter}]{GallowaySprietsma2026}
Galloway-Sprietsma, M., Bae, J., Phillips, T., {et~al.} 2026, \href{http://dx.doi.org/10.3847/1538-3881/ae8bae}{\color{blue}The Astronomical Journal}, 172, 172

\bibitem[{{Garufi} {et~al.}(2018){Garufi}, {Benisty}, {Pinilla}, {Tazzari}, {Dominik}, {Ginski}, {Henning}, {Kral}, {Langlois}, {M{\'e}nard}, {Stolker}, {Szulagyi}, {Villenave}, \& {van der Plas}}]{Garufi2018}
{Garufi}, A., {Benisty}, M., {Pinilla}, P., {et~al.} 2018, \href{http://dx.doi.org/10.1051/0004-6361/201833872}{\color{blue}\aap}, \href{https://ui.adsabs.harvard.edu/abs/2018A&A...620A..94G}{620, A94}

\bibitem[{{Garufi} {et~al.}(2026){Garufi}, {Ginski}, {Benisty}, {Vioque}, {Winter}, {Huang}, {Manara}, \& {Dominik}}]{Garufi2026}
{Garufi}, A., {Ginski}, C., {Benisty}, M., {et~al.} 2026, \href{https://ui.adsabs.harvard.edu/abs/2026arXiv260301703G}{\href{http://dx.doi.org/10.48550/arXiv.2603.01703}{\color{blue}arXiv e-prints}, arXiv:2603.01703}

\bibitem[{{Garufi} {et~al.}(2022){Garufi}, {Podio}, {Codella}, {Segura-Cox}, {Vander Donckt}, {Mercimek}, {Bacciotti}, {Fedele}, {Kasper}, {Pineda}, {Humphreys}, \& {Testi}}]{Garufi2022}
{Garufi}, A., {Podio}, L., {Codella}, C., {et~al.} 2022, \href{http://dx.doi.org/10.1051/0004-6361/202141264}{\color{blue}\aap}, \href{https://ui.adsabs.harvard.edu/abs/2022A&A...658A.104G}{658, A104}

\bibitem[{{Ginsburg} {et~al.}(2019){Ginsburg}, {Sip{\H{o}}cz}, {Brasseur}, {Cowperthwaite}, {Craig}, {Deil}, {Guillochon}, {Guzman}, {Liedtke}, {Lian Lim}, {Lockhart}, {Mommert}, {Morris}, {Norman}, {Parikh}, {Persson}, {Robitaille}, {Segovia}, {Singer}, {Tollerud}, {de Val-Borro}, {Valtchanov}, {Woillez}, {Astroquery Collaboration}, \& {a subset of astropy Collaboration}}]{Ginsburg2019}
{Ginsburg}, A., {Sip{\H{o}}cz}, B.~M., {Brasseur}, C.~E., {et~al.} 2019, \href{http://dx.doi.org/10.3847/1538-3881/aafc33}{\color{blue}\aj}, \href{https://ui.adsabs.harvard.edu/abs/2019AJ....157...98G}{157, 98}

\bibitem[{{Ginski} {et~al.}(2021){Ginski}, {Facchini}, {Huang}, {Benisty}, {Vaendel}, {Stapper}, {Dominik}, {Bae}, {M{\'e}nard}, {Muro-Arena}, {Hogerheijde}, {McClure}, {van Holstein}, {Birnstiel}, {Boehler}, {Bohn}, {Flock}, {Mamajek}, {Manara}, {Pinilla}, {Pinte}, \& {Ribas}}]{Ginski2021}
{Ginski}, C., {Facchini}, S., {Huang}, J., {et~al.} 2021, \href{http://dx.doi.org/10.3847/2041-8213/abdf57}{\color{blue}\apjl}, \href{https://ui.adsabs.harvard.edu/abs/2021ApJ...908L..25G}{908, L25}

\bibitem[{{Grant} {et~al.}(2022){Grant}, {Espaillat}, {Brittain}, {Scott-Joseph}, \& {Calvet}}]{Grant2022}
{Grant}, S.~L., {Espaillat}, C.~C., {Brittain}, S., {Scott-Joseph}, C., \& {Calvet}, N. 2022, \href{http://dx.doi.org/10.3847/1538-4357/ac450a}{\color{blue}\apj}, \href{https://ui.adsabs.harvard.edu/abs/2022ApJ...926..229G}{926, 229}

\bibitem[{{Grant} {et~al.}(2023){Grant}, {Stapper}, {Hogerheijde}, {van Dishoeck}, {Brittain}, \& {Vioque}}]{GrantStapper2023}
{Grant}, S.~L., {Stapper}, L.~M., {Hogerheijde}, M.~R., {et~al.} 2023, \href{http://dx.doi.org/10.3847/1538-3881/acf128}{\color{blue}\aj}, \href{https://ui.adsabs.harvard.edu/abs/2023AJ....166..147G}{166, 147}

\bibitem[{{Gupta} {et~al.}(2024){Gupta}, {Miotello}, {Williams}, {Birnstiel}, {Kuffmeier}, \& {Yen}}]{Gupta2024}
{Gupta}, A., {Miotello}, A., {Williams}, J.~P., {et~al.} 2024, \href{http://dx.doi.org/10.1051/0004-6361/202348007}{\color{blue}\aap}, \href{https://ui.adsabs.harvard.edu/abs/2024A&A...683A.133G}{683, A133}

\bibitem[{{Guzm{\'a}n-D{\'\i}az} {et~al.}(2021){Guzm{\'a}n-D{\'\i}az}, {Mendigut{\'\i}a}, {Montesinos}, {Oudmaijer}, {Vioque}, {Rodrigo}, {Solano}, {Meeus}, \& {Marcos-Arenal}}]{GuzmanDiaz2021}
{Guzm{\'a}n-D{\'\i}az}, J., {Mendigut{\'\i}a}, I., {Montesinos}, B., {et~al.} 2021, \href{http://dx.doi.org/10.1051/0004-6361/202039519}{\color{blue}\aap}, \href{https://ui.adsabs.harvard.edu/abs/2021A&A...650A.182G}{650, A182}

\bibitem[{{Guzm{\'a}n-D{\'\i}az} {et~al.}(2023){Guzm{\'a}n-D{\'\i}az}, {Montesinos}, {Mendigut{\'\i}a}, {Kama}, {Meeus}, {Vioque}, {Oudmaijer}, \& {Villaver}}]{GuzmanDiaz2023}
{Guzm{\'a}n-D{\'\i}az}, J., {Montesinos}, B., {Mendigut{\'\i}a}, I., {et~al.} 2023, \href{http://dx.doi.org/10.1051/0004-6361/202245427}{\color{blue}\aap}, \href{https://ui.adsabs.harvard.edu/abs/2023A&A...671A.140G}{671, A140}

\bibitem[{{Hammond} {et~al.}(2023){Hammond}, {Christiaens}, {Price}, {Toci}, {Pinte}, {Juillard}, \& {Garg}}]{Hammond2023}
{Hammond}, I., {Christiaens}, V., {Price}, D.~J., {et~al.} 2023, \href{http://dx.doi.org/10.1093/mnrasl/slad027}{\color{blue}\mnras}, \href{https://ui.adsabs.harvard.edu/abs/2023MNRAS.522L..51H}{522, L51}

\bibitem[{Harris {et~al.}(2020)Harris, Millman, van~der Walt, Gommers, Virtanen, Cournapeau, Wieser, Taylor, Berg, Smith, Kern, Picus, Hoyer, van Kerkwijk, Brett, Haldane, del R{\'{i}}o, Wiebe, Peterson, G{\'{e}}rard-Marchant, Sheppard, Reddy, Weckesser, Abbasi, Gohlke, \& Oliphant}]{Harris2020}
Harris, C.~R., Millman, K.~J., van~der Walt, S.~J., {et~al.} 2020, \href{http://dx.doi.org/10.1038/s41586-020-2649-2}{\color{blue}Nature}, 585, 585

\bibitem[{{Hartmann} {et~al.}(2016){Hartmann}, {Herczeg}, \& {Calvet}}]{Hartmann2016}
{Hartmann}, L., {Herczeg}, G., \& {Calvet}, N. 2016, \href{http://dx.doi.org/10.1146/annurev-astro-081915-023347}{\color{blue}\araa}, \href{https://ui.adsabs.harvard.edu/abs/2016ARA&A..54..135H}{54, 135}

\bibitem[{{Herbig}(1960)}]{Herbig1960}
{Herbig}, G.~H. 1960, \href{http://dx.doi.org/10.1086/190050}{\color{blue}\apjs}, \href{https://ui.adsabs.harvard.edu/abs/1960ApJS....4..337H}{4, 337}

\bibitem[{{Hildebrand}(1983)}]{Hildebrand1983}
{Hildebrand}, R.~H. 1983, \qjras, \href{https://ui.adsabs.harvard.edu/abs/1983QJRAS..24..267H}{24, 267}

\bibitem[{{Honda} {et~al.}(2015){Honda}, {Maaskant}, {Okamoto}, {Kataza}, {Yamashita}, {Miyata}, {Sako}, {Fujiyoshi}, {Sakon}, {Fujiwara}, {Kamizuka}, {Mulders}, {Lopez-Rodriguez}, {Packham}, \& {Onaka}}]{Honda2015}
{Honda}, M., {Maaskant}, K., {Okamoto}, Y.~K., {et~al.} 2015, \href{http://dx.doi.org/10.1088/0004-637X/804/2/143}{\color{blue}\apj}, \href{https://ui.adsabs.harvard.edu/abs/2015ApJ...804..143H}{804, 143}

\bibitem[{{Huang} {et~al.}(2021){Huang}, {Bergin}, {{\"O}berg}, {Andrews}, {Teague}, {Law}, {Kalas}, {Aikawa}, {Bae}, {Bergner}, {Booth}, {Bosman}, {Calahan}, {Cataldi}, {Cleeves}, {Czekala}, {Ilee}, {Le Gal}, {Guzm{\'a}n}, {Long}, {Loomis}, {M{\'e}nard}, {Nomura}, {Qi}, {Schwarz}, {Tsukagoshi}, {van't Hoff}, {Walsh}, {Wilner}, {Yamato}, \& {Zhang}}]{Huang2021}
{Huang}, J., {Bergin}, E.~A., {{\"O}berg}, K.~I., {et~al.} 2021, \href{http://dx.doi.org/10.3847/1538-4365/ac143e}{\color{blue}\apjs}, \href{https://ui.adsabs.harvard.edu/abs/2021ApJS..257...19H}{257, 19}

\bibitem[{{H{\"u}hn} \& {Dullemond}(2025)}]{Huhn2025}
{H{\"u}hn}, L.-A. \& {Dullemond}, C.~P. 2025, \href{http://dx.doi.org/10.1051/0004-6361/202556203}{\color{blue}\aap}, \href{https://ui.adsabs.harvard.edu/abs/2025A&A...704A.222H}{704, A222}

\bibitem[{Hunter(2007)}]{Hunter2007}
Hunter, J.~D. 2007, \href{http://dx.doi.org/10.1109/MCSE.2007.55}{\color{blue}Computing in Science \& Engineering}, 9, 9

\bibitem[{{Iglesias} {et~al.}(2023){Iglesias}, {Pani{\'c}}, {van den Ancker}, {Petr-Gotzens}, {Siess}, {Vioque}, {Pascucci}, {Oudmaijer}, \& {Miley}}]{Iglesias2022}
{Iglesias}, D.~P., {Pani{\'c}}, O., {van den Ancker}, M., {et~al.} 2023, \href{http://dx.doi.org/10.1093/mnras/stac3619}{\color{blue}\mnras}, \href{https://ui.adsabs.harvard.edu/abs/2023MNRAS.519.3958I}{519, 3958}

\bibitem[{{Isella} {et~al.}(2016){Isella}, {Guidi}, {Testi}, {Liu}, {Li}, {Li}, {Weaver}, {Boehler}, {Carperter}, {De Gregorio-Monsalvo}, {Manara}, {Natta}, {P{\'e}rez}, {Ricci}, {Sargent}, {Tazzari}, \& {Turner}}]{Isella2016}
{Isella}, A., {Guidi}, G., {Testi}, L., {et~al.} 2016, \href{http://dx.doi.org/10.1103/PhysRevLett.117.251101}{\color{blue}\prl}, \href{https://ui.adsabs.harvard.edu/abs/2016PhRvL.117y1101I}{117, 251101}

\bibitem[{{Izquierdo} {et~al.}(2022){Izquierdo}, {Facchini}, {Rosotti}, {van Dishoeck}, \& {Testi}}]{Izquierdo2022}
{Izquierdo}, A.~F., {Facchini}, S., {Rosotti}, G.~P., {van Dishoeck}, E.~F., \& {Testi}, L. 2022, \href{http://dx.doi.org/10.3847/1538-4357/ac474d}{\color{blue}\apj}, \href{https://ui.adsabs.harvard.edu/abs/2022ApJ...928....2I}{928, 2}

\bibitem[{{Jiang} {et~al.}(2026){Jiang}, {Semenov}, {Benisty}, {Pi{\'e}tu}, {Henning}, {Rivi{\`e}re-Marichalar}, {Stapper}, \& {Chapillon}}]{Jiang2026}
{Jiang}, H., {Semenov}, D., {Benisty}, M., {et~al.} 2026, \href{https://ui.adsabs.harvard.edu/abs/2026arXiv260718683J}{\href{http://dx.doi.org/10.48550/arXiv.2607.18683}{\color{blue}arXiv e-prints}, arXiv:2607.18683}

\bibitem[{{Johnson} {et~al.}(2010){Johnson}, {Aller}, {Howard}, \& {Crepp}}]{Johnson2010}
{Johnson}, J.~A., {Aller}, K.~M., {Howard}, A.~W., \& {Crepp}, J.~R. 2010, \href{http://dx.doi.org/10.1086/655775}{\color{blue}\pasp}, \href{https://ui.adsabs.harvard.edu/abs/2010PASP..122..905J}{122, 905}

\bibitem[{{Johnston} {et~al.}(2026){Johnston}, {Panic}, {Reffert}, {Liu}, \& {Ma}}]{Johnston2026}
{Johnston}, H.~F., {Panic}, O., {Reffert}, S., {Liu}, B., \& {Ma}, X. 2026, \href{https://ui.adsabs.harvard.edu/abs/2026arXiv260303014J}{\href{http://dx.doi.org/10.48550/arXiv.2603.03014}{\color{blue}arXiv e-prints}, arXiv:2603.03014}

\bibitem[{{Kaeufer} {et~al.}(2023){Kaeufer}, {Woitke}, {Min}, {Kamp}, \& {Pinte}}]{Kaeufer2023}
{Kaeufer}, T., {Woitke}, P., {Min}, M., {Kamp}, I., \& {Pinte}, C. 2023, \href{https://ui.adsabs.harvard.edu/abs/2023arXiv230204629K}{\href{http://dx.doi.org/10.48550/arXiv.2302.04629}{\color{blue}arXiv e-prints}, arXiv:2302.04629}

\bibitem[{{Kama} {et~al.}(2015){Kama}, {Folsom}, \& {Pinilla}}]{Kama2015}
{Kama}, M., {Folsom}, C.~P., \& {Pinilla}, P. 2015, \href{http://dx.doi.org/10.1051/0004-6361/201527094}{\color{blue}\aap}, \href{https://ui.adsabs.harvard.edu/abs/2015A&A...582L..10K}{582, L10}

\bibitem[{{Kelly}(2007)}]{Kelly2007}
{Kelly}, B.~C. 2007, \href{http://dx.doi.org/10.1086/519947}{\color{blue}\apj}, \href{https://ui.adsabs.harvard.edu/abs/2007ApJ...665.1489K}{665, 1489}

\bibitem[{{Kennicutt}(1998)}]{Kennicutt_ea_1998}
{Kennicutt}, Jr., R.~C. 1998, \href{http://dx.doi.org/10.1086/305588}{\color{blue}\apj}, \href{https://ui.adsabs.harvard.edu/abs/1998ApJ...498..541K}{498, 541}

\bibitem[{{Kim} {et~al.}(2020){Kim}, {Takahashi}, {Nomura}, {Tsukagoshi}, {Lee}, {Muto}, {Dong}, {Hasegawa}, {Hashimoto}, {Kanagawa}, {Kataoka}, {Konishi}, {Liu}, {Momose}, {Sitko}, \& {Tomida}}]{Kim2020}
{Kim}, S., {Takahashi}, S., {Nomura}, H., {et~al.} 2020, \href{http://dx.doi.org/10.3847/1538-4357/ab5d2b}{\color{blue}\apj}, \href{https://ui.adsabs.harvard.edu/abs/2020ApJ...888...72K}{888, 72}

\bibitem[{{K{\"o}hler} {et~al.}(2016){K{\"o}hler}, {Kasper}, {Herbst}, {Ratzka}, \& {Bertrang}}]{Kohler2016}
{K{\"o}hler}, R., {Kasper}, M., {Herbst}, T.~M., {Ratzka}, T., \& {Bertrang}, G.~H.~M. 2016, \href{http://dx.doi.org/10.1051/0004-6361/201527125}{\color{blue}\aap}, \href{https://ui.adsabs.harvard.edu/abs/2016A&A...587A..35K}{587, A35}

\bibitem[{{Kraus} {et~al.}(2008){Kraus}, {Hofmann}, {Benisty}, {Berger}, {Chesneau}, {Isella}, {Malbet}, {Meilland}, {Nardetto}, {Natta}, {Preibisch}, {Schertl}, {Smith}, {Stee}, {Tatulli}, {Testi}, \& {Weigelt}}]{Kraus2008}
{Kraus}, S., {Hofmann}, K.~H., {Benisty}, M., {et~al.} 2008, \href{http://dx.doi.org/10.1051/0004-6361:200809946}{\color{blue}\aap}, \href{https://ui.adsabs.harvard.edu/abs/2008A&A...489.1157K}{489, 1157}

\bibitem[{{Kraus} {et~al.}(2017){Kraus}, {Kreplin}, {Fukugawa}, {Muto}, {Sitko}, {Young}, {Bate}, {Grady}, {Harries}, {Monnier}, {Willson}, \& {Wisniewski}}]{Kraus2017}
{Kraus}, S., {Kreplin}, A., {Fukugawa}, M., {et~al.} 2017, \href{http://dx.doi.org/10.3847/2041-8213/aa8edc}{\color{blue}\apjl}, \href{https://ui.adsabs.harvard.edu/abs/2017ApJ...848L..11K}{848, L11}

\bibitem[{{Kraus} {et~al.}(2020){Kraus}, {Kreplin}, {Young}, {Bate}, {Monnier}, {Harries}, {Avenhaus}, {Kluska}, {Laws}, {Rich}, {Willson}, {Aarnio}, {Adams}, {Andrews}, {Anugu}, {Bae}, {ten Brummelaar}, {Calvet}, {Cur{\'e}}, {Davies}, {Ennis}, {Espaillat}, {Gardner}, {Hartmann}, {Hinkley}, {Labdon}, {Lanthermann}, {LeBouquin}, {Schaefer}, {Setterholm}, {Wilner}, \& {Zhu}}]{Kraus2020}
{Kraus}, S., {Kreplin}, A., {Young}, A.~K., {et~al.} 2020, \href{http://dx.doi.org/10.1126/science.aba4633}{\color{blue}Science}, \href{https://ui.adsabs.harvard.edu/abs/2020Sci...369.1233K}{369, 1233}

\bibitem[{{Kuffmeier} {et~al.}(2023){Kuffmeier}, {Jensen}, \& {Haugb{\o}lle}}]{Kuffmeier2023}
{Kuffmeier}, M., {Jensen}, S.~S., \& {Haugb{\o}lle}, T. 2023, \href{http://dx.doi.org/10.1140/epjp/s13360-023-03880-y}{\color{blue}European Physical Journal Plus}, \href{https://ui.adsabs.harvard.edu/abs/2023EPJP..138..272K}{138, 272}

\bibitem[{{Kuhn} {et~al.}(2021){Kuhn}, {de Souza}, {Krone-Martins}, {Castro-Ginard}, {Ishida}, {Povich}, {Hillenbrand}, \& {COIN Collaboration}}]{Kuhn2021}
{Kuhn}, M.~A., {de Souza}, R.~S., {Krone-Martins}, A., {et~al.} 2021, \href{http://dx.doi.org/10.3847/1538-4365/abe465}{\color{blue}\apjs}, \href{https://ui.adsabs.harvard.edu/abs/2021ApJS..254...33K}{254, 33}

\bibitem[{{Kunitomo} {et~al.}(2021){Kunitomo}, {Ida}, {Takeuchi}, {Pani{\'c}}, {Miley}, \& {Suzuki}}]{Kunitomo2021}
{Kunitomo}, M., {Ida}, S., {Takeuchi}, T., {et~al.} 2021, \href{http://dx.doi.org/10.3847/1538-4357/abdb2a}{\color{blue}\apj}, \href{https://ui.adsabs.harvard.edu/abs/2021ApJ...909..109K}{909, 109}

\bibitem[{{Lada}(1987)}]{Lada1987}
{Lada}, C.~J. 1987, in IAU Symposium, Vol. 115, Star Forming Regions, ed. M.~{Peimbert} \& J.~{Jugaku}, \href{https://ui.adsabs.harvard.edu/abs/1987IAUS..115....1L}{1}

\bibitem[{{Lagrange} {et~al.}(2010){Lagrange}, {Bonnefoy}, {Chauvin}, {Apai}, {Ehrenreich}, {Boccaletti}, {Gratadour}, {Rouan}, {Mouillet}, {Lacour}, \& {Kasper}}]{Lagrange2010}
{Lagrange}, A.~M., {Bonnefoy}, M., {Chauvin}, G., {et~al.} 2010, \href{http://dx.doi.org/10.1126/science.1187187}{\color{blue}Science}, \href{https://ui.adsabs.harvard.edu/abs/2010Sci...329...57L}{329, 57}

\bibitem[{{Latour} {et~al.}(2026){Latour}, {Christiaens}, {Absil}, {Bonse}, {Savonet}, {Juillard}, {Hammond}, {Casassus}, {Cieza}, {Cugno}, {Desgrange}, {Lacour}, {Mawet}, {Montesinos}, {Perez}, {Pinte}, {Reggiani}, {Stolker}, {van der Marel}, \& {Zurlo}}]{Latour2026}
{Latour}, J., {Christiaens}, V., {Absil}, O., {et~al.} 2026, \href{http://dx.doi.org/10.1051/0004-6361/202558458}{\color{blue}\aap}, \href{https://ui.adsabs.harvard.edu/abs/2026A&A...711A.292L}{711, A292}

\bibitem[{{Leemker} {et~al.}(2024){Leemker}, {Booth}, {van Dishoeck}, {W{\"o}lfer}, \& {Dent}}]{Leemker2024}
{Leemker}, M., {Booth}, A.~S., {van Dishoeck}, E.~F., {W{\"o}lfer}, L., \& {Dent}, B. 2024, \href{http://dx.doi.org/10.1051/0004-6361/202349072}{\color{blue}\aap}, \href{https://ui.adsabs.harvard.edu/abs/2024A&A...687A.299L}{687, A299}

\bibitem[{{Leemker} {et~al.}(2025){Leemker}, {Tobin}, {Facchini}, {Curone}, {Booth}, {Furuya}, \& {van't Hoff}}]{Leemker2025}
{Leemker}, M., {Tobin}, J.~J., {Facchini}, S., {et~al.} 2025, \href{http://dx.doi.org/10.1038/s41550-025-02663-y}{\color{blue}Nature Astronomy}, \href{https://ui.adsabs.harvard.edu/abs/2025NatAs...9.1486L}{9, 1486}

\bibitem[{{Liu} {et~al.}(2026){Liu}, {Liu}, {Yao}, {Cheng}, {Qu}, {Liu}, {Cui}, \& {Fang}}]{Liu2026}
{Liu}, J., {Liu}, J., {Yao}, J., {et~al.} 2026, \href{http://dx.doi.org/10.3847/1538-3881/ae2c4f}{\color{blue}\aj}, \href{https://ui.adsabs.harvard.edu/abs/2026AJ....171...93L}{171, 93}

\bibitem[{{Liu} {et~al.}(2019){Liu}, {Dipierro}, {Ragusa}, {Lodato}, {Herczeg}, {Long}, {Harsono}, {Boehler}, {Menard}, {Johnstone}, {Pascucci}, {Pinilla}, {Salyk}, {van der Plas}, {Cabrit}, {Fischer}, {Hendler}, {Manara}, {Nisini}, {Rigliaco}, {Avenhaus}, {Banzatti}, \& {Gully-Santiago}}]{Liu2019}
{Liu}, Y., {Dipierro}, G., {Ragusa}, E., {et~al.} 2019, \href{http://dx.doi.org/10.1051/0004-6361/201834157}{\color{blue}\aap}, \href{https://ui.adsabs.harvard.edu/abs/2019A&A...622A..75L}{622, A75}

\bibitem[{{Liu} {et~al.}(2022){Liu}, {Linz}, {Fang}, {Henning}, {Wolf}, {Flock}, {Rosotti}, {Wang}, \& {Li}}]{Liu2022}
{Liu}, Y., {Linz}, H., {Fang}, M., {et~al.} 2022, \href{http://dx.doi.org/10.1051/0004-6361/202244505}{\color{blue}\aap}, \href{https://ui.adsabs.harvard.edu/abs/2022A&A...668A.175L}{668, A175}

\bibitem[{{Longarini} {et~al.}(2025{\natexlab{a}}){Longarini}, {Lodato}, {Rosotti}, {Andrews}, {Winter}, {Stadler}, {Izquierdo}, {Galloway-Sprietsma}, {Facchini}, {Curone}, {Benisty}, {Teague}, {Bae}, {Barraza-Alfaro}, {Cataldi}, {Czekala}, {Cuello}, {Fasano}, {Flock}, {Fukagawa}, {Garg}, {Hall}, {Hammond}, {Hardiman}, {Hilder}, {Huang}, {Ilee}, {Isella}, {Kanagawa}, {Lesur}, {Loomis}, {M{\'e}nard}, {Orihara}, {Pinte}, {Price}, {Testi}, {Fernandez}, {W{\"o}lfer}, {Yen}, {Yoshida}, \& {Zawadzki}}]{Longarini2025}
{Longarini}, C., {Lodato}, G., {Rosotti}, G., {et~al.} 2025{\natexlab{a}}, \href{http://dx.doi.org/10.3847/2041-8213/adc431}{\color{blue}\apjl}, \href{https://ui.adsabs.harvard.edu/abs/2025ApJ...984L..17L}{984, L17}

\bibitem[{{Longarini} {et~al.}(2025{\natexlab{b}}){Longarini}, {Price}, {Kratter}, {Lodato}, \& {Clarke}}]{Longarini2025_infall}
{Longarini}, C., {Price}, D.~J., {Kratter}, K.~M., {Lodato}, G., \& {Clarke}, C.~J. 2025{\natexlab{b}}, \href{http://dx.doi.org/10.1093/mnras/staf1018}{\color{blue}\mnras}, \href{https://ui.adsabs.harvard.edu/abs/2025MNRAS.541.1145L}{541, 1145}

\bibitem[{{Luhman}(2022)}]{Luhman2022}
{Luhman}, K.~L. 2022, \href{http://dx.doi.org/10.3847/1538-3881/ac35e2}{\color{blue}\aj}, \href{https://ui.adsabs.harvard.edu/abs/2022AJ....163...24L}{163, 24}

\bibitem[{{Maaskant} {et~al.}(2013){Maaskant}, {Honda}, {Waters}, {Tielens}, {Dominik}, {Min}, {Verhoeff}, {Meeus}, \& {van den Ancker}}]{Maaskant2013}
{Maaskant}, K.~M., {Honda}, M., {Waters}, L.~B.~F.~M., {et~al.} 2013, \href{http://dx.doi.org/10.1051/0004-6361/201321300}{\color{blue}\aap}, \href{https://ui.adsabs.harvard.edu/abs/2013A&A...555A..64M}{555, A64}

\bibitem[{{Malfait} {et~al.}(1998){Malfait}, {Bogaert}, \& {Waelkens}}]{Malfait1998}
{Malfait}, K., {Bogaert}, E., \& {Waelkens}, C. 1998, \aap, \href{https://ui.adsabs.harvard.edu/abs/1998A&A...331..211M}{331, 211}

\bibitem[{{Manara} {et~al.}(2023){Manara}, {Ansdell}, {Rosotti}, {Hughes}, {Armitage}, {Lodato}, \& {Williams}}]{Manara2023}
{Manara}, C.~F., {Ansdell}, M., {Rosotti}, G.~P., {et~al.} 2023, in Astronomical Society of the Pacific Conference Series, Vol. 534, Protostars and Planets VII, ed. S.~{Inutsuka}, Y.~{Aikawa}, T.~{Muto}, K.~{Tomida}, \& M.~{Tamura}, \href{https://ui.adsabs.harvard.edu/abs/2023ASPC..534..539M}{539}

\bibitem[{{Manara} {et~al.}(2018){Manara}, {Morbidelli}, \& {Guillot}}]{Manara2018}
{Manara}, C.~F., {Morbidelli}, A., \& {Guillot}, T. 2018, \href{http://dx.doi.org/10.1051/0004-6361/201834076}{\color{blue}\aap}, \href{https://ui.adsabs.harvard.edu/abs/2018A&A...618L...3M}{618, L3}

\bibitem[{{Manara} {et~al.}(2016){Manara}, {Rosotti}, {Testi}, {Natta}, {Alcal{\'a}}, {Williams}, {Ansdell}, {Miotello}, {van der Marel}, {Tazzari}, {Carpenter}, {Guidi}, {Mathews}, {Oliveira}, {Prusti}, \& {van Dishoeck}}]{Manara2016}
{Manara}, C.~F., {Rosotti}, G., {Testi}, L., {et~al.} 2016, \href{http://dx.doi.org/10.1051/0004-6361/201628549}{\color{blue}\aap}, \href{https://ui.adsabs.harvard.edu/abs/2016A&A...591L...3M}{591, L3}

\bibitem[{{Marcos-Arenal} {et~al.}(2021){Marcos-Arenal}, {Mendigut{\'\i}a}, {Koumpia}, {Oudmaijer}, {Vioque}, {Guzm{\'a}n-D{\'\i}az}, {Wichittanakom}, {de Wit}, {Montesinos}, \& {Ilee}}]{2021A&A...652A..68M}
{Marcos-Arenal}, P., {Mendigut{\'\i}a}, I., {Koumpia}, E., {et~al.} 2021, \href{http://dx.doi.org/10.1051/0004-6361/202140724}{\color{blue}\aap}, \href{https://ui.adsabs.harvard.edu/abs/2021A&A...652A..68M}{652, A68}

\bibitem[{{Marino} {et~al.}(2026){Marino}, {Matr{\`a}}, {Hughes}, {Ehrhardt}, {Kennedy}, {del Burgo}, {Brennan}, {Han}, {Jankovic}, {Lovell}, {Mac Manamon}, {Milli}, {Weber}, {Zawadzki}, {Bendahan-West}, {Fehr}, {Mansell}, {Olofsson}, {Pearce}, {Bayo}, {Matthews}, {L{\"o}hne}, {Wyatt}, {{\'A}brah{\'a}m}, {Bonduelle}, {Booth}, {Cataldi}, {Carpenter}, {Chiang}, {Ertel}, {Hales}, {Henning}, {K{\'o}sp{\'a}l}, {Krivov}, {Luppe}, {MacGregor}, {Marshall}, {Mo{\'o}r}, {P{\'e}rez}, {Sefilian}, {Sepulveda}, \& {Wilner}}]{Marino2026}
{Marino}, S., {Matr{\`a}}, L., {Hughes}, A.~M., {et~al.} 2026, \href{http://dx.doi.org/10.1051/0004-6361/202556489}{\color{blue}\aap}, \href{https://ui.adsabs.harvard.edu/abs/2026A&A...705A.195M}{705, A195}

\bibitem[{{Marois} {et~al.}(2008){Marois}, {Macintosh}, {Barman}, {Zuckerman}, {Song}, {Patience}, {Lafreni{\`e}re}, \& {Doyon}}]{Marois2008}
{Marois}, C., {Macintosh}, B., {Barman}, T., {et~al.} 2008, \href{http://dx.doi.org/10.1126/science.1166585}{\color{blue}Science}, \href{https://ui.adsabs.harvard.edu/abs/2008Sci...322.1348M}{322, 1348}

\bibitem[{{Marois} {et~al.}(2010){Marois}, {Zuckerman}, {Konopacky}, {Macintosh}, \& {Barman}}]{Marois2010}
{Marois}, C., {Zuckerman}, B., {Konopacky}, Q.~M., {Macintosh}, B., \& {Barman}, T. 2010, \href{http://dx.doi.org/10.1038/nature09684}{\color{blue}\nat}, \href{https://ui.adsabs.harvard.edu/abs/2010Natur.468.1080M}{468, 1080}

\bibitem[{{Mason} {et~al.}(2001){Mason}, {Wycoff}, {Hartkopf}, {Douglass}, \& {Worley}}]{Mason2001}
{Mason}, B.~D., {Wycoff}, G.~L., {Hartkopf}, W.~I., {Douglass}, G.~G., \& {Worley}, C.~E. 2001, \href{http://dx.doi.org/10.1086/323920}{\color{blue}\aj}, \href{https://ui.adsabs.harvard.edu/abs/2001AJ....122.3466M}{122, 3466}

\bibitem[{{Meeus} {et~al.}(2001){Meeus}, {Waters}, {Bouwman}, {van den Ancker}, {Waelkens}, \& {Malfait}}]{Meeus2001}
{Meeus}, G., {Waters}, L.~B.~F.~M., {Bouwman}, J., {et~al.} 2001, \href{http://dx.doi.org/10.1051/0004-6361:20000144}{\color{blue}\aap}, \href{https://ui.adsabs.harvard.edu/abs/2001A&A...365..476M}{365, 476}

\bibitem[{{M{\'e}nard} {et~al.}(2020){M{\'e}nard}, {Cuello}, {Ginski}, {van der Plas}, {Villenave}, {Gonzalez}, {Pinte}, {Benisty}, {Boccaletti}, {Price}, {Boehler}, {Chripko}, {de Boer}, {Dominik}, {Garufi}, {Gratton}, {Hagelberg}, {Henning}, {Langlois}, {Maire}, {Pinilla}, {Ruane}, {Schmid}, {van Holstein}, {Vigan}, {Zurlo}, {Hubin}, {Pavlov}, {Rochat}, {Sauvage}, \& {Stadler}}]{Menard2020}
{M{\'e}nard}, F., {Cuello}, N., {Ginski}, C., {et~al.} 2020, \href{http://dx.doi.org/10.1051/0004-6361/202038356}{\color{blue}\aap}, \href{https://ui.adsabs.harvard.edu/abs/2020A&A...639L...1M}{639, L1}

\bibitem[{{Mendigut{\'\i}a}(2020)}]{Mendigutia2020}
{Mendigut{\'\i}a}, I. 2020, \href{http://dx.doi.org/10.3390/galaxies8020039}{\color{blue}Galaxies}, \href{https://ui.adsabs.harvard.edu/abs/2020Galax...8...39M}{8, 39}

\bibitem[{{Mendigut{\'\i}a} {et~al.}(2011){Mendigut{\'\i}a}, {Calvet}, {Montesinos}, {Mora}, {Muzerolle}, {Eiroa}, {Oudmaijer}, \& {Mer{\'\i}n}}]{2011A&A...535A..99M}
{Mendigut{\'\i}a}, I., {Calvet}, N., {Montesinos}, B., {et~al.} 2011, \href{http://dx.doi.org/10.1051/0004-6361/201117444}{\color{blue}\aap}, \href{https://ui.adsabs.harvard.edu/abs/2011A&A...535A..99M}{535, A99}

\bibitem[{{Mendigut{\'\i}a} {et~al.}(2026){Mendigut{\'\i}a}, {Campbell-White}, {Montesinos}, {Maldonado}, {Fullana-Garc{\'\i}a}, {Mirouh}, {Meeus}, {Vioque}, {Sicilia-Aguilar}, {Zapatero-Osorio}, {Villaver}, \& {Kahar}}]{Mendigutia2026}
{Mendigut{\'\i}a}, I., {Campbell-White}, J., {Montesinos}, B., {et~al.} 2026, \href{https://ui.adsabs.harvard.edu/abs/2026arXiv260405040M}{arXiv e-prints, arXiv:2604.05040}

\bibitem[{{Mendigut{\'\i}a} {et~al.}(2024){Mendigut{\'\i}a}, {Lillo-Box}, {Vioque}, {Maldonado}, {Montesinos}, {Hu{\'e}lamo}, \& {Wang}}]{2024A&A...686L...1M}
{Mendigut{\'\i}a}, I., {Lillo-Box}, J., {Vioque}, M., {et~al.} 2024, \href{http://dx.doi.org/10.1051/0004-6361/202449368}{\color{blue}\aap}, \href{https://ui.adsabs.harvard.edu/abs/2024A&A...686L...1M}{686, L1}

\bibitem[{{Mendigut{\'\i}a} {et~al.}(2012){Mendigut{\'\i}a}, {Mora}, {Montesinos}, {Eiroa}, {Meeus}, {Mer{\'\i}n}, \& {Oudmaijer}}]{Mendigutia2012}
{Mendigut{\'\i}a}, I., {Mora}, A., {Montesinos}, B., {et~al.} 2012, \href{http://dx.doi.org/10.1051/0004-6361/201219110}{\color{blue}\aap}, \href{https://ui.adsabs.harvard.edu/abs/2012A&A...543A..59M}{543, A59}

\bibitem[{{Mendigut{\'\i}a} {et~al.}(2017){Mendigut{\'\i}a}, {Oudmaijer}, {Garufi}, {Lumsden}, {Hu{\'e}lamo}, {Cheetham}, {de Wit}, {Norris}, {Olguin}, \& {Tuthill}}]{Mendigutia2017}
{Mendigut{\'\i}a}, I., {Oudmaijer}, R.~D., {Garufi}, A., {et~al.} 2017, \href{http://dx.doi.org/10.1051/0004-6361/201731131}{\color{blue}\aap}, \href{https://ui.adsabs.harvard.edu/abs/2017A&A...608A.104M}{608, A104}

\bibitem[{{Mesa} {et~al.}(2022){Mesa}, {Ginski}, {Gratton}, {Ertel}, {Wagner}, {Bonavita}, {Fedele}, {Meyer}, {Henning}, {Langlois}, {Garufi}, {Antoniucci}, {Claudi}, {Defr{\`e}re}, {Desidera}, {Janson}, {Pawellek}, {Rigliaco}, {Squicciarini}, {Zurlo}, {Boccaletti}, {Bonnefoy}, {Cantalloube}, {Chauvin}, {Feldt}, {Hagelberg}, {Hugot}, {Lagrange}, {Lazzoni}, {Maurel}, {Perrot}, {Petit}, {Rouan}, \& {Vigan}}]{Mesa2022}
{Mesa}, D., {Ginski}, C., {Gratton}, R., {et~al.} 2022, \href{http://dx.doi.org/10.1051/0004-6361/202142219}{\color{blue}\aap}, \href{https://ui.adsabs.harvard.edu/abs/2022A&A...658A..63M}{658, A63}

\bibitem[{{Miley} {et~al.}(2025){Miley}, {Kennedy}, {Ribas}, {Macias}, {Carpenter}, {Vioque}, {Luhman}, {Haworth}, {Weber}, {Perez}, \& {Zurlo}}]{2025A&A...703A.235M}
{Miley}, J.~M., {Kennedy}, G.~M., {Ribas}, {\'A}., {et~al.} 2025, \href{http://dx.doi.org/10.1051/0004-6361/202554463}{\color{blue}\aap}, \href{https://ui.adsabs.harvard.edu/abs/2025A&A...703A.235M}{703, A235}

\bibitem[{{Monnier} {et~al.}(2019){Monnier}, {Harries}, {Bae}, {Setterholm}, {Laws}, {Aarnio}, {Adams}, {Andrews}, {Calvet}, {Espaillat}, {Hartmann}, {Kraus}, {McClure}, {Miller}, {Oppenheimer}, {Wilner}, \& {Zhu}}]{Monnier2019}
{Monnier}, J.~D., {Harries}, T.~J., {Bae}, J., {et~al.} 2019, \href{http://dx.doi.org/10.3847/1538-4357/aafe87}{\color{blue}\apj}, \href{https://ui.adsabs.harvard.edu/abs/2019ApJ...872..122M}{872, 122}

\bibitem[{{Mo{\'o}r} {et~al.}(2017){Mo{\'o}r}, {Cur{\'e}}, {K{\'o}sp{\'a}l}, {{\'A}brah{\'a}m}, {Csengeri}, {Eiroa}, {Gunawan}, {Henning}, {Hughes}, {Juh{\'a}sz}, {Pawellek}, \& {Wyatt}}]{2017ApJ...849..123M}
{Mo{\'o}r}, A., {Cur{\'e}}, M., {K{\'o}sp{\'a}l}, {\'A}., {et~al.} 2017, \href{http://dx.doi.org/10.3847/1538-4357/aa8e4e}{\color{blue}\apj}, \href{https://ui.adsabs.harvard.edu/abs/2017ApJ...849..123M}{849, 123}

\bibitem[{{Mulders} {et~al.}(2021){Mulders}, {Pascucci}, {Ciesla}, \& {Fernandes}}]{Mulders2021}
{Mulders}, G.~D., {Pascucci}, I., {Ciesla}, F.~J., \& {Fernandes}, R.~B. 2021, \href{http://dx.doi.org/10.3847/1538-4357/ac178e}{\color{blue}\apj}, \href{https://ui.adsabs.harvard.edu/abs/2021ApJ...920...66M}{920, 66}

\bibitem[{{Mulders} {et~al.}(2017){Mulders}, {Pascucci}, {Manara}, {Testi}, {Herczeg}, {Henning}, {Mohanty}, \& {Lodato}}]{Mulders2017}
{Mulders}, G.~D., {Pascucci}, I., {Manara}, C.~F., {et~al.} 2017, \href{http://dx.doi.org/10.3847/1538-4357/aa8906}{\color{blue}\apj}, \href{https://ui.adsabs.harvard.edu/abs/2017ApJ...847...31M}{847, 31}

\bibitem[{{Nakatani} {et~al.}(2023){Nakatani}, {Turner}, {Hasegawa}, {Cataldi}, {Aikawa}, {Marino}, \& {Kobayashi}}]{Nakatani2023}
{Nakatani}, R., {Turner}, N.~J., {Hasegawa}, Y., {et~al.} 2023, \href{http://dx.doi.org/10.3847/2041-8213/ad0ed8}{\color{blue}\apjl}, \href{https://ui.adsabs.harvard.edu/abs/2023ApJ...959L..28N}{959, L28}

\bibitem[{{Natta} {et~al.}(2006){Natta}, {Testi}, \& {Randich}}]{Natta2006}
{Natta}, A., {Testi}, L., \& {Randich}, S. 2006, \href{http://dx.doi.org/10.1051/0004-6361:20054706}{\color{blue}\aap}, \href{https://ui.adsabs.harvard.edu/abs/2006A&A...452..245N}{452, 245}

\bibitem[{{Nguyen} {et~al.}(2022){Nguyen}, {Costa}, {Girardi}, {Volpato}, {Bressan}, {Chen}, {Marigo}, {Fu}, \& {Goudfrooij}}]{2022A&A...665A.126N}
{Nguyen}, C.~T., {Costa}, G., {Girardi}, L., {et~al.} 2022, \href{http://dx.doi.org/10.1051/0004-6361/202244166}{\color{blue}\aap}, \href{https://ui.adsabs.harvard.edu/abs/2022A&A...665A.126N}{665, A126}

\bibitem[{{Nidhi} {et~al.}(2023){Nidhi}, {Mathew}, {Shridharan}, {Arun}, {Anusha}, \& {Kartha}}]{2023MNRAS.524.5166N}
{Nidhi}, S., {Mathew}, B., {Shridharan}, B., {et~al.} 2023, \href{http://dx.doi.org/10.1093/mnras/stad2067}{\color{blue}\mnras}, \href{https://ui.adsabs.harvard.edu/abs/2023MNRAS.524.5166N}{524, 5166}

\bibitem[{{Nielsen} {et~al.}(2019){Nielsen}, {De Rosa}, {Macintosh}, {Wang}, {Ruffio}, {Chiang}, {Marley}, {Saumon}, {Savransky}, {Ammons}, {Bailey}, {Barman}, {Blain}, {Bulger}, {Burrows}, {Chilcote}, {Cotten}, {Czekala}, {Doyon}, {Duch{\^e}ne}, {Esposito}, {Fabrycky}, {Fitzgerald}, {Follette}, {Fortney}, {Gerard}, {Goodsell}, {Graham}, {Greenbaum}, {Hibon}, {Hinkley}, {Hirsch}, {Hom}, {Hung}, {Dawson}, {Ingraham}, {Kalas}, {Konopacky}, {Larkin}, {Lee}, {Lin}, {Maire}, {Marchis}, {Marois}, {Metchev}, {Millar-Blanchaer}, {Morzinski}, {Oppenheimer}, {Palmer}, {Patience}, {Perrin}, {Poyneer}, {Pueyo}, {Rafikov}, {Rajan}, {Rameau}, {Rantakyr{\"o}}, {Ren}, {Schneider}, {Sivaramakrishnan}, {Song}, {Soummer}, {Tallis}, {Thomas}, {Ward-Duong}, \& {Wolff}}]{Nielsen2019}
{Nielsen}, E.~L., {De Rosa}, R.~J., {Macintosh}, B., {et~al.} 2019, \href{http://dx.doi.org/10.3847/1538-3881/ab16e9}{\color{blue}\aj}, \href{https://ui.adsabs.harvard.edu/abs/2019AJ....158...13N}{158, 13}

\bibitem[{{{\"O}berg} {et~al.}(2015){{\"O}berg}, {Guzm{\'a}n}, {Furuya}, {Qi}, {Aikawa}, {Andrews}, {Loomis}, \& {Wilner}}]{Oberg2015}
{{\"O}berg}, K.~I., {Guzm{\'a}n}, V.~V., {Furuya}, K., {et~al.} 2015, \href{http://dx.doi.org/10.1038/nature14276}{\color{blue}\nat}, \href{https://ui.adsabs.harvard.edu/abs/2015Natur.520..198O}{520, 198}

\bibitem[{{{\"O}berg} {et~al.}(2021){{\"O}berg}, {Guzm{\'a}n}, {Walsh}, {Aikawa}, {Bergin}, {Law}, {Loomis}, {Alarc{\'o}n}, {Andrews}, {Bae}, {Bergner}, {Boehler}, {Booth}, {Bosman}, {Calahan}, {Cataldi}, {Cleeves}, {Czekala}, {Furuya}, {Huang}, {Ilee}, {Kurtovic}, {Le Gal}, {Liu}, {Long}, {M{\'e}nard}, {Nomura}, {P{\'e}rez}, {Qi}, {Schwarz}, {Sierra}, {Teague}, {Tsukagoshi}, {Yamato}, {van't Hoff}, {Waggoner}, {Wilner}, \& {Zhang}}]{Oberg2021}
{{\"O}berg}, K.~I., {Guzm{\'a}n}, V.~V., {Walsh}, C., {et~al.} 2021, \href{http://dx.doi.org/10.3847/1538-4365/ac1432}{\color{blue}\apjs}, \href{https://ui.adsabs.harvard.edu/abs/2021ApJS..257....1O}{257, 1}

\bibitem[{{Offner} {et~al.}(2023){Offner}, {Moe}, {Kratter}, {Sadavoy}, {Jensen}, \& {Tobin}}]{Offner2023}
{Offner}, S.~S.~R., {Moe}, M., {Kratter}, K.~M., {et~al.} 2023, in Astronomical Society of the Pacific Conference Series, Vol. 534, Protostars and Planets VII, ed. S.~{Inutsuka}, Y.~{Aikawa}, T.~{Muto}, K.~{Tomida}, \& M.~{Tamura}, \href{https://ui.adsabs.harvard.edu/abs/2023ASPC..534..275O}{275}

\bibitem[{{Orcajo} {et~al.}(2025){Orcajo}, {Cieza}, {Guilera}, {P{\'e}rez}, {Rannou}, {Gonz{\'a}lez-Ruilova}, {Batalla-Falcon}, {Bhowmik}, {Chavan}, {Casassus}, {Dasgupta}, {Diaz}, {Gomez}, {Hales}, {Miley}, {Miller Bertolami}, {Nogueira}, {Ronco}, {Ruiz-Rodriguez}, {Sierra}, {Venturini}, {Weber}, {Williams}, \& {Zurlo}}]{Orcajo2025}
{Orcajo}, S., {Cieza}, L.~A., {Guilera}, O., {et~al.} 2025, \href{http://dx.doi.org/10.3847/2041-8213/adcd58}{\color{blue}\apjl}, \href{https://ui.adsabs.harvard.edu/abs/2025ApJ...984L..57O}{984, L57}

\bibitem[{{Padoan} {et~al.}(2025){Padoan}, {Pan}, {Pelkonen}, {Haugb{\o}lle}, \& {Nordlund}}]{Padoan2025}
{Padoan}, P., {Pan}, L., {Pelkonen}, V.-M., {Haugb{\o}lle}, T., \& {Nordlund}, {\AA}. 2025, \href{http://dx.doi.org/10.1038/s41550-025-02529-3}{\color{blue}Nature Astronomy}, \href{https://ui.adsabs.harvard.edu/abs/2025NatAs...9..862P}{9, 862}

\bibitem[{{Painter} {et~al.}(2025){Painter}, {Andrews}, {Chandler}, {Ueda}, {Wilner}, {Long}, {Macias}, {Carrasco-Gonzalez}, {Chung}, {Liu}, {Birnstiel}, \& {Hughes}}]{Painter2025}
{Painter}, C., {Andrews}, S.~M., {Chandler}, C.~J., {et~al.} 2025, \href{http://dx.doi.org/10.33232/001c.144268}{\color{blue}The Open Journal of Astrophysics}, \href{https://ui.adsabs.harvard.edu/abs/2025OJAp....8E.134P}{8, 134}

\bibitem[{pandas~development team(2020)}]{reback2020pandas}
pandas~development team, T. 2020, pandas-dev/pandas: Pandas

\bibitem[{{Pani{\'c}} {et~al.}(2021){Pani{\'c}}, {Haworth}, {Petr-Gotzens}, {Miley}, {van den Ancker}, {Vioque}, {Siess}, {Parker}, {Clarke}, {Kamp}, {Kennedy}, {Oudmaijer}, {Pascucci}, {Richards}, {Ratzka}, \& {Qi}}]{Panic2021}
{Pani{\'c}}, O., {Haworth}, T.~J., {Petr-Gotzens}, M.~G., {et~al.} 2021, \href{http://dx.doi.org/10.1093/mnras/staa3834}{\color{blue}\mnras}, \href{https://ui.adsabs.harvard.edu/abs/2021MNRAS.501.4317P}{501, 4317}

\bibitem[{{Pascual} {et~al.}(2016){Pascual}, {Montesinos}, {Meeus}, {Marshall}, {Mendigut{\'\i}a}, \& {Sandell}}]{Pascual2016}
{Pascual}, N., {Montesinos}, B., {Meeus}, G., {et~al.} 2016, \href{http://dx.doi.org/10.1051/0004-6361/201526605}{\color{blue}\aap}, \href{https://ui.adsabs.harvard.edu/abs/2016A&A...586A...6P}{586, A6}

\bibitem[{{Pascucci} {et~al.}(2016){Pascucci}, {Testi}, {Herczeg}, {Long}, {Manara}, {Hendler}, {Mulders}, {Krijt}, {Ciesla}, {Henning}, {Mohanty}, {Drabek-Maunder}, {Apai}, {Sz{\H{u}}cs}, {Sacco}, \& {Olofsson}}]{Pascucci2016}
{Pascucci}, I., {Testi}, L., {Herczeg}, G.~J., {et~al.} 2016, \href{http://dx.doi.org/10.3847/0004-637X/831/2/125}{\color{blue}\apj}, \href{https://ui.adsabs.harvard.edu/abs/2016ApJ...831..125P}{831, 125}

\bibitem[{{Pecaut} \& {Mamajek}(2013)}]{Pecaut2013}
{Pecaut}, M.~J. \& {Mamajek}, E.~E. 2013, \href{http://dx.doi.org/10.1088/0067-0049/208/1/9}{\color{blue}\apjs}, \href{https://ui.adsabs.harvard.edu/abs/2013ApJS..208....9P}{208, 9}

\bibitem[{{P{\'e}rez} {et~al.}(2019){P{\'e}rez}, {Casassus}, {Baruteau}, {Dong}, {Hales}, \& {Cieza}}]{Perez2019}
{P{\'e}rez}, S., {Casassus}, S., {Baruteau}, C., {et~al.} 2019, \href{http://dx.doi.org/10.3847/1538-3881/ab1f88}{\color{blue}\aj}, \href{https://ui.adsabs.harvard.edu/abs/2019AJ....158...15P}{158, 15}

\bibitem[{{Pfalzner} {et~al.}(2022){Pfalzner}, {Dehghani}, \& {Michel}}]{Pfalzner_ea_2022}
{Pfalzner}, S., {Dehghani}, S., \& {Michel}, A. 2022, \href{http://dx.doi.org/10.3847/2041-8213/ac9839}{\color{blue}\apjl}, \href{https://ui.adsabs.harvard.edu/abs/2022ApJ...939L..10P}{939, L10}

\bibitem[{{Pineda} {et~al.}(2019){Pineda}, {Szul{\'a}gyi}, {Quanz}, {van Dishoeck}, {Garufi}, {Meru}, {Mulders}, {Testi}, {Meyer}, \& {Reggiani}}]{Pineda2019}
{Pineda}, J.~E., {Szul{\'a}gyi}, J., {Quanz}, S.~P., {et~al.} 2019, \href{http://dx.doi.org/10.3847/1538-4357/aaf389}{\color{blue}\apj}, \href{https://ui.adsabs.harvard.edu/abs/2019ApJ...871...48P}{871, 48}

\bibitem[{{Ping} {et~al.}(2026){Ping}, {Anania}, {Pinilla}, \& {Vioque}}]{2026arXiv260222050P}
{Ping}, J., {Anania}, R., {Pinilla}, P., \& {Vioque}, M. 2026, \href{https://ui.adsabs.harvard.edu/abs/2026arXiv260222050P}{\href{http://dx.doi.org/10.48550/arXiv.2602.22050}{\color{blue}arXiv e-prints}, arXiv:2602.22050}

\bibitem[{{Pinilla} {et~al.}(2022{\natexlab{a}}){Pinilla}, {Benisty}, {Kurtovic}, {Bae}, {Dong}, {Zhu}, {Andrews}, {Carpenter}, {Ginski}, {Huang}, {Isella}, {P{\'e}rez}, {Ricci}, {Rosotti}, {Villenave}, \& {Wilner}}]{Pinilla2022b}
{Pinilla}, P., {Benisty}, M., {Kurtovic}, N.~T., {et~al.} 2022{\natexlab{a}}, \href{http://dx.doi.org/10.1051/0004-6361/202243704}{\color{blue}\aap}, \href{https://ui.adsabs.harvard.edu/abs/2022A&A...665A.128P}{665, A128}

\bibitem[{{Pinilla} {et~al.}(2022{\natexlab{b}}){Pinilla}, {Garufi}, \& {G{\'a}rate}}]{Pinilla2022a}
{Pinilla}, P., {Garufi}, A., \& {G{\'a}rate}, M. 2022{\natexlab{b}}, \href{http://dx.doi.org/10.1051/0004-6361/202243637}{\color{blue}\aap}, \href{https://ui.adsabs.harvard.edu/abs/2022A&A...662L...8P}{662, L8}

\bibitem[{{Pinilla} {et~al.}(2020){Pinilla}, {Pascucci}, \& {Marino}}]{Pinilla2020}
{Pinilla}, P., {Pascucci}, I., \& {Marino}, S. 2020, \href{http://dx.doi.org/10.1051/0004-6361/201937003}{\color{blue}\aap}, \href{https://ui.adsabs.harvard.edu/abs/2020A&A...635A.105P}{635, A105}

\bibitem[{{Pinte} {et~al.}(2018){Pinte}, {Price}, {M{\'e}nard}, {Duch{\^e}ne}, {Dent}, {Hill}, {de Gregorio-Monsalvo}, {Hales}, \& {Mentiplay}}]{Pinte2018}
{Pinte}, C., {Price}, D.~J., {M{\'e}nard}, F., {et~al.} 2018, \href{http://dx.doi.org/10.3847/2041-8213/aac6dc}{\color{blue}\apjl}, \href{https://ui.adsabs.harvard.edu/abs/2018ApJ...860L..13P}{860, L13}

\bibitem[{{Polnitzky} {et~al.}(2025){Polnitzky}, {Ratzenb{\"o}ck}, {Gro{\ss}schedl}, \& {Alves}}]{Polnitzky2025}
{Polnitzky}, F.~A., {Ratzenb{\"o}ck}, S., {Gro{\ss}schedl}, J.~E., \& {Alves}, J. 2025, \href{https://ui.adsabs.harvard.edu/abs/2025arXiv251206873P}{arXiv e-prints, arXiv:2512.06873}

\bibitem[{{Poorta} {et~al.}(2025){Poorta}, {Hogerheijde}, {de Koter}, {Kaper}, {Backs}, {Ram{\'\i}rez Tannus}, {McClure}, {Hygate}, {Rab}, {Klaassen}, \& {Derkink}}]{Poorta2025}
{Poorta}, J., {Hogerheijde}, M., {de Koter}, A., {et~al.} 2025, \href{http://dx.doi.org/10.1051/0004-6361/202451042}{\color{blue}\aap}, \href{https://ui.adsabs.harvard.edu/abs/2025A&A...694A.295P}{694, A295}

\bibitem[{{Price} \& {Bate}(2009)}]{Price2009}
{Price}, D.~J. \& {Bate}, M.~R. 2009, \href{http://dx.doi.org/10.1111/j.1365-2966.2009.14969.x}{\color{blue}\mnras}, \href{https://ui.adsabs.harvard.edu/abs/2009MNRAS.398...33P}{398, 33}

\bibitem[{{Price} {et~al.}(2018){Price}, {Cuello}, {Pinte}, {Mentiplay}, {Casassus}, {Christiaens}, {Kennedy}, {Cuadra}, {Sebastian Perez}, {Marino}, {Armitage}, {Zurlo}, {Juhasz}, {Ragusa}, {Laibe}, \& {Lodato}}]{Price2018}
{Price}, D.~J., {Cuello}, N., {Pinte}, C., {et~al.} 2018, \href{http://dx.doi.org/10.1093/mnras/sty647}{\color{blue}\mnras}, \href{https://ui.adsabs.harvard.edu/abs/2018MNRAS.477.1270P}{477, 1270}

\bibitem[{{Quintana} {et~al.}(2025){Quintana}, {Wright}, \& {Mart{\'\i}nez Garc{\'\i}a}}]{Quintana_ea_2025}
{Quintana}, A.~L., {Wright}, N.~J., \& {Mart{\'\i}nez Garc{\'\i}a}, J. 2025, \href{http://dx.doi.org/10.1093/mnras/staf083}{\color{blue}\mnras}, \href{https://ui.adsabs.harvard.edu/abs/2025MNRAS.538.1367Q}{538, 1367}

\bibitem[{{Radley} {et~al.}(2026){Radley}, {Ilee}, {Busquet}, {Liu}, {Pontoppidan}, {Ribas}, {Audard}, {Bianchi}, {Bourke}, {Codella}, {Coutens}, {Girart}, {Hoare}, {Jim{\'e}nez-Serra}, {Johnstone}, {Loinard}, {Pani{\'c}}, {Pineda}, {Podio}, {Tobin}, \& {Wilner}}]{Radley2026}
{Radley}, I.~C., {Ilee}, J.~D., {Busquet}, G., {et~al.} 2026, \href{https://ui.adsabs.harvard.edu/abs/2026arXiv260715468R}{\href{http://dx.doi.org/10.48550/arXiv.2607.15468}{\color{blue}arXiv e-prints}, arXiv:2607.15468}

\bibitem[{{Ragusa} {et~al.}(2025){Ragusa}, {Lodato}, {Cuello}, {Vioque}, {Manara}, \& {Toci}}]{2025A&A...698A.102R}
{Ragusa}, E., {Lodato}, G., {Cuello}, N., {et~al.} 2025, \href{http://dx.doi.org/10.1051/0004-6361/202554462}{\color{blue}\aap}, \href{https://ui.adsabs.harvard.edu/abs/2025A&A...698A.102R}{698, A102}

\bibitem[{{Reffert} {et~al.}(2015){Reffert}, {Bergmann}, {Quirrenbach}, {Trifonov}, \& {K{\"u}nstler}}]{Reffert2015}
{Reffert}, S., {Bergmann}, C., {Quirrenbach}, A., {Trifonov}, T., \& {K{\"u}nstler}, A. 2015, \href{http://dx.doi.org/10.1051/0004-6361/201322360}{\color{blue}\aap}, \href{https://ui.adsabs.harvard.edu/abs/2015A&A...574A.116R}{574, A116}

\bibitem[{{Ren} {et~al.}(2020){Ren}, {Dong}, {van Holstein}, {Ruffio}, {Calvin}, {Girard}, {Benisty}, {Boccaletti}, {Esposito}, {Choquet}, {Mawet}, {Pueyo}, {Stolker}, {Chiang}, {Boer}, {Debes}, {Garufi}, {Grady}, {Hines}, {Maire}, {M{\'e}nard}, {Millar-Blanchaer}, {Perrin}, {Poteet}, \& {Schneider}}]{Ren2020}
{Ren}, B., {Dong}, R., {van Holstein}, R.~G., {et~al.} 2020, \href{http://dx.doi.org/10.3847/2041-8213/aba43e}{\color{blue}\apjl}, \href{https://ui.adsabs.harvard.edu/abs/2020ApJ...898L..38R}{898, L38}

\bibitem[{{Ren} {et~al.}(2024){Ren}, {Xie}, {Benisty}, {Dong}, {Bae}, {Stolker}, {van Holstein}, {Debes}, {Garufi}, {Ginski}, \& {Kraus}}]{Ren2024}
{Ren}, B.~B., {Xie}, C., {Benisty}, M., {et~al.} 2024, \href{http://dx.doi.org/10.1051/0004-6361/202348114}{\color{blue}\aap}, \href{https://ui.adsabs.harvard.edu/abs/2024A&A...681L...2R}{681, L2}

\bibitem[{{Ribas} {et~al.}(2015){Ribas}, {Bouy}, \& {Mer{\'\i}n}}]{Ribas_ea_2015}
{Ribas}, {\'A}., {Bouy}, H., \& {Mer{\'\i}n}, B. 2015, \href{http://dx.doi.org/10.1051/0004-6361/201424846}{\color{blue}\aap}, \href{https://ui.adsabs.harvard.edu/abs/2015A&A...576A..52R}{576, A52}

\bibitem[{{Ronco} {et~al.}(2024){Ronco}, {Schreiber}, {Villaver}, {Guilera}, \& {Miller Bertolami}}]{Ronco2024}
{Ronco}, M.~P., {Schreiber}, M.~R., {Villaver}, E., {Guilera}, O.~M., \& {Miller Bertolami}, M.~M. 2024, \href{http://dx.doi.org/10.1051/0004-6361/202347762}{\color{blue}\aap}, \href{https://ui.adsabs.harvard.edu/abs/2024A&A...682A.155R}{682, A155}

\bibitem[{{Rosotti} {et~al.}(2020){Rosotti}, {Benisty}, {Juh{\'a}sz}, {Teague}, {Clarke}, {Dominik}, {Dullemond}, {Klaassen}, {Matr{\`a}}, \& {Stolker}}]{Rosotti2020}
{Rosotti}, G.~P., {Benisty}, M., {Juh{\'a}sz}, A., {et~al.} 2020, \href{http://dx.doi.org/10.1093/mnras/stz3090}{\color{blue}\mnras}, \href{https://ui.adsabs.harvard.edu/abs/2020MNRAS.491.1335R}{491, 1335}

\bibitem[{{Rosotti} {et~al.}(2017){Rosotti}, {Clarke}, {Manara}, \& {Facchini}}]{Rosotti2017}
{Rosotti}, G.~P., {Clarke}, C.~J., {Manara}, C.~F., \& {Facchini}, S. 2017, \href{http://dx.doi.org/10.1093/mnras/stx595}{\color{blue}\mnras}, \href{https://ui.adsabs.harvard.edu/abs/2017MNRAS.468.1631R}{468, 1631}

\bibitem[{{Rota} {et~al.}(2022){Rota}, {Manara}, {Miotello}, {Lodato}, {Facchini}, {Koutoulaki}, {Herczeg}, {Long}, {Tazzari}, {Cabrit}, {Harsono}, {M{\'e}nard}, {Pinilla}, {van der Plas}, {Ragusa}, \& {Yen}}]{Rota2022}
{Rota}, A.~A., {Manara}, C.~F., {Miotello}, A., {et~al.} 2022, \href{http://dx.doi.org/10.1051/0004-6361/202141035}{\color{blue}\aap}, \href{https://ui.adsabs.harvard.edu/abs/2022A&A...662A.121R}{662, A121}

\bibitem[{{Sanchis} {et~al.}(2020){Sanchis}, {Testi}, {Natta}, {Manara}, {Ercolano}, {Preibisch}, {Henning}, {Facchini}, {Miotello}, {de Gregorio-Monsalvo}, {Lopez}, {Mu{\v{z}}i{\'c}}, {Pascucci}, {Santamar{\'\i}a-Miranda}, {Scholz}, {Tazzari}, {van Terwisga}, \& {Williams}}]{Sanchis2020}
{Sanchis}, E., {Testi}, L., {Natta}, A., {et~al.} 2020, \href{http://dx.doi.org/10.1051/0004-6361/201936913}{\color{blue}\aap}, \href{https://ui.adsabs.harvard.edu/abs/2020A&A...633A.114S}{633, A114}

\bibitem[{{Sellek} {et~al.}(2020){Sellek}, {Booth}, \& {Clarke}}]{Sellek2020}
{Sellek}, A.~D., {Booth}, R.~A., \& {Clarke}, C.~J. 2020, \href{http://dx.doi.org/10.1093/mnras/staa2519}{\color{blue}\mnras}, \href{https://ui.adsabs.harvard.edu/abs/2020MNRAS.498.2845S}{498, 2845}

\bibitem[{{Semenov} {et~al.}(2021){Semenov}, {Kravtsov}, \& {Gnedin}}]{Semenov_ea_2021}
{Semenov}, V.~A., {Kravtsov}, A.~V., \& {Gnedin}, N.~Y. 2021, \href{http://dx.doi.org/10.3847/1538-4357/ac0a77}{\color{blue}\apj}, \href{https://ui.adsabs.harvard.edu/abs/2021ApJ...918...13S}{918, 13}

\bibitem[{{Shridharan} {et~al.}(2021){Shridharan}, {Mathew}, {Nidhi}, {Anusha}, {Arun}, {Kartha}, \& {Kumar}}]{Shridharan2021}
{Shridharan}, B., {Mathew}, B., {Nidhi}, S., {et~al.} 2021, \href{http://dx.doi.org/10.1088/1674-4527/21/11/288}{\color{blue}Research in Astronomy and Astrophysics}, \href{https://ui.adsabs.harvard.edu/abs/2021RAA....21..288S}{21, 288}

\bibitem[{{Somigliana} {et~al.}(2024){Somigliana}, {Testi}, {Rosotti}, {Toci}, {Lodato}, {Anania}, {Tabone}, {Tazzari}, {Klessen}, {Lebreuilly}, {Hennebelle}, \& {Molinari}}]{Somigliana2024}
{Somigliana}, A., {Testi}, L., {Rosotti}, G., {et~al.} 2024, \href{http://dx.doi.org/10.1051/0004-6361/202450744}{\color{blue}\aap}, \href{https://ui.adsabs.harvard.edu/abs/2024A&A...689A.285S}{689, A285}

\bibitem[{{Speedie} {et~al.}(2025){Speedie}, {Dong}, {Teague}, {Segura-Cox}, {Pineda}, {Calcino}, {Longarini}, {Hall}, {Tang}, {Hashimoto}, {Paneque-Carre{\~n}o}, {Lodato}, \& {Veronesi}}]{Speedie2025}
{Speedie}, J., {Dong}, R., {Teague}, R., {et~al.} 2025, \href{http://dx.doi.org/10.3847/2041-8213/adb7d5}{\color{blue}\apjl}, \href{https://ui.adsabs.harvard.edu/abs/2025ApJ...981L..30S}{981, L30}

\bibitem[{{Squicciarini} {et~al.}(2025){Squicciarini}, {Mazoyer}, {Lagrange}, {Chomez}, {Delorme}, {Flasseur}, {Kiefer}, {Bergeon}, {Albert}, \& {Meunier}}]{Squicciarini2025}
{Squicciarini}, V., {Mazoyer}, J., {Lagrange}, A.-M., {et~al.} 2025, \href{http://dx.doi.org/10.1051/0004-6361/202452310}{\color{blue}\aap}, \href{https://ui.adsabs.harvard.edu/abs/2025A&A...693A..54S}{693, A54}

\bibitem[{{Stadler} {et~al.}(2026){Stadler}, {Benisty}, {Zagaria}, {Izquierdo}, {Speedie}, {Winter}, {W{\"o}lfer}, {Bae}, {Facchini}, {Fasano}, {Kurtovic}, \& {Teague}}]{Stadler2026}
{Stadler}, J., {Benisty}, M., {Zagaria}, F., {et~al.} 2026, \href{https://ui.adsabs.harvard.edu/abs/2026arXiv260115262S}{\href{http://dx.doi.org/10.48550/arXiv.2601.15262}{\color{blue}arXiv e-prints}, arXiv:2601.15262}

\bibitem[{{Stapper} {et~al.}(2025{\natexlab{a}}){Stapper}, {Hogerheijde}, {van Dishoeck}, {Booth}, {Grant}, \& {van Terwisga}}]{Stapper2025a}
{Stapper}, L.~M., {Hogerheijde}, M.~R., {van Dishoeck}, E.~F., {et~al.} 2025{\natexlab{a}}, \href{http://dx.doi.org/10.1051/0004-6361/202450678}{\color{blue}\aap}, \href{https://ui.adsabs.harvard.edu/abs/2025A&A...693A..49S}{693, A49}

\bibitem[{{Stapper} {et~al.}(2024){Stapper}, {Hogerheijde}, {van Dishoeck}, {Lin}, {Ahmadi}, {Booth}, {Grant}, {Immer}, {Leemker}, \& {P{\'e}rez-S{\'a}nchez}}]{Stapper2024}
{Stapper}, L.~M., {Hogerheijde}, M.~R., {van Dishoeck}, E.~F., {et~al.} 2024, \href{http://dx.doi.org/10.1051/0004-6361/202347271}{\color{blue}\aap}, \href{https://ui.adsabs.harvard.edu/abs/2024A&A...682A.149S}{682, A149}

\bibitem[{{Stapper} {et~al.}(2022){Stapper}, {Hogerheijde}, {van Dishoeck}, \& {Mentel}}]{Stapper2022}
{Stapper}, L.~M., {Hogerheijde}, M.~R., {van Dishoeck}, E.~F., \& {Mentel}, R. 2022, \href{http://dx.doi.org/10.1051/0004-6361/202142164}{\color{blue}\aap}, \href{https://ui.adsabs.harvard.edu/abs/2022A&A...658A.112S}{658, A112}

\bibitem[{{Stapper} {et~al.}(2025{\natexlab{b}}){Stapper}, {Hogerheijde}, {van Dishoeck}, {Vioque}, {Williams}, \& {Ginski}}]{Stapper2025b}
{Stapper}, L.~M., {Hogerheijde}, M.~R., {van Dishoeck}, E.~F., {et~al.} 2025{\natexlab{b}}, \href{http://dx.doi.org/10.1051/0004-6361/202450260}{\color{blue}\aap}, \href{https://ui.adsabs.harvard.edu/abs/2025A&A...693A.286S}{693, A286}

\bibitem[{{Stolker} {et~al.}(2016){Stolker}, {Dominik}, {Avenhaus}, {Min}, {de Boer}, {Ginski}, {Schmid}, {Juhasz}, {Bazzon}, {Waters}, {Garufi}, {Augereau}, {Benisty}, {Boccaletti}, {Henning}, {Langlois}, {Maire}, {M{\'e}nard}, {Meyer}, {Pinte}, {Quanz}, {Thalmann}, {Beuzit}, {Carbillet}, {Costille}, {Dohlen}, {Feldt}, {Gisler}, {Mouillet}, {Pavlov}, {Perret}, {Petit}, {Pragt}, {Rochat}, {Roelfsema}, {Salasnich}, {Soenke}, \& {Wildi}}]{Stolker2016}
{Stolker}, T., {Dominik}, C., {Avenhaus}, H., {et~al.} 2016, \href{http://dx.doi.org/10.1051/0004-6361/201528039}{\color{blue}\aap}, \href{https://ui.adsabs.harvard.edu/abs/2016A&A...595A.113S}{595, A113}

\bibitem[{{Stolker} {et~al.}(2025){Stolker}, {Samland}, {Waters}, {van den Ancker}, {Balmer}, {Lacour}, {Sitko}, {Wang}, {Nowak}, {Maire}, {Kammerer}, {Otten}, {Abuter}, {Amorim}, {Benisty}, {Berger}, {Beust}, {Blunt}, {Boccaletti}, {Bonnefoy}, {Bonnet}, {Bordoni}, {Bourdarot}, {Brandner}, {Cantalloube}, {Caselli}, {Charnay}, {Chauvin}, {Chavez}, {Chomez}, {Choquet}, {Christiaens}, {Cl{\'e}net}, {Coud{\'e} du Foresto}, {Cridland}, {Davies}, {Dembet}, {Dexter}, {Dominik}, {Drescher}, {Duvert}, {Eckart}, {Eisenhauer}, {F{\"o}rster Schreiber}, {Garcia}, {Garcia Lopez}, {Gardner}, {Gendron}, {Genzel}, {Gillessen}, {Girard}, {Grant}, {Haubois}, {Hei{\ss}el}, {Henning}, {Hinkley}, {Hippler}, {Houll{\'e}}, {Hubert}, {Jocou}, {Keppler}, {Kervella}, {Kreidberg}, {Kurtovic}, {Lagrange}, {Lapeyr{\`e}re}, {Le Bouquin}, {Lutz}, {Mang}, {Marleau}, {M{\'e}rand}, {Min}, {Molli{\`e}re}, {Monnier}, {Mordasini}, {Mouillet}, {Nasedkin}, {Ott}, {Paladini}, {Paumard}, {Perraut}, {Perrin}, {Pfuhl}, {Pourr{\'e}}, {Pueyo}, {Quanz},
  {Ribeiro}, {Rickman}, {Rustamkulov}, {Shangguan}, {Shimizu}, {Sing}, {Stadler}, {Straub}, {Straubmeier}, {Sturm}, {Tacconi}, {van Dishoeck}, {Vigan}, {Vincent}, {von Fellenberg}, {Widmann}, {Winterhalder}, {Woillez}, \& {Yazici}}]{Stolker2025}
{Stolker}, T., {Samland}, M., {Waters}, L.~B.~F.~M., {et~al.} 2025, \href{http://dx.doi.org/10.1051/0004-6361/202555064}{\color{blue}\aap}, \href{https://ui.adsabs.harvard.edu/abs/2025A&A...700A..21S}{700, A21}

\bibitem[{{Tabone} {et~al.}(2022){Tabone}, {Rosotti}, {Lodato}, {Armitage}, {Cridland}, \& {van Dishoeck}}]{Tabone2022}
{Tabone}, B., {Rosotti}, G.~P., {Lodato}, G., {et~al.} 2022, \href{http://dx.doi.org/10.1093/mnrasl/slab124}{\color{blue}\mnras}, \href{https://ui.adsabs.harvard.edu/abs/2022MNRAS.512L..74T}{512, L74}

\bibitem[{{Tabone} {et~al.}(2025){Tabone}, {Rosotti}, {Trapman}, {Pinilla}, {Pascucci}, {Somigliana}, {Alexander}, {Vioque}, {Anania}, {Kuznetsova}, {Zhang}, {P{\'e}rez}, {Cieza}, {Carpenter}, {Deng}, {Agurto-Gangas}, {Ruiz-Rodriguez}, {Sierra}, {Kurtovic}, {Miley}, {Gonz{\'a}lez-Ruilova}, {TorresVillanueva}, {Hogerheijde}, {Schwarz}, {Toci}, {Testi}, \& {Lodato}}]{Tabone2025}
{Tabone}, B., {Rosotti}, G.~P., {Trapman}, L., {et~al.} 2025, \href{http://dx.doi.org/10.3847/1538-4357/adc7b1}{\color{blue}\apj}, \href{https://ui.adsabs.harvard.edu/abs/2025ApJ...989....7T}{989, 7}

\bibitem[{{Tang} {et~al.}(2017){Tang}, {Guilloteau}, {Dutrey}, {Muto}, {Shen}, {Gu}, {Inutsuka}, {Momose}, {Pietu}, {Fukagawa}, {Chapillon}, {Ho}, {di Folco}, {Corder}, {Ohashi}, \& {Hashimoto}}]{Tang2017}
{Tang}, Y.-W., {Guilloteau}, S., {Dutrey}, A., {et~al.} 2017, \href{http://dx.doi.org/10.3847/1538-4357/aa6af7}{\color{blue}\apj}, \href{https://ui.adsabs.harvard.edu/abs/2017ApJ...840...32T}{840, 32}

\bibitem[{{Teague} {et~al.}(2025){Teague}, {Benisty}, {Facchini}, {Fukagawa}, {Pinte}, {Andrews}, {Bae}, {Barraza-Alfaro}, {Cataldi}, {Cuello}, {Curone}, {Czekala}, {Fasano}, {Flock}, {Galloway-Sprietsma}, {Garg}, {Hall}, {Hammond}, {Hilder}, {Huang}, {Ilee}, {Izquierdo}, {Kanagawa}, {Lesur}, {Lodato}, {Longarini}, {Loomis}, {Masset}, {Menard}, {Orihara}, {Price}, {Rosotti}, {Stadler}, {Testi}, {Yen}, {Wafflard-Fernandez}, {Wilner}, {Winter}, {W{\"o}lfer}, {Yoshida}, \& {Zawadzki}}]{Teague2025}
{Teague}, R., {Benisty}, M., {Facchini}, S., {et~al.} 2025, \href{http://dx.doi.org/10.3847/2041-8213/adc43b}{\color{blue}\apjl}, \href{https://ui.adsabs.harvard.edu/abs/2025ApJ...984L...6T}{984, L6}

\bibitem[{{Temmink} {et~al.}(2023){Temmink}, {Booth}, {van der Marel}, \& {van Dishoeck}}]{Temmink2023}
{Temmink}, M., {Booth}, A.~S., {van der Marel}, N., \& {van Dishoeck}, E.~F. 2023, \href{http://dx.doi.org/10.1051/0004-6361/202346272}{\color{blue}\aap}, \href{https://ui.adsabs.harvard.edu/abs/2023A&A...675A.131T}{675, A131}

\bibitem[{Temmink {et~al.}(2026)Temmink, van Dishoeck, Booth, van~der Marel, Benisty, \& Hogerheijde}]{Temmink2026}
Temmink, M., van Dishoeck, E.~F., Booth, A.~S., {et~al.} 2026, The asymmetric carbon-rich chemistry of the planet-forming disk of HD 142527 triggered by late infall

\bibitem[{{Testi} {et~al.}(2022{\natexlab{a}}){Testi}, {Natta}, {Manara}, {de Gregorio Monsalvo}, {Lodato}, {Lopez}, {Muzic}, {Pascucci}, {Sanchis}, {Miranda}, {Scholz}, {De Simone}, \& {Williams}}]{Testi2022}
{Testi}, L., {Natta}, A., {Manara}, C.~F., {et~al.} 2022{\natexlab{a}}, \href{http://dx.doi.org/10.1051/0004-6361/202141380}{\color{blue}\aap}, \href{https://ui.adsabs.harvard.edu/abs/2022A&A...663A..98T}{663, A98}

\bibitem[{{Testi} {et~al.}(2022{\natexlab{b}}){Testi}, {Natta}, {Manara}, {de Gregorio Monsalvo}, {Lodato}, {Lopez}, {Muzic}, {Pascucci}, {Sanchis}, {Miranda}, {Scholz}, {De Simone}, \& {Williams}}]{2022A&A...663A..98T}
{Testi}, L., {Natta}, A., {Manara}, C.~F., {et~al.} 2022{\natexlab{b}}, \href{http://dx.doi.org/10.1051/0004-6361/202141380}{\color{blue}\aap}, \href{https://ui.adsabs.harvard.edu/abs/2022A&A...663A..98T}{663, A98}

\bibitem[{{Th{\'e}} {et~al.}(1994){Th{\'e}}, {de Winter}, \& {Perez}}]{The1994}
{Th{\'e}}, P.~S., {de Winter}, D., \& {Perez}, M.~R. 1994, \aaps, \href{https://ui.adsabs.harvard.edu/abs/1994A&AS..104..315T}{104, 315}

\bibitem[{{Thomas} {et~al.}(2023){Thomas}, {Rodgers}, {van der Bliek}, {Doppmann}, {Bouvier}, {Salvo}, {Beuzit}, \& {Rigaut}}]{2023AJ....165..135T}
{Thomas}, S.~J., {Rodgers}, B., {van der Bliek}, N.~S., {et~al.} 2023, \href{http://dx.doi.org/10.3847/1538-3881/aca803}{\color{blue}\aj}, \href{https://ui.adsabs.harvard.edu/abs/2023AJ....165..135T}{165, 135}

\bibitem[{{Uyama} {et~al.}(2020){Uyama}, {Muto}, {Mawet}, {Christiaens}, {Hashimoto}, {Kudo}, {Kuzuhara}, {Ruane}, {Beichman}, {Absil}, {Akiyama}, {Bae}, {Bottom}, {Choquet}, {Currie}, {Dong}, {Follette}, {Fukagawa}, {Guidi}, {Huby}, {Kwon}, {Mayama}, {Meshkat}, {Reggiani}, {Ricci}, {Serabyn}, {Tamura}, {Testi}, {Wallack}, {Williams}, \& {Zhu}}]{Uyama2020}
{Uyama}, T., {Muto}, T., {Mawet}, D., {et~al.} 2020, \href{http://dx.doi.org/10.3847/1538-3881/ab7006}{\color{blue}\aj}, \href{https://ui.adsabs.harvard.edu/abs/2020AJ....159..118U}{159, 118}

\bibitem[{{Vacca} \& {Sandell}(2022)}]{2022ApJ...941..189V}
{Vacca}, W.~D. \& {Sandell}, G. 2022, \href{http://dx.doi.org/10.3847/1538-4357/ac94c4}{\color{blue}\apj}, \href{https://ui.adsabs.harvard.edu/abs/2022ApJ...941..189V}{941, 189}

\bibitem[{{Valeg{\r{a}}rd} {et~al.}(2022){Valeg{\r{a}}rd}, {Ginski}, {Dominik}, {Bae}, {Benisty}, {Birnstiel}, {Facchini}, {Garufi}, {Hogerheijde}, {van Holstein}, {Langlois}, {Manara}, {Pinilla}, {Rab}, {Ribas}, {Waters}, \& {Williams}}]{Valegard2022}
{Valeg{\r{a}}rd}, P.-G., {Ginski}, C., {Dominik}, C., {et~al.} 2022, \href{http://dx.doi.org/10.1051/0004-6361/202244001}{\color{blue}\aap}, \href{https://ui.adsabs.harvard.edu/abs/2022A&A...668A..25V}{668, A25}

\bibitem[{{Valeg{\r{a}}rd} {et~al.}(2021){Valeg{\r{a}}rd}, {Waters}, \& {Dominik}}]{Valegard2021}
{Valeg{\r{a}}rd}, P.~G., {Waters}, L.~B.~F.~M., \& {Dominik}, C. 2021, \href{http://dx.doi.org/10.1051/0004-6361/202039802}{\color{blue}\aap}, \href{https://ui.adsabs.harvard.edu/abs/2021A&A...652A.133V}{652, A133}

\bibitem[{{van den Ancker} {et~al.}(2021){van den Ancker}, {Gentile Fusillo}, {Haworth}, {Manara}, {Miles-P{\'a}ez}, {Oudmaijer}, {Pani{\'c}}, {Petit dit de la Roche}, {Petr-Gotzens}, \& {Vioque}}]{2021A&A...651L..11V}
{van den Ancker}, M.~E., {Gentile Fusillo}, N.~P., {Haworth}, T.~J., {et~al.} 2021, \href{http://dx.doi.org/10.1051/0004-6361/202141070}{\color{blue}\aap}, \href{https://ui.adsabs.harvard.edu/abs/2021A&A...651L..11V}{651, L11}

\bibitem[{{van der Marel} \& {Mulders}(2021)}]{vanderMarel2021}
{van der Marel}, N. \& {Mulders}, G.~D. 2021, \href{http://dx.doi.org/10.3847/1538-3881/ac0255}{\color{blue}\aj}, \href{https://ui.adsabs.harvard.edu/abs/2021AJ....162...28V}{162, 28}

\bibitem[{{van der Plas} {et~al.}(2017{\natexlab{a}}){van der Plas}, {M{\'e}nard}, {Canovas}, {Avenhaus}, {Casassus}, {Pinte}, {Caceres}, \& {Cieza}}]{vanderPlas2017a}
{van der Plas}, G., {M{\'e}nard}, F., {Canovas}, H., {et~al.} 2017{\natexlab{a}}, \href{http://dx.doi.org/10.1051/0004-6361/201731392}{\color{blue}\aap}, \href{https://ui.adsabs.harvard.edu/abs/2017A&A...607A..55V}{607, A55}

\bibitem[{{van der Plas} {et~al.}(2016){van der Plas}, {M{\'e}nard}, {Ward-Duong}, {Bulger}, {Harvey}, {Pinte}, {Patience}, {Hales}, \& {Casassus}}]{vanderPlas2016}
{van der Plas}, G., {M{\'e}nard}, F., {Ward-Duong}, K., {et~al.} 2016, \href{http://dx.doi.org/10.3847/0004-637X/819/2/102}{\color{blue}\apj}, \href{https://ui.adsabs.harvard.edu/abs/2016ApJ...819..102V}{819, 102}

\bibitem[{{van der Plas} {et~al.}(2017{\natexlab{b}}){van der Plas}, {Wright}, {M{\'e}nard}, {Casassus}, {Canovas}, {Pinte}, {Maddison}, {Maaskant}, {Avenhaus}, {Cieza}, {Perez}, \& {Ubach}}]{vanderPlas2017b}
{van der Plas}, G., {Wright}, C.~M., {M{\'e}nard}, F., {et~al.} 2017{\natexlab{b}}, \href{http://dx.doi.org/10.1051/0004-6361/201629523}{\color{blue}\aap}, \href{https://ui.adsabs.harvard.edu/abs/2017A&A...597A..32V}{597, A32}

\bibitem[{{van der Velden}(2020)}]{cmasher}
{van der Velden}, E. 2020, \href{http://dx.doi.org/10.21105/joss.02004}{\color{blue}The Journal of Open Source Software}, \href{https://ui.adsabs.harvard.edu/abs/2020JOSS....5.2004V}{5, 2004}

\bibitem[{{van Kempen} {et~al.}(2010){van Kempen}, {Green}, {Evans}, {van Dishoeck}, {Kristensen}, {Herczeg}, {Mer{\'\i}n}, {Lee}, {J{\o}rgensen}, {Bouwman}, {Acke}, {Adamkovics}, {Augereau}, {Bergin}, {Blake}, {Brown}, {Carr}, {Chen}, {Cieza}, {Dominik}, {Dullemond}, {Dunham}, {Glassgold}, {G{\"u}del}, {Harvey}, {Henning}, {Hogerheijde}, {Jaffe}, {Kim}, {Knez}, {Lacy}, {Maret}, {Meeus}, {Meijerink}, {Mulders}, {Mundy}, {Najita}, {Olofsson}, {Pontoppidan}, {Salyk}, {Sturm}, {Visser}, {Waters}, {Waelkens}, \& {Y{\i}ld{\i}z}}]{vanKempen2010}
{van Kempen}, T.~A., {Green}, J.~D., {Evans}, N.~J., {et~al.} 2010, \href{http://dx.doi.org/10.1051/0004-6361/201014686}{\color{blue}\aap}, \href{https://ui.adsabs.harvard.edu/abs/2010A&A...518L.128V}{518, L128}

\bibitem[{{van Terwisga} {et~al.}(2022){van Terwisga}, {Hacar}, {van Dishoeck}, {Oonk}, \& {Portegies Zwart}}]{vanTerwisga2022}
{van Terwisga}, S.~E., {Hacar}, A., {van Dishoeck}, E.~F., {Oonk}, R., \& {Portegies Zwart}, S. 2022, \href{http://dx.doi.org/10.1051/0004-6361/202141913}{\color{blue}\aap}, \href{https://ui.adsabs.harvard.edu/abs/2022A&A...661A..53V}{661, A53}

\bibitem[{{Vides} {et~al.}(2023){Vides}, {Sallum}, {Eisner}, {Skemer}, \& {Murray-Clay}}]{2023ApJ...958..123V}
{Vides}, C.~L., {Sallum}, S., {Eisner}, J., {Skemer}, A., \& {Murray-Clay}, R. 2023, \href{http://dx.doi.org/10.3847/1538-4357/acfda6}{\color{blue}\apj}, \href{https://ui.adsabs.harvard.edu/abs/2023ApJ...958..123V}{958, 123}

\bibitem[{{Vieira} {et~al.}(2011){Vieira}, {Gregorio-Hetem}, {Hetem}, {Stasi{\'n}ska}, \& {Szczerba}}]{2011A&A...526A..24V}
{Vieira}, R.~G., {Gregorio-Hetem}, J., {Hetem}, A., {Stasi{\'n}ska}, G., \& {Szczerba}, R. 2011, \href{http://dx.doi.org/10.1051/0004-6361/201015592}{\color{blue}\aap}, \href{https://ui.adsabs.harvard.edu/abs/2011A&A...526A..24V}{526, A24}

\bibitem[{{Vieira} {et~al.}(2003){Vieira}, {Corradi}, {Alencar}, {Mendes}, {Torres}, {Quast}, {Guimar{\~a}es}, \& {da Silva}}]{Vieira2003}
{Vieira}, S.~L.~A., {Corradi}, W.~J.~B., {Alencar}, S.~H.~P., {et~al.} 2003, \href{http://dx.doi.org/10.1086/379553}{\color{blue}\aj}, \href{https://ui.adsabs.harvard.edu/abs/2003AJ....126.2971V}{126, 2971}

\bibitem[{{Vioque} {et~al.}(2023){Vioque}, {Cavieres}, {Pantaleoni Gonz{\'a}lez}, {Ribas}, {Oudmaijer}, {Mendigut{\'\i}a}, {Kilian}, {C{\'a}novas}, \& {Kuhn}}]{Vioque2023}
{Vioque}, M., {Cavieres}, M., {Pantaleoni Gonz{\'a}lez}, M., {et~al.} 2023, \href{http://dx.doi.org/10.3847/1538-3881/acf75f}{\color{blue}\aj}, \href{https://ui.adsabs.harvard.edu/abs/2023AJ....166..183V}{166, 183}

\bibitem[{{Vioque} {et~al.}(2018){Vioque}, {Oudmaijer}, {Baines}, {Mendigut{\'\i}a}, \& {P{\'e}rez-Mart{\'\i}nez}}]{Vioque2018}
{Vioque}, M., {Oudmaijer}, R.~D., {Baines}, D., {Mendigut{\'\i}a}, I., \& {P{\'e}rez-Mart{\'\i}nez}, R. 2018, \href{http://dx.doi.org/10.1051/0004-6361/201832870}{\color{blue}\aap}, \href{https://ui.adsabs.harvard.edu/abs/2018A&A...620A.128V}{620, A128}

\bibitem[{{Vioque} {et~al.}(2020){Vioque}, {Oudmaijer}, {Schreiner}, {Mendigut{\'\i}a}, {Baines}, {Mowlavi}, \& {P{\'e}rez-Mart{\'\i}nez}}]{Vioque2020}
{Vioque}, M., {Oudmaijer}, R.~D., {Schreiner}, M., {et~al.} 2020, \href{http://dx.doi.org/10.1051/0004-6361/202037731}{\color{blue}\aap}, \href{https://ui.adsabs.harvard.edu/abs/2020A&A...638A..21V}{638, A21}

\bibitem[{{Vioque} {et~al.}(2022){Vioque}, {Oudmaijer}, {Wichittanakom}, {Mendigut{\'\i}a}, {Baines}, {Pani{\'c}}, {Iglesias}, {Miley}, \& {P{\'e}rez-Mart{\'\i}nez}}]{Vioque2022}
{Vioque}, M., {Oudmaijer}, R.~D., {Wichittanakom}, C., {et~al.} 2022, \href{http://dx.doi.org/10.3847/1538-4357/ac5c46}{\color{blue}\apj}, \href{https://ui.adsabs.harvard.edu/abs/2022ApJ...930...39V}{930, 39}

\bibitem[{Virtanen {et~al.}(2020)Virtanen, Gommers, Oliphant, Haberland, Reddy, Cournapeau, Burovski, Peterson, Weckesser, Bright, {van der Walt}, Brett, Wilson, Millman, Mayorov, Nelson, Jones, Kern, Larson, Carey, Polat, Feng, Moore, {VanderPlas}, Laxalde, Perktold, Cimrman, Henriksen, Quintero, Harris, Archibald, Ribeiro, Pedregosa, {van Mulbregt}, \& {SciPy 1.0 Contributors}}]{2020SciPy-NMeth}
Virtanen, P., Gommers, R., Oliphant, T.~E., {et~al.} 2020, \href{http://dx.doi.org/10.1038/s41592-019-0686-2}{\color{blue}Nature Methods}, \href{https://rdcu.be/b08Wh}{17, 261}

\bibitem[{{Waters} \& {Waelkens}(1998)}]{Waters1998}
{Waters}, L.~B.~F.~M. \& {Waelkens}, C. 1998, \href{http://dx.doi.org/10.1146/annurev.astro.36.1.233}{\color{blue}\araa}, \href{https://ui.adsabs.harvard.edu/abs/1998ARA&A..36..233W}{36, 233}

\bibitem[{{Wheelwright} {et~al.}(2010){Wheelwright}, {Oudmaijer}, \& {Goodwin}}]{Wheelwright2010}
{Wheelwright}, H.~E., {Oudmaijer}, R.~D., \& {Goodwin}, S.~P. 2010, \href{http://dx.doi.org/10.1111/j.1365-2966.2009.15708.x}{\color{blue}\mnras}, \href{https://ui.adsabs.harvard.edu/abs/2010MNRAS.401.1199W}{401, 1199}

\bibitem[{{Wichittanakom} {et~al.}(2020){Wichittanakom}, {Oudmaijer}, {Fairlamb}, {Mendigut{\'\i}a}, {Vioque}, \& {Ababakr}}]{Wichittanakom2020}
{Wichittanakom}, C., {Oudmaijer}, R.~D., {Fairlamb}, J.~R., {et~al.} 2020, \href{http://dx.doi.org/10.1093/mnras/staa169}{\color{blue}\mnras}, \href{https://ui.adsabs.harvard.edu/abs/2020MNRAS.493..234W}{493, 234}

\bibitem[{{Williams} {et~al.}(2019){Williams}, {Cieza}, {Hales}, {Ansdell}, {Ruiz-Rodriguez}, {Casassus}, {Perez}, \& {Zurlo}}]{Williams2019}
{Williams}, J.~P., {Cieza}, L., {Hales}, A., {et~al.} 2019, \href{http://dx.doi.org/10.3847/2041-8213/ab1338}{\color{blue}\apjl}, \href{https://ui.adsabs.harvard.edu/abs/2019ApJ...875L...9W}{875, L9}

\bibitem[{{Wilner} {et~al.}(2026){Wilner}, {Lovell}, {Andrews}, {Vioque}, {Long}, \& {Matr{\`a}}}]{Wilner2026}
{Wilner}, D.~J., {Lovell}, J.~B., {Andrews}, S.~M., {et~al.} 2026, \href{http://dx.doi.org/10.3847/1538-4357/ae6f01}{\color{blue}\apj}, \href{https://ui.adsabs.harvard.edu/abs/2026ApJ..1005...41W}{1005, 41}

\bibitem[{{Winter} {et~al.}(2024){Winter}, {Benisty}, {Manara}, \& {Gupta}}]{Winter2024}
{Winter}, A.~J., {Benisty}, M., {Manara}, C.~F., \& {Gupta}, A. 2024, \href{http://dx.doi.org/10.1051/0004-6361/202452120}{\color{blue}\aap}, \href{https://ui.adsabs.harvard.edu/abs/2024A&A...691A.169W}{691, A169}

\bibitem[{{Wittenmyer} {et~al.}(2020){Wittenmyer}, {Butler}, {Horner}, {Clark}, {Tinney}, {Carter}, {Wang}, {Johnson}, \& {Collins}}]{Wittenmyer2020}
{Wittenmyer}, R.~A., {Butler}, R.~P., {Horner}, J., {et~al.} 2020, \href{http://dx.doi.org/10.1093/mnras/stz3378}{\color{blue}\mnras}, \href{https://ui.adsabs.harvard.edu/abs/2020MNRAS.491.5248W}{491, 5248}

\bibitem[{{W{\"o}lfer} {et~al.}(2021){W{\"o}lfer}, {Facchini}, {Kurtovic}, {Teague}, {van Dishoeck}, {Benisty}, {Ercolano}, {Lodato}, {Miotello}, {Rosotti}, {Testi}, \& {Ubeira Gabellini}}]{Wolfer2021}
{W{\"o}lfer}, L., {Facchini}, S., {Kurtovic}, N.~T., {et~al.} 2021, \href{http://dx.doi.org/10.1051/0004-6361/202039469}{\color{blue}\aap}, \href{https://ui.adsabs.harvard.edu/abs/2021A&A...648A..19W}{648, A19}

\bibitem[{{Wolthoff} {et~al.}(2022){Wolthoff}, {Reffert}, {Quirrenbach}, {Jones}, {Wittenmyer}, \& {Jenkins}}]{Wolthoff2022}
{Wolthoff}, V., {Reffert}, S., {Quirrenbach}, A., {et~al.} 2022, \href{http://dx.doi.org/10.1051/0004-6361/202142501}{\color{blue}\aap}, \href{https://ui.adsabs.harvard.edu/abs/2022A&A...661A..63W}{661, A63}

\bibitem[{{Yang} {et~al.}(2023){Yang}, {Liu}, {Muto}, {Hashimoto}, {Dong}, {Kanagawa}, {Momose}, {Akiyama}, {Hasegawa}, {Tsukagoshi}, {Konishi}, \& {Tamura}}]{Yang2023}
{Yang}, Y., {Liu}, H.~B., {Muto}, T., {et~al.} 2023, \href{http://dx.doi.org/10.3847/1538-4357/acc325}{\color{blue}\apj}, \href{https://ui.adsabs.harvard.edu/abs/2023ApJ...948..110Y}{948, 110}

\bibitem[{{Yasui} {et~al.}(2014){Yasui}, {Kobayashi}, {Tokunaga}, \& {Saito}}]{Yasui_ea_2014}
{Yasui}, C., {Kobayashi}, N., {Tokunaga}, A.~T., \& {Saito}, M. 2014, \href{http://dx.doi.org/10.1093/mnras/stu1013}{\color{blue}\mnras}, \href{https://ui.adsabs.harvard.edu/abs/2014MNRAS.442.2543Y}{442, 2543}

\bibitem[{{Zagaria} {et~al.}(2022){Zagaria}, {Clarke}, {Rosotti}, \& {Manara}}]{Zagaria2022}
{Zagaria}, F., {Clarke}, C.~J., {Rosotti}, G.~P., \& {Manara}, C.~F. 2022, \href{http://dx.doi.org/10.1093/mnras/stac621}{\color{blue}\mnras}, \href{https://ui.adsabs.harvard.edu/abs/2022MNRAS.512.3538Z}{512, 3538}

\bibitem[{{Zagaria} {et~al.}(2023){Zagaria}, {Rosotti}, {Alexander}, \& {Clarke}}]{Zagaria2023}
{Zagaria}, F., {Rosotti}, G.~P., {Alexander}, R.~D., \& {Clarke}, C.~J. 2023, \href{http://dx.doi.org/10.1140/epjp/s13360-022-03616-4}{\color{blue}European Physical Journal Plus}, \href{https://ui.adsabs.harvard.edu/abs/2023EPJP..138...25Z}{138, 25}

\bibitem[{{Zallio} {et~al.}(2026){Zallio}, {Vioque}, {Andrews}, {Empey}, {Rosotti}, {Miotello}, {Manara}, {Carpenter}, {Deng}, {Kurtovic}, {Law}, {Longarini}, {Paneque-Carreno}, {Teague}, {Villenave}, {Yen}, \& {Zagaria}}]{Zallio2026}
{Zallio}, L., {Vioque}, M., {Andrews}, S.~M., {et~al.} 2026, \href{https://ui.adsabs.harvard.edu/abs/2026arXiv260303422Z}{\href{http://dx.doi.org/10.48550/arXiv.2603.03422}{\color{blue}arXiv e-prints}, arXiv:2603.03422}

\bibitem[{{Zari} {et~al.}(2018){Zari}, {Hashemi}, {Brown}, {Jardine}, \& {de Zeeuw}}]{Zari2018}
{Zari}, E., {Hashemi}, H., {Brown}, A.~G.~A., {Jardine}, K., \& {de Zeeuw}, P.~T. 2018, \href{http://dx.doi.org/10.1051/0004-6361/201834150}{\color{blue}\aap}, \href{https://ui.adsabs.harvard.edu/abs/2018A&A...620A.172Z}{620, A172}

\bibitem[{{Zhang} \& {Yuan}(2023)}]{2023ApJS..264...14Z}
{Zhang}, R. \& {Yuan}, H. 2023, \href{http://dx.doi.org/10.3847/1538-4365/ac9dfa}{\color{blue}\apjs}, \href{https://ui.adsabs.harvard.edu/abs/2023ApJS..264...14Z}{264, 14}

\bibitem[{{Zhang} {et~al.}(2021){Zhang}, {Snellen}, {Bohn}, {Molli{\`e}re}, {Ginski}, {Hoeijmakers}, {Kenworthy}, {Mamajek}, {Meshkat}, {Reggiani}, \& {Snik}}]{ZhangY2021}
{Zhang}, Y., {Snellen}, I. A.~G., {Bohn}, A.~J., {et~al.} 2021, \href{http://dx.doi.org/10.1038/s41586-021-03616-x}{\color{blue}\nat}, \href{https://ui.adsabs.harvard.edu/abs/2021Natur.595..370Z}{595, 370}

\bibitem[{{Zhang} {et~al.}(2022){Zhang}, {Hou}, {Luo}, {Li}, {Qin}, {Lu}, {Li}, {Chen}, \& {Zhao}}]{2022ApJS..259...38Z}
{Zhang}, Y.-J., {Hou}, W., {Luo}, A.-L., {et~al.} 2022, \href{http://dx.doi.org/10.3847/1538-4365/ac4964}{\color{blue}\apjs}, \href{https://ui.adsabs.harvard.edu/abs/2022ApJS..259...38Z}{259, 38}

\bibitem[{Zhao {et~al.}(2026)Zhao, Lau, Drążkowska, Birnstiel, \& Stammler}]{Zhao2026}
Zhao, H., Lau, T. C.~H., Drążkowska, J., Birnstiel, T., \& Stammler, S.~M. 2026, Late-infall-induced formation of giant planets, multi-generational planetesimals, and disk substructures

\end{thebibliography}

\appendix
\section{Comments on individual sources}\label{app:individual_sources}

While everything indicates a pre-main sequence nature, upon inspection, 25 sources appear in unexpected locations of the HR diagram. Here we therefore revise the stellar parameters for these sources. Nineteen are to the left of the ZAMS (and hence are not compatible with the pre-main sequence phase) and six, at the top-right of the diagram, are so massive and young it is surprising they were detected by the \textit{Gaia} passbands. 

Those 19 to the left of the ZAMS are (for 5 objects we change stellar parameters after checking their original position in the HR diagram):

\begin{itemize}
    \item 2MASS J05065551-0321132. It has a high \textit{Gaia} RUWE, probably indicating an inaccurate \textit{Gaia} parallax. We take the distance and luminosity of \citet{Fairlamb2015} instead.
    
    \item 2MASS J05424769-0947222. It has both uncertain stellar parameters and high \textit{Gaia} RUWE. It is close enough to the pre-main sequence tracks to get a stellar mass and age assigned within uncertainties.
    
    \item 2MASS J20242954+4214019. A binary with high \textit{Gaia} RUWE, probably indicating an inaccurate \textit{Gaia} parallax. It falls outside of the pre-main sequence tracks and no mass or age could be derived for this source.
    
    \item Bad 16. It has a high \textit{Gaia} RUWE, probably indicating an inaccurate \textit{Gaia} parallax. It is close enough to the pre-main sequence tracks to get a stellar mass and age assigned within uncertainties.
    
    \item CD-25 11111A.  It has a high \textit{Gaia} RUWE, probably indicating an inaccurate \textit{Gaia} parallax. We take the distance and luminosity of \citet{Fairlamb2015} instead.
    
    \item  LkHa 25 (or V590 Mon). \citet{Fairlamb2015} assigns it a much larger distance of $\sim1.7$ kpc than the 694 pc \textit{Gaia} distance. Because assuming the \textit{Gaia} distance it appears close enough to the pre-main sequence tracks to get a stellar mass and age assigned within uncertainties, we stick to 694 pc for this source, but with the caveat that it could be much more massive.
    
    \item GSC 05360-01033.  A binary with uncertain stellar parameters. It falls outside of the pre-main sequence tracks and no mass or age could be derived for this source.
    
    \item HBC 324.  It has a high \textit{Gaia} RUWE, probably indicating an inaccurate \textit{Gaia} parallax. It is close enough to the pre-main sequence tracks to get a stellar mass and age assigned within uncertainties.
    
    \item IRAS 09245-5228. A binary with an unclear nature (\citealp{2011A&A...526A..24V}). It is close enough to the pre-main sequence tracks to get a stellar mass and age assigned within uncertainties.
    
    \item IRAS 13445-3624. Very massive O9 star, it has a high \textit{Gaia} RUWE, probably indicating an inaccurate \textit{Gaia} parallax. It falls outside of the pre-main sequence tracks and no mass or age could be derived for this source.
    
    \item IRAS 18528+0400. Very massive O9 star, it has a high \textit{Gaia} RUWE, probably indicating an inaccurate \textit{Gaia} parallax. It falls outside of the pre-main sequence tracks and no mass or age could be derived for this source.
    
    \item KK Oph. It is a binary (\citealp{Pascual2016, Panic2021}). It has a high \textit{Gaia} RUWE, probably indicating an inaccurate \textit{Gaia} parallax. We take the distance and luminosity of \citet{Fairlamb2015} instead.
    
    \item R CrA. It has a very uncertain spectral type in the literature. However, it is close enough to the pre-main sequence tracks to get a stellar mass and age assigned within uncertainties.
    
    \item SS 43. Clearly hot and massive, it is close enough to the pre-main sequence tracks to get a stellar mass and age assigned within uncertainties.
    
    \item PV Cep. Surrounded by a cloud, it has a high \textit{Gaia} RUWE, probably indicating an inaccurate \textit{Gaia} parallax. It falls outside of the pre-main sequence tracks and no mass or age could be derived for this source.
    
    \item UY Ori. It has a high \textit{Gaia} RUWE, probably indicating an inaccurate \textit{Gaia} parallax. We take the distance and luminosity of \citet{Fairlamb2015} instead.
    
    \item V1012 Ori. It has a high \textit{Gaia} RUWE, probably indicating an inaccurate \textit{Gaia} parallax. We take the distance and luminosity of \citet{Fairlamb2015} instead.
    
    \item V892 Tau. A binary, it has a high \textit{Gaia} RUWE, probably indicating an inaccurate \textit{Gaia} parallax. It falls outside of the pre-main sequence tracks and no mass or age could be derived for this source (but it is a known 6 M$_{\odot}$ total equal mass binary, \citealp{2023ApJ...958..123V}).
    
    \item VOS 42. Perhaps inaccurate effective temperature. It is close enough to the pre-main sequence tracks to get a stellar mass and age assigned within uncertainties.
\end{itemize}

Those six massive and young ones are (for one source, MWC~297, we change stellar parameters after evaluating the information available):

\begin{itemize}
    \item * 17 Lep (8.3~\Msun). Described as a symbiotic binary star which perhaps introduces an inaccurate \textit{Gaia} parallax. However, the \textit{Gaia} RUWE is low. No changes were made.
    \item HD 190073 (6.0~\Msun). Effective temperature, distance and luminosity seem correct. No changes were made.
    \item HD 95881 (6.4~\Msun). Effective temperature, distance and luminosity seem correct. No changes were made.
    \item HD 98922 (7.1~\Msun). Perhaps it has an inaccurate distance from the comparison to \citet{Fairlamb2015}, but \textit{Gaia} RUWE is low. No changes were made.
    \item MWC 297 (10.6~\Msun). We find this source has incorrect stellar parameters and luminosities in the considered literature (Sect. \ref{sec:target_selection}). We adopt the values of the detailed analysis of \citet{2022ApJ...941..189V} to place it on a reasonable location of the pre-main sequence tracks.
    \item Z CMa (9.4~\Msun). The distance to this source is very uncertain, and indeed its \textit{Gaia} RUWE is very high, probably indicating an inaccurate \textit{Gaia} parallax. From the HR diagram perspective, it would be more natural for this source to be closer than its current Gaia distance of $\sim640$ pc. However, previous studies have placed it at even larger distances, which would mean it needs to be even more luminous and massive than what is reported here. In the absence of evidence, we use the \textit{Gaia} parallax and resulting luminosity for this source, but we raise caution about the mass and age provided.
\end{itemize}

\renewcommand{\thetable}{B.1}
\begin{table*}
\tiny\centering
\caption{Overview of ALMA and ACA observations.}
\begin{tabularx}{\textwidth}{l|cXXXc}
\hline\hline
\makecell{Project Code \\ \hspace{1mm}} & \makecell{\# Targets \\ \hspace{1mm}} & \makecell{Dates \\ \hspace{1mm}} & \makecell{Int. Time \\ (min)} & \makecell{Calibrators \\ \hspace{1mm}} & \makecell{Array \\ \hspace{1mm}} \\
\hline2021.2.00005.S & 45 & 23/07/22, 31/07/22, 08/08/22, 12/08/22, 22/08/22, 28/08/22, 06/09/22, 09/09/22, 10/09/22, 11/09/22, 28/09/22, 22/04/23, 23/04/23, 26/05/23, 01/06/23, 17/06/23, 25/06/23, 03/07/23, 05/07/23 & median=4.0\newline min=3.2\newline max=12.3 & J0210-5101, J0241-0815, J0319+4130, J0423-0120, J0501-0159, J0516-6207, J0521+2112, J0532+0732, J0538-4405, J0609-1542, J0648-1744, J0811-4929, J0828-3731, J0854+2006, J1037-2934, J1427-4206, J1524-5903, J1830+0619, J1851+0035, J1924-2914, J1925+2106, J2055-1234, J2232+1143, J2258-2758 & ACA \\
\hline2022.1.01155.S & 27 & 06/04/23, 08/04/23, 09/04/23, 15/04/23, 17/04/23 & median=1.6\newline min=1.6\newline max=9.5 & J0423-0120, J0541-0541, J0726-4728, J1713-2658, J1851+0035, J1924+1540, J1924-2914, J1925-3401 & 12m \\
\hline2022.1.01460.S & 5 & 25/04/23, 09/05/23, 24/05/23, 09/06/23, 01/07/23 & median=4.9\newline min=4.9\newline max=9.7 & J0643+0857, J0854+2006, J1147-6753, J1256-0547, J1427-4206, J1554-2704, J1700-2610, J1924-2914 & ACA \\
\hline2023.1.00561.S & 22 & 31/05/24, 02/06/24, 08/06/24, 30/06/24, 07/07/24, 09/07/24, 18/07/24, 01/08/24, 02/08/24, 05/08/24, 07/08/24 & median=6.9\newline min=6.9\newline max=20.0 & J0303-6211, J0336+3218, J0423-0120, J0438+3004, J0440+1437, J0529-0519, J0530+1331, J0538-4405 & 12m \\
\hline2023.1.00937.S & 109 & 08/10/23, 11/10/23, 28/10/23, 29/10/23, 31/10/23, 16/11/23, /11/23, 22/11/23, 28/11/23, 30/11/23, 01/12/23, 02/12/23, 03/12/23, 17/12/23, 18/12/23, 08/03/24, 16/03/24, 17/03/24, 23/04/24, 28/04/24, 01/05/24, 06/05/24, 07/05/24, 08/05/24, 13/05/24, 21/05/24, 25/05/24, 27/05/24, 28/05/24, 31/05/24, 02/06/24 & median=4.5\newline min=2.5\newline max=13.1 & J0238+1636, J0336+3218, J0418+3801, J0423-0120, J0510+1800, J0516-6207, J0530+1331, J0532+0732, J0538-4405, J0607-0834, J0609-1542, J0648-1744, J0700+1709, J0726-4728, J0828-3731, J0854+2006, J0904-5735, J1037-2934, J1058+0133, J1058-8003, J1103-5357, J1256-0547, J1427-4206, J1534-3526, J1625-2527, J1832-2039, J1834-0301, J1851+0035, J1924-2914, J1925+2106, J1925-3401, J1935+2031, J2232+1143, J2258-2758, J2323-0317 & ACA \\
\hline2024.1.00408.S & 29 & 01/10/24, 02/10/24, 04/10/24, 05/10/24, 11/10/24, 12/10/24, 13/10/24, 14/10/24, 15/10/24, 24/10/24, 26/10/24, 06/11/24, 19/11/24 & median=40.4\newline min=22.8\newline max=97.8 & J0423-0120, J0521+2112, J0538-4405, J0550+2326, J0607-0834, J0643+0857, J0648-1744, J0828-3731, J0904-5735, J0922-3959, J1058+0133, J1107-4449, J1145-6954, J1147-6753, J1427-4206, J1517-2422, J1617-5848, J1720-3552, J1733-1304, J1802-3940, J1838+0404, J1924-2914, J1935+2031, J2002+1501, J2015+3710, J2023-0123, J2232+1143 & ACA \\
\hline
\end{tabularx}\\
\label{tab:ALMA_obs}
\end{table*}

\section{New millimeter observations overview and data reduction}
\label{app:new_mm_data}
This section gives an overview of the data taken with the ACA, ALMA, the SMA, and NOEMA. In addition, the steps taken for the imaging of the data are set out, and the resulting images are presented. Lastly, the method for determining the fluxes is explained. For the errors on the flux, we report the rms noise in Table~\ref{tab:herbig_fluxes}. Additionally, for ALMA/ACA\footnote{\url{https://almascience.eso.org/proposing/documents-and-tools/cycle12/alma-proposers-guide}} we use an absolute flux calibration error of 10\%, and for NOEMA\footnote{\url{https://www.iram.fr/IRAMFR/GILDAS/doc/html/pdbi-cookbook-html/node17.html}} and the SMA\footnote{\url{https://lweb.cfa.harvard.edu/sma/mir/sma-calibration-example5.pdf}} we use 20\%.

\begin{figure}[h!]
    \centering
    \includegraphics[width=0.5\textwidth]{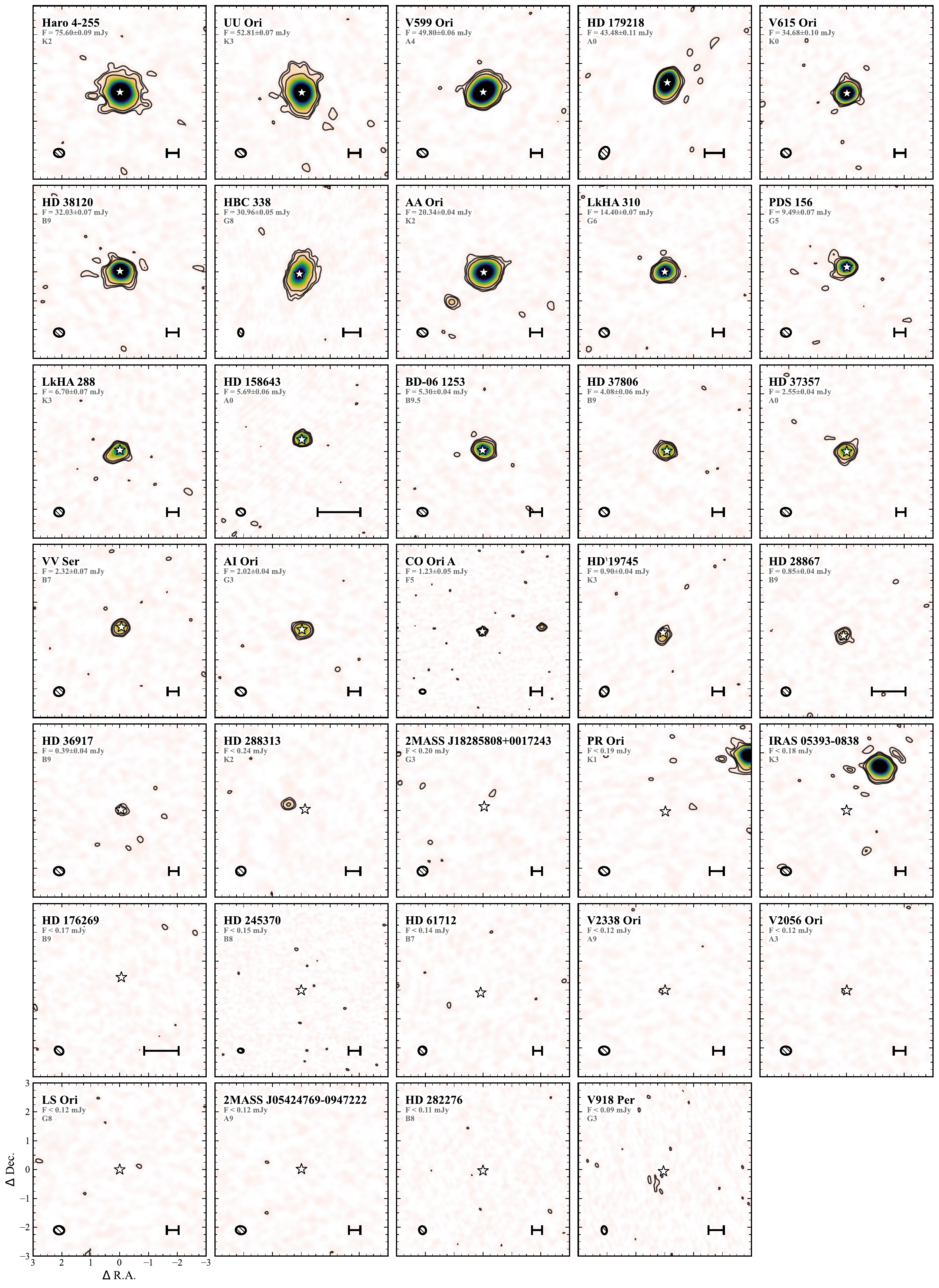}
    \caption{The images of all ALMA targets ordered by integrated flux (from top left to bottom right). The position of the Herbig star is indicated by the star. The scalebar indicates a size of 200~au. The color scale goes from $-1\times$rms to $100\times$rms. The location of the Herbig is highlighted with the star. The contours indicate 3, 5, and 10 sigma levels. Below the name of the target, the integrated flux and the spectral type are shown. The size of the beam is shown in the bottom left.}
    \label{fig:ALMA_gallery}
\end{figure}

\begin{figure*}[h!]
    \centering
    \includegraphics[width=0.9\textwidth]{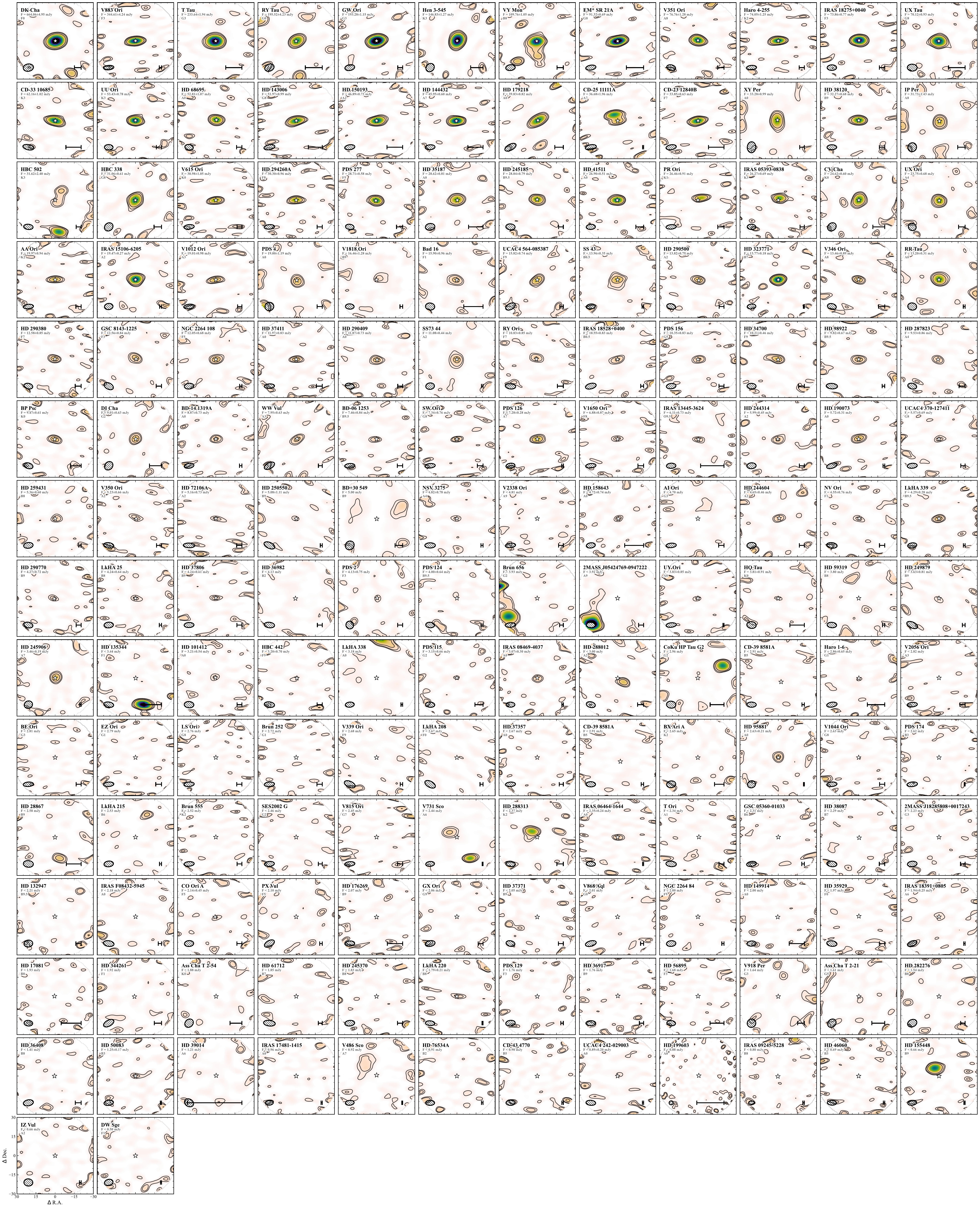}
    \caption{The images of all ACA targets ordered by integrated flux (from top left to bottom right). The position of the Herbig star is indicated by the star. The scalebar indicates a size of 2000~au. The color scale goes from $-1\times$rms to $100\times$rms. The contours indicate 3, 5, and 10 sigma levels. Below the name of the target, the integrated flux and the spectral type are shown. The size of the beam is shown in the bottom left.}
    \label{fig:ACA_gallery}
\end{figure*}

\subsection{ACA and ALMA data reduction}
The new ACA and ALMA data were taken between 23 July 2022 and 19 November 2024. On-source times mostly range from about 1.5 to 10 minutes per source. Specifically for the sources observed in  project 2024.1.00408.S, integration times were much longer ranging from 22.8 to 97.8 minutes with a median of 40.4 minutes. See Table~\ref{tab:ALMA_obs} for observing dates, calibrators, and integration times per project code. Due to the large number of targets, observing conditions ranged from very good (pwv of 0.24~mm) to bad (pwv of 3.3~mm) with typically a precipitable water vapor column of 1.2~mm. All data were tuned at a reference frequency of 225~GHz such that there is a continuum window, and the three most common CO isotopologues ($^{12}$CO, $^{13}$CO, and C$^{18}$O) were covered. The latter will be part of an upcoming work.

\renewcommand{\thetable}{B.2}
\begin{table*}[t]
\caption{Overview of SMA observations.}
\centering
\begin{tabular}{c c c c c c c c} 
\hline
Date & Program & Config & \# Ants & LO & Source$^{a}$ & \multicolumn{2}{c}{Calibrators}  \\
~    & Code    &        & ~       & (GHz) & Group        & Passband & Flux \\ 
\hline
30 May 2024 & 2024A-S016 & COM & 6 & 225.5,225.5 & 1 & 3C279 & Ceres,Pallas \\
09 Jun 2024 & 2024A-H001 & COM & 6 & 225.5,225.5 & 2 & 3C345 & Neptune \\
25 Jul 2024 & 2024A-H001 & SUB & 7 & 225.6,225.6 & 1 & 3C279 & Callisto,Uranus\\
05 Aug 2024 & 2024A-H001 & SUB & 7 & 225.6,225.6 & 4 & 3C84 & Uranus \\
29 Sep 2024 & 2024A-H001 & COM & 5 & 225.5,225.5 & 3 & 0423-013 & Neptune \\
11 Oct 2024 & 2024A-S016 & COM & 6 & 225.0,236.0 & 2 & Uranus & Uranus \\
12 Oct 2024 & 2024A-S016 & COM & 6 & 225.0,236.0 & 3 & 3C84 & Callisto,Uranus\\
17 Oct 2024 & 2024A-S016 & COM & 6 & 225.5,225.5 & 2 & 3C345 & Callisto,Uranus\\
04 Nov 2024 & 2024A-S016 & COM & 5 & 225.5,225.5 & 3 & 3C84 & Callisto,Uranus\\
\hline
\end{tabular}
\tablefoot{$^a$ {\em Group 1}: HBC 694, HD 150193, IRAS 18528+040, IRAS 20144+352, LkH$\alpha$ 224, 
MSX6C G034.0, PX Vul, UCAC4 365-124575, UCAC4 370-127411, UCAC4 564-085387, V522 Cyg, V868 Aql; 
{\em Group 2}: 22-115, BD+46 3471, BD+61 154, DI Cep, HD 235495, IRAS 21183+464, LkH$\alpha$ 324, 
MQ Cas, TYC 4062-230-1, CAC4 677-0999320, UCAC4 685-099740, VX Cas; 
{\em Group 3}: 2MASS J0506, 2MASS J0528, brun 252, brun 555, brun 656, EX Ori, GX Ori, 
HD 245370, UY Ori, V1044 Ori, V1650 Ori, V815 Ori; 
{\em Group 4}: BD+73-1031, GGR 195, HBC 234, LkH$\alpha$ 259, PV Cep, UCAC4 780-045562
}
\label{tab:SMA_obs}
\end{table*}

After obtaining the pipeline calibrated data, the reduction of the ACA and ALMA data were done using the \texttt{Common Astronomy Software Applications} (CASA) software package version 6.7.0.31 \citep{CASA2022}. In a couple of cases multiple ACA or ALMA observations were taken as part of different project codes. First these overlapping ACA data (17 targets) or ALMA data (15 targets) were combined using the \texttt{concat} task after shifting them toward a common phasecenter using the \texttt{phaseshift} task in CASA. The executions for HD~101412 and HD~199603 in project 2021.2.00005.S were excluded due to an incorrect declination specified for both targets. Coincidentally, both were still covered in other projects.

After combining duplicate observations, the ACA and ALMA data were cleaned and imaged. Using the \texttt{tclean} task in CASA, the data were cleaned down to 2$\sigma$ as set by the \texttt{nsigma} parameter. The cleaning masks were automatically generated using the \texttt{auto-multithresh} option. The standard values\footnote{\url{https://casaguides.nrao.edu/index.php/Automasking_Guide_CASA_6.5.4}} for the automasking parameters were used, except that the \texttt{noisethreshold} was lowered to 4.0 (or in some low signal-to-noise cases to 3.0) for the ACA data, as some emission was not picked up for the fainter sources in the sample. Additionally, the \texttt{growiterations} parameter was lowered to only 2 iterations, as the standard value of 75 iterations resulted in excessively large masks, particularly for the low S/N targets. To obtain the cleaned continuum images multi-frequency synthesis was used on the combined continuum windows present in each data set. The \texttt{Hogbom} deconvolve algorithm was used for the ACA data, while \texttt{multiscale} was used for the ALMA observations. In the latter case the scales were set to 0 (point source), 1, 2, and 5 times the size of the beam in pixels. For all data the beam size was set to six pixels along its minor axis. For the imaging Briggs weighting was used with a robust value of 0.5 and the image size was set such that the complete field of view of the telescope was imaged. The resulting continuum images are presented in Figs.~\ref{fig:ACA_gallery} and \ref{fig:ALMA_gallery}.

The integrated fluxes were calculated using two different methods. For the unresolved ACA data, a Gaussian model was fit using the \texttt{imfit} method in CASA. The resulting integrated flux of the fitted Gaussian we report in this work. The corresponding rms noise of the flux was calculating by using the inverted cleaning mask with the \texttt{imstat} task in CASA. For three sources (CD-25~11111A, HBC~502, and V731~Sco) there is additional emission close to the Herbig resulting in an incorrect Gaussian fit. For these sources the fluxes were determined using an aperture that was placed by hand which isolated the contribution of the source as much as possible from the additional emission around it. For V731~Sco specifically, we report an upper limit equal to the integrated flux inside a beam-sized aperture centered on the target, as there is strong contamination from a nearby source and the integrated flux cannot accurately be determined. For the ALMA data, as some are (marginally) resolved, we use a circular aperture curve-of-growth method. The apertures were centered on the peak of the continuum emission. We report the flux at the first peak in the curve-of-growth. If the radius at the position of the peak flux was smaller than 2 times the size of the beam minor axis the flux from the Gaussian fit is reported. No significant differences were found for the two methods. For eight disks resolved emission is detected out of the in total 22 detected targets with ALMA data, see the first eight disks in Fig.~\ref{fig:ALMA_gallery}.

\begin{figure}[b!]
    \centering
    \includegraphics[width=0.5\textwidth]{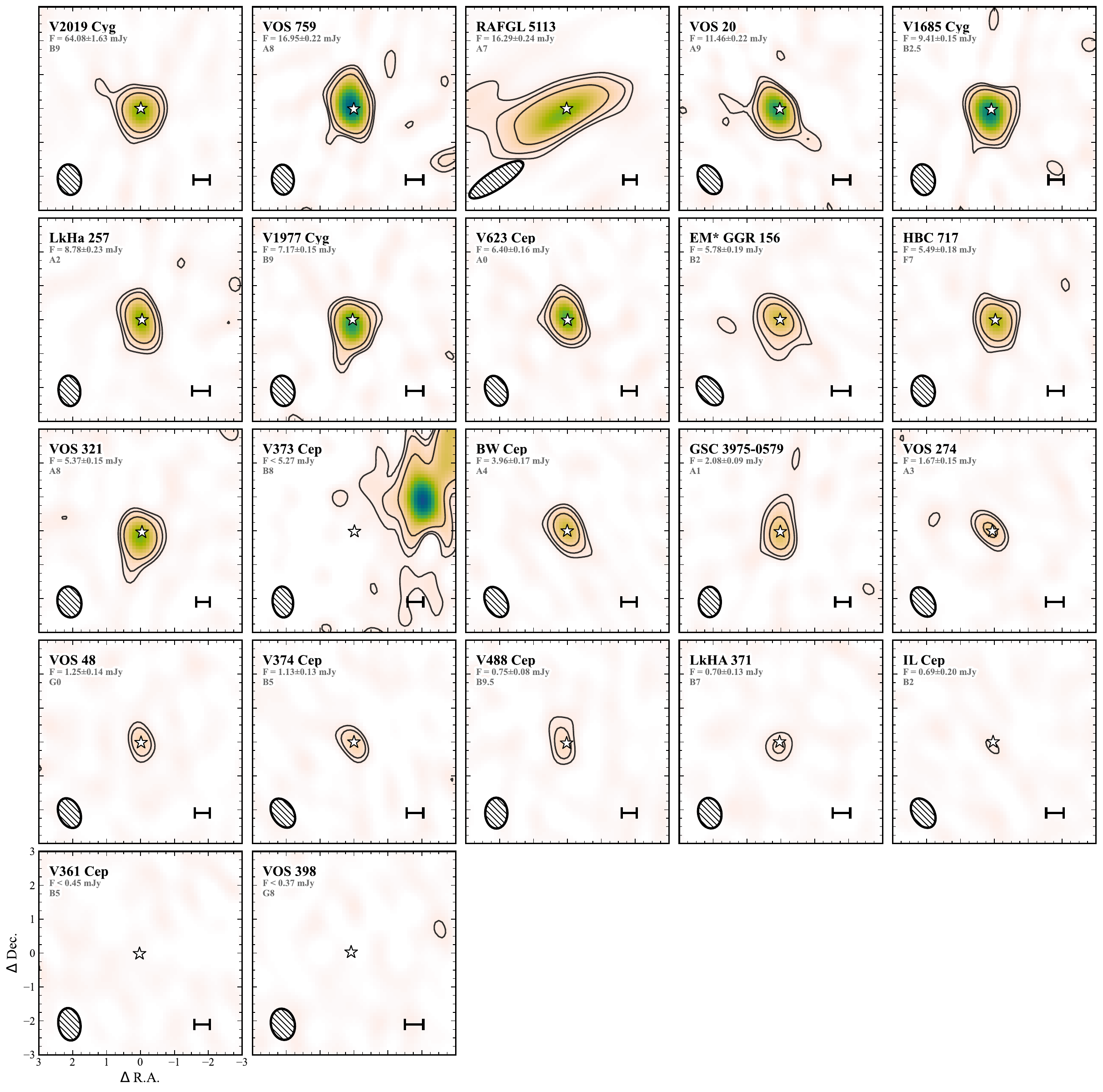}
    \caption{The images of all NOEMA targets ordered by integrated flux (from top left to bottom right). The position of the Herbig star is indicated by the star. The scalebar indicates a size of 500~au. The color scale goes from $-1\times$rms to $100\times$rms. The contours indicate 3, 5, and 10 sigma levels. Below the name of the target, the integrated flux and the spectral type are shown. The size of the beam is shown in the bottom left.}
    \label{fig:NOEMA_gallery}
\end{figure}

There are several cases where there is emission in the ACA data detected away from the center of the image. To ensure that this emission originates from the disk around the Herbig star, the \textit{Gaia} DR3 coordinates with propagated proper motions were compared to the position of these detections. If the peak emission did not fall within the beam centered on the Gaia position, we deem the source of the emission to not originate from the Herbig. In those cases we report a $3\times\text{rms}$ upper limit. In cases with no clear detection (if the peak flux was lower than $3$ or $5\sigma$ for the ACA or ALMA observations respectively) we also report a $3\times\text{rms}$ upper limit. The resulting fluxes are presented in Table~\ref{tab:herbig_fluxes}.

\subsection{SMA data reduction}
We used the SMA on Maunakea, Hawaii over 9 tracks in 2024 May through November to observe 42 Herbig stars at 1.3~mm under project codes 2024A-S016 and 2024A-H001. Table~\ref{tab:SMA_obs} provides a brief log of these observations. With the goal of obtaining photometry for compact sources, these observations were performed with 5 to 7 antennas in either the COM configuration (baselines lengths 8 to 70~m) or SUB configuration (baseline lengths 8 to 30~m) that provided synthesized beam sizes of $4''$ to $6''$ (fwhm). The ``230'' and ``240'' receivers, each with an IF range of $4-16$~GHz in two sidebands, were tuned to LO frequencies near 225~GHz. The SWARM digital backend processed a total continuum bandwidth of 24 GHz in each of two polarizations. 

Observations of the target Herbig stars were divided into 4 groups by sky location, 3 of which included 12 sources, and 1 with 6 sources. In each track, observations of all sources in a group were interleaved with nearby calibrators, with observations of flux and bandpass calibrators obtained before and/or after the source observations. The observing sequence typically consisted of $3\times30$~second integrations on two gain calibrators followed by $10\times30$~second integrations on a source. This strategy resulted in approximately 30 minutes of integration time on each source per night. The observations were made under a wide variety of weather conditions, and observations of the source groups were repeated as needed in an effort to obtain an rms noise near 1~mJy for each source. The SMA primary beam size is $55''$ (fwhm) at 1.3~mm and provides an ample field of view for these compact sources. 

The data were processed using the SMA data reduction pipeline \texttt{COMPASS} (Calibrator Observations for Measuring the Performance of Array Sensitivity and Stability)\footnote{Keating, G. Submillimeter Array Newsletter No. 39, pp. 11-12, January 2025}, which provided automated spectral flagging, production of gain, flux, and bandpass calibration solutions from all of the available calibrators, and fiducial CLEAN continuum images with basic image statistics. Given the limited $(u,v)$ coverage per source, images of some fields were limited in dynamic range to about 20 by the presence of bright sources, resulting in higher rms noise levels. From the basic image statistics the integrated fluxes were taken. These are presented in Table~\ref{tab:herbig_fluxes}.

\renewcommand{\thetable}{B.3}
\begin{table*}[t]
\caption{Overview of NOEMA observations.}
\centering
\begin{tabular}{c c c c c c} 
\hline
Date & \# Ants & Source$^{a}$ & \multicolumn{3}{c}{Calibrators}  \\
~    & ~       & Setup        & Passband & Flux & Ph. \& Amp. \\ 
\hline
14 Dec 2025 & 12 & 2 & 3C454.3 & MWC349 & J2146+608 \\
17 Dec 2025 & 12 & 1 & 3C454.3 & MWC349 & J2005+403 \\
26 Dec 2025 & 12 & 3 & J2200+420 & MWC349 & J2037+511 \\
27 Dec 2025 & 11 & 4 & 2200+420 & MWC349 & J0300+470 \\
\hline
\end{tabular}
\tablefoot{$^a$ {\em Setup 1}: V1685~Cyg, VOS~321, VOS~398, LkH$\alpha$~371, V1977~Cyg, V2019~Cyg, HBC~717; 
{\em Setup 2}: V623~Cep, VOS~48, BW~Cep, VOS~20, IL~Cep, VOS~274, V374~Cep, EM*~GGR~156, V361~Cep, V373~Cep; 
{\em Setup 3}: V488~Cep, GSC~3975-0579, VOS~759, LkH$\alpha$~257; 
{\em Setup 4}: RAFGL~5113
}
\label{tab:NOEMA_obs}
\end{table*}

\renewcommand{\thefigure}{C.1}
\begin{figure*}[t]
 \centering
    \includegraphics[width=\textwidth]{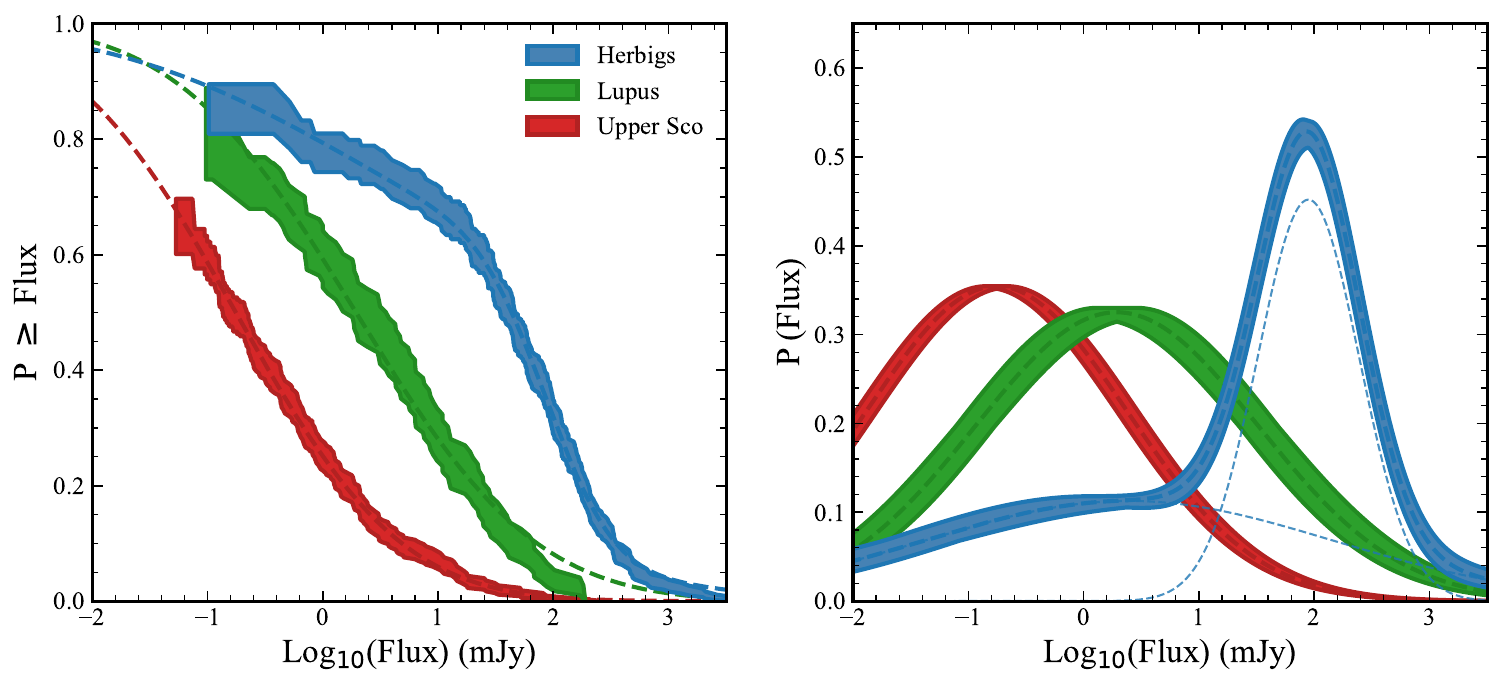}
    \caption{The same as Fig.~\ref{fig:cdf_Mdust}, but using the fluxes instead of disk dust masses. The fluxes have been scaled to a common distance of 160~pc and frequency of 230~GHz using a spectral index of 2.}
    \label{fig:cdf_flux}
\end{figure*}

\subsection{NOEMA data reduction}
The NOEMA data were taken over four tracks on the 14th (setup 2), 17th (setup 1), 26th (setup 3), and 27th (setup 4) of December 2025 for a total of 22 Herbig stars. The data were taken with the C configuration, with baselines ranging from 20 to 370 meters. The resulting synthesized beam has a size of 1$''$ $\times$ 0.6$''$ (except for RAFGL~5113, see Fig.~\ref{fig:NOEMA_gallery}). The observations were performed with all twelve antennas for setups~1-3 and eleven antennas for setup~4. The observations started with the bandpass and flux calibrators, after which the targets were observed with the phase and amplitude calibrator in between. This resulted in a typical total on-source time of 12 to 20 minutes. An overview of the observations can be found in Table~\ref{tab:NOEMA_obs}. The observations made use of the same spectral setup as \citet{Stapper2025a} used. The wide-band correlator PolyFiX was tuned such that the lower side-band (LSB) and upper side-band (USB) ranged from 203.8 to 211.7 GHz and 219.3 to 227.2 GHz, respectively, with a channel width of 2 MHz ($\sim$2.7 km~s$^{-1}$). These were used to obtain the continuum fluxes. In addition, 13 high resolution spectral windows with a width of 62.5 kHz ($\sim$86 m~s$^{-1}$) were centered on several molecular emission lines, in particular the $J=2-1$ transition of \ce{^13CO}, \ce{C^18O}, and \ce{C^17O}. Apart from \ce{^13CO} in V2019~Sco, no other lines have been detected. 

The standard data calibration pipeline was done with the Continuum and Line Interferometer Calibration (\texttt{CLIC}) program of the Grenoble Image and Line Data Analysis Software (\texttt{GILDAS}\footnote{\url{https://www.iram.fr/IRAMFR/GILDAS}}). Due to problems with the receiver for basebands 7 and 8 on antenna~3 during the observations on the 26th and 27th of December (setups 3 and 4), additional flagging was done during the calibration process. For all observations the polarizations were averaged during the amplitude calibration, for the phase calibration this was only done for setups 3 and 4. After calibrating the data, the data were exported to \texttt{uvfits} files.

The data were imaged using the task \texttt{tclean} in CASA. The data were cleaned down to $2\sigma$ using the \texttt{nsigma} parameter. The \texttt{Hogbom} deconvolve algorithm was used with a robust value of 0.5. The resulting continuum images are presented in Fig.~\ref{fig:NOEMA_gallery}. Using the \texttt{imfit} task, a Gaussian model was fit to the image. The resulting integrated flux is presented in Table~\ref{tab:herbig_fluxes}. The noise was determined by an annulus around the source. In cases of a non-detection we report a $3\sigma$ upper limit.

\renewcommand{\thefigure}{D.1}
\begin{figure*}[b]
    \centering
    \includegraphics[width=\textwidth]{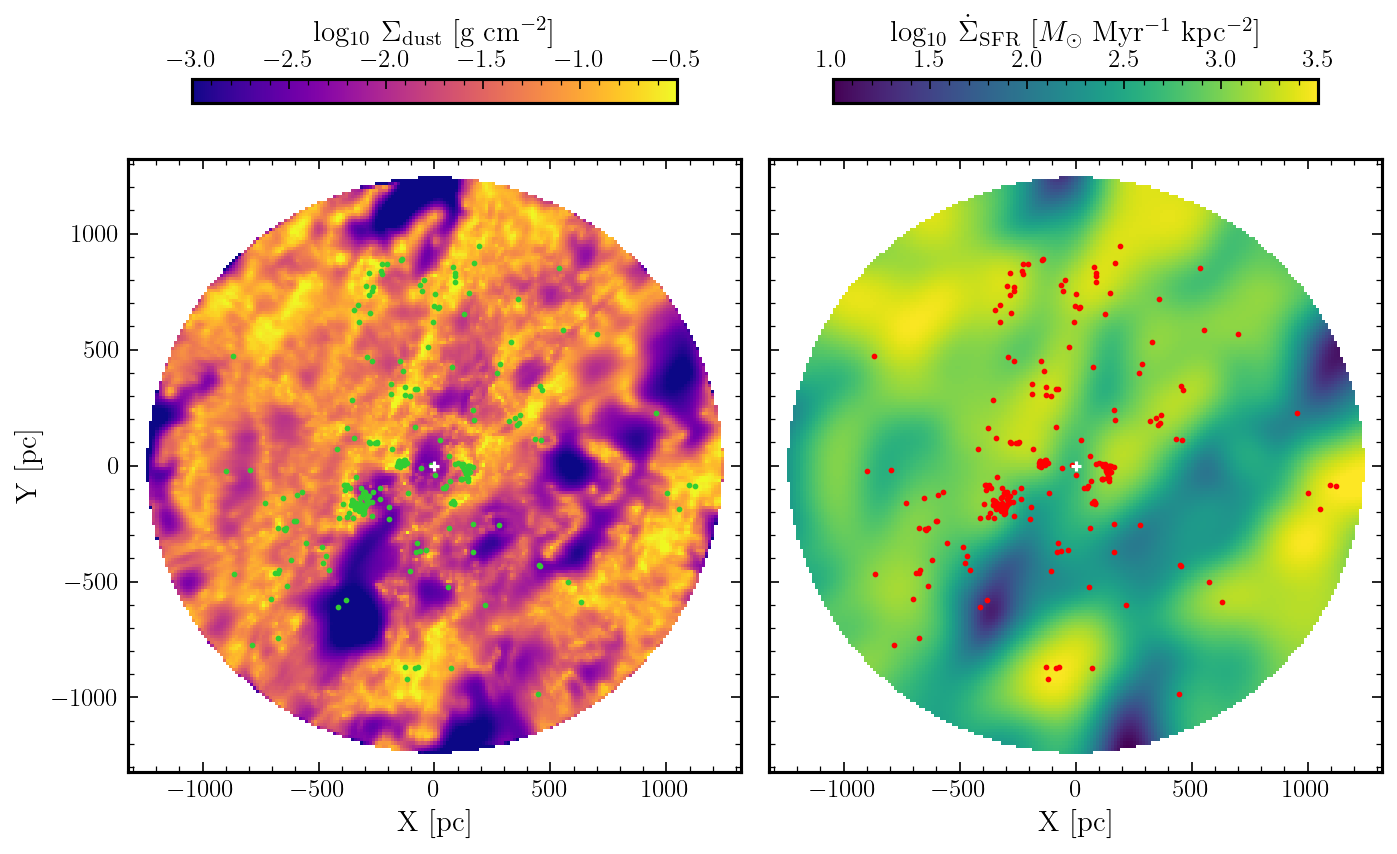}
    \caption{The obtained dust surface density and the estimated star-formation rate plotted with the Sun at the center of the images. The Herbig stars are indicated by the markers.}
    \label{fig:sfr}
\end{figure*}

\section{Flux distributions}
\label{app:flux_distributions}
In Fig.~\ref{fig:cdf_flux} we present the cumulative distributions similar to those in Fig.~\ref{fig:cdf_Mdust}, but now using the fluxes instead of the dust masses. The fluxes were scaled to a common distance of 160~pc and frequency of 230~GHz using a spectral index of 2. The low ``disk mass'' tail is still present for the distribution of the Herbig disks, again making a bimodal distribution the best fit to the data. We find that the Herbig disk flux distribution consists of two populations, one with a mean flux of $89.02^{+5.77}_{-5.77}$~mJy and one with a mean flux of $2.46^{+2.51}_{-1.36}$~mJy. The respective standard deviations are $0.44^{+0.01}_{-0.01}$ and $1.77^{+0.08}_{-0.11}$ in units of log$_{10}$(Flux/mJy). For Lupus and Upper~Scorpius we find mean fluxes of $1.94^{+0.93}_{-0.62}$~mJy and $0.18^{+0.04}_{-0.03}$~mJy, and standard deviations of $1.23^{+0.02}_{-0.03}$ and $1.13^{+0.01}_{-0.01}$.

\section{From 3D dust extinction to a star-formation rate map}
\label{app:sfr_from_dust}
To be able to estimate the number of Herbig stars within 1~kpc, we first need to determine the star-formation rate. To do that, we start by constructing a Sun–centered map of the star-formation rate surface density
$\dot{\Sigma}_{\rm SFR}(x,y)$ in the Galactic plane starting from the
three-dimensional dust map of \citet{Edenhofer_ea_2024}, using the
the \texttt{dustmaps} package. We translate that extinction map into a star formation rate as follows.

\subsection{Geometry and extinction slab}

We work in a Sun–centred Cartesian frame $(x,y,z)$, with $x$ and $y$ in the
Galactic plane and $z$ perpendicular to it.  At each grid point $(x,y,z)$ we
convert to Galactic coordinates $(l,b,d)$ via
\begin{align}
  d &= \sqrt{x^2 + y^2 + z^2}, \\
  l &= \arctan2(y,x), \\
  b &= \arcsin\!\left(\frac{z}{d}\right).
\end{align}
We compute the extinction per unit path length $A_V$:
\begin{equation}
  \frac{{\rm d}A_V}{{\rm d}s}(x,y,z)
  \equiv A_V'(x,y,z).
\end{equation}
For a given $(x,y)$, we integrate $A_V'(x,y,z)$ along the vertical direction
within a symmetric slab $|z|\leq z_{\max}$:
\begin{equation}
  A_V^{\rm slab}(x,y)
  = \int_{-z_{\max}}^{+z_{\max}} A_V'(x,y,z) \, {\rm d}z.
\end{equation}
This yields the total $V$-band extinction through the local dust layer
above and below the mid-plane at that $(x,y)$.

\subsection{Gas surface density and log-space smoothing}

We convert the extinction slab to a gas surface density using a constant
conversion factor,
\begin{equation}
  \Sigma_{\rm gas}^{(0)}(x,y)
  = \alpha_{A_V} \, A_V^{\rm slab}(x,y),
\end{equation}
with fiducial calibration
$\alpha_{A_V} \approx 1.8\times 10^{21} \mu m_\mathrm{H} \mathrm{mag}^{-1} \approx 21.3~{\rm M}_\odot~{\rm pc}^{-2}~{\rm mag}^{-1}$, where $\mu$ is the mean molecular weight.

Below a certain length scale, young star locations are decorrelated with gas due to feedback and other effects \citep{Semenov_ea_2021}. To estimate the star formation rate, we therefore smooth $\Sigma_{\rm gas}^{(0)}$ in the plane using a Gaussian kernel in surface density. We apply an isotropic Gaussian filter of full width at half maximum
$R_{\rm FWHM}=200$~pc to the mean-subtracted surface density, adopting a Gaussian kernel with standard deviation
$\sigma = R_{\rm FWHM}/2.355$ in the plane. Our choice of FWHM is somewhat arbitrary, but approximately corresponds to the galactic gas scale height and the largest scales of individual molecular clouds. 

\subsection{Kennicutt–Schmidt law and birth surface density}

We convert the smoothed gas map to an SFR surface density via a
Kennicutt--Schmidt--type relation:
\begin{equation}
  \dot{\Sigma}_{\rm SFR}^{\rm (raw)}(x,y)
  = A_{\rm KS} \, \Sigma_{\rm gas}(x,y)^{N_{\rm KS}},
  \label{eq:ks_law}
\end{equation}
where we adopt fiducial values
$A_{\rm KS} = 2.5\times10^{-4}~{\rm M}_\odot~{\rm yr}^{-1}~{\rm kpc}^{-2}$
(for $\Sigma_{\rm gas}$ in ${\rm M}_\odot~{\rm pc}^{-2}$) and
$N_{\rm KS} = 1.4$ \citep{Kennicutt_ea_1998}.
This yields a ``raw'' SFR map $\dot{\Sigma}_{\rm SFR}^{\rm (raw)}(x,y)$
defined on a uniform Cartesian grid in $x$ and $y$.

The overall normalisation of equation~(\ref{eq:ks_law}) is uncertain,
because both the extinction--to--gas conversion and the small--scale
star--formation law are approximate.  We therefore {renormalise}
the entire map so that the total SFR within a cylindrical aperture of
radius $R_{\rm lim} = 1~{\rm kpc}$ matches the value inferred by
\citet{Quintana_ea_2025},
$\mathrm{SFR}_{\rm lim}(<R_{\rm lim}) = 2896~{\rm M}_\odot~{\rm Myr}^{-1}$.
On the Cartesian grid we identify all finite, positive SFR pixels inside
$R \le R_{\rm lim}$ and compute their mean raw SFR surface density
$\langle \dot{\Sigma}_{\rm SFR}^{\rm (raw)} \rangle_{R\le R_{\rm lim}}$.
The required mean surface density to reproduce
$\mathrm{SFR}_{\rm lim}$ is
\begin{equation}
  \langle \dot{\Sigma}_{\rm SFR} \rangle_{\rm req}
  =
  \frac{\mathrm{SFR}_{\rm lim}}
       {\pi (R_{\rm lim}/{\rm kpc})^2 },
\end{equation}
We then define a single multiplicative scale factor
\begin{equation}
  s_{\rm Q} =
  \frac{\langle \dot{\Sigma}_{\rm SFR} \rangle_{\rm req}}
       {\langle \dot{\Sigma}_{\rm SFR}^{\rm (raw)} \rangle_{R\le R_{\rm lim}}},
\end{equation}
and rescale the entire map,
\begin{equation}
  \dot{\Sigma}_{\rm SFR}(x,y)
  = s_{\rm Q}\,\dot{\Sigma}_{\rm SFR}^{\rm (raw)}(x,y).
\end{equation}The resultant surface density and star formation rate maps are shown in Figure~\ref{fig:sfr}.

\subsection{Star formation rate to number of stars}
To convert the SFR surface density to a birth surface density of stellar
systems per unit area and per unit time we divide by the mean stellar
mass $\bar{M}$ of a \citet{Chabrier_ea_2003} IMF $\xi(M_*)$, normalised over
$[M_{\min},M_{\max}] = [0.08,120]~{\rm M}_\odot$:
\begin{equation}
  \dot{N}_{\rm birth}(x,y)
  =
    \frac{\dot{\Sigma}_{\rm SFR}(x,y)}
         {\bar{M}}.
\end{equation}
In the radial completeness model of
Section~\ref{sec:completeness} we allow for a small residual
global scaling through the factor $f_\mathrm{SFR}$, for which we assume a prior corresponding to the fractional uncertainties quoted by \citet{Quintana_ea_2025}. We truncate this prior below $1$, since the \citet{Quintana_ea_2025} SFR is effectively a lower limit.

In practice we tabulate the IMF on a fine logarithmic mass grid, $M_k\in[0.08,120]~\mathrm{M_\odot}$, and compute: \begin{align} \xi(M_k) &\propto \text{Chabrier2003}(M_k), \\ f_j &= \frac{\displaystyle \sum_{k:\,M_{j,\min}\le M_k<M_{j,\max}} \xi(M_k)\,\Delta M_k} {\displaystyle \sum_{k}\xi(M_k)\,\Delta M_k}, \end{align} where $f_j$ is the fraction of all stars formed in mass bin $j$. The IMF--averaged stellar mass is \begin{equation} \bar{M} = \frac{\displaystyle \sum_k M_k\,\xi(M_k)\,\Delta M_k} {\displaystyle \sum_k \xi(M_k)\,\Delta M_k}. \end{equation} This $\bar{M}$ is used to convert between SFR (in $\mathrm{M_\odot\,yr^{-1}}$) and total formation rate of stellar systems (in $\mathrm{yr^{-1}}$).

\subsection{Completeness model fit}
\label{app:corner_plot}
Here we present the fit of our completeness model, see Fig.~\ref{fig:corner}. By inverting eq.~(\ref{eq:logit_fdet}), we estimate the completeness of the catalog as follows:
\begin{equation}
    f_\mathrm{det} = \frac{e^x}{1+e^x} 
\end{equation}where

\begin{equation}
\begin{aligned}
    x &= a_0
          + a_R \ln\!\left(\frac{R}{R_0}\right)
          + a_M \ln\!\left(\frac{M}{M_0}\right) \\
      &\approx -2.6 - 3.7 \ln \left( \frac{R}{500 \,\rm{pc}}\right) + 1.9 \ln\!\left(\frac{M}{1 \,M_\odot}\right).
      \label{eq:logit_fdet_filled_in}
\end{aligned}
\end{equation}

\renewcommand{\thefigure}{D.2}
\begin{figure}[h!]
    \centering
    \includegraphics[width=\linewidth]{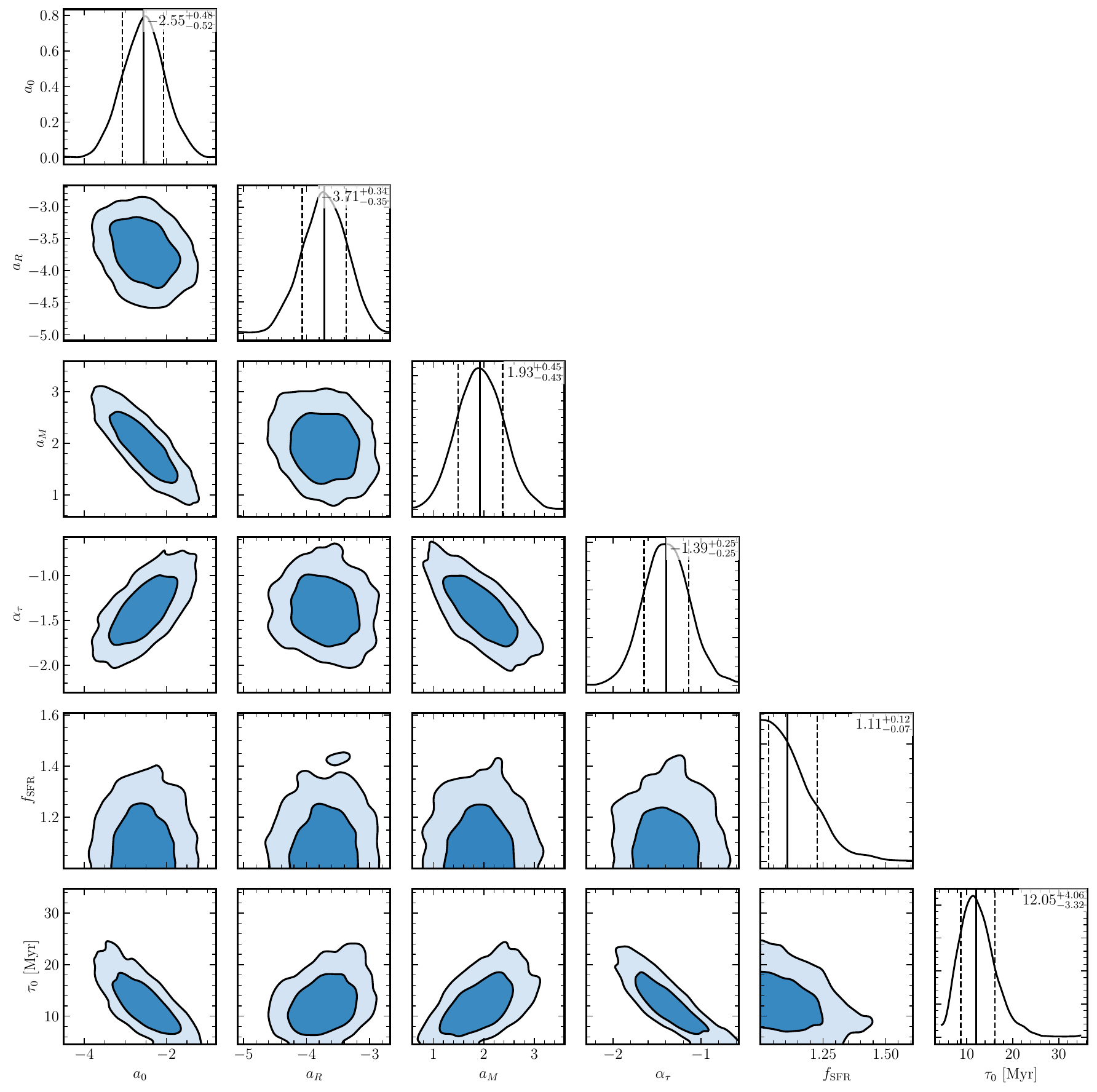}
    \caption{Corner plot of the posterior distributions of our fitted completeness model.}
    \label{fig:corner}
\end{figure}

\end{document}